\documentclass{aa}  

\usepackage{natbib,appendix,epsfig,graphicx,color}
\usepackage[varg]{txfonts}
\definecolor{Blue}{rgb}{0.06,0.20,0.93}
\bibpunct{(}{)}{;}{a}{}{,}
\def\Ha {H$_\alpha$\,}

\def\Hgama {H$_\gamma$\,}
\def\Hdelta {H$_\delta$\,}

\def\Hdelta{H$_\delta$\,}

\def\heiid {He II $\lambda$4686\,}

\def\niiia{N~III $\lambda\lambda$4634, 4640, 4642\,}

\def\niva{N~IV $\lambda$4058\,}

\def\nv{N~V $\lambda\lambda$4604, 4620\,}

\def\Mdot {$\dot M$\,}
\def\MV {M$_{\rm V}$\,}
\def\Yhe {$Y_{\rm He}$} 
\def\kms {km~s$^{\rm -1}$\,} 
\def\Rstar {$R_\star$\,}

\def\Mspec {$M_{\rm spec}$\,}
\def\Mevol {$M_{\rm evol}$\,}
\def\Minit {$M_{\rm init}$\,}
\def\Teff {$T_{\rm eff}$\,}
\def\vinit {$v_{init}$\,}
\def\vinf {$v_\infty$\,}

\def\vsini {$v \sin i$\,}
\def\vcrit {$v_{crit}$\,} 
\def\logg {$\log g$\,}
\def\loggc {$\log g_{\rm c}$\,}
\def\Vr {$V_{\rm r}$\,}

\def\vrot{$V_{\rm rot}$\,}

\def\logl {$\log L/L_\odot$\,}
\def\vmac {$v_{\rm mac}$\,}
\def\vmic {$v_{\rm mic}$\,}

\def\Msun {$M_\odot$\,}
\def\Rsun {$R_\odot$\,}
\def\Lsun {$L_\odot$\,}

\def \Rstare {R_\star}

\def \Teffe {T_{\rm eff}}

\def \beq{\begin{equation}}
\def \eeq{\end{equation}}
\def \ben{\begin{enumerate}} 
\def \een{\end{enumerate}} 
\def \beqa{\begin{eqnarray}}
\def \eeqa{\end{eqnarray}}
\begin{document}

   \title{Spectroscopic and physical parameters of Galactic O-type stars}

   \subtitle{III. Mass discrepancy and rotational mixing}

\author{N. Markova\inst{1}, J. Puls\inst{2} and N. Langer\inst{3}}

\offprints{N. Markova,\\ \email{nmarkova@astro.bas.bg}}

\institute{Institute of Astronomy, 
        National Astronomical Observatory,
        Bulgarian Academy of Sciences, 
        P.O. Box 136, 4700 Smolyan, Bulgaria\\ 
        \email{nmarkova@astro.bas.bg}
\and LMU M\"{u}nchen, Universit\"{a}ts-Sternwarte,
        Scheinerstrasse 1, D-81679 M\"unchen, Germany\\
        \email{uh101aw@usm.uni-muenchen.de}
\and
Argelander-Institut f\"r Astronomie, Bonn University, Auf dem H\"ugel 71,
53121 Bonn, Germany
}
\date{Received; Accepted }

\abstract
{Massive stars play a key role in the evolution of the Universe.} 
{Our goal is to compare observed and predicted properties of single Galactic O stars 
to identify and constrain uncertain physical parameters and processes in stellar evolution 
and atmosphere models. }
{We used a sample of 53 objects with spectral types from O3 to O9.7. 
For 30 of these, we determined the main photospheric and wind parameters, using optical 
spectroscopy and applying the FASTWIND code. For the remaining objects, literature data, 
obtained by means of the CMFGEN code, were used instead. The properties of our sample were 
compared to published predictions based on two grids of evolution models that include 
rotationally induced mixing.}
{Within each luminosity class, we find a close correlation of N surface abundance 
and luminosity, and a stronger N enrichment in more massive and evolved O stars. 
Additionally, a correlation of the surface nitrogen and helium abundances is observed. 
The large number of nitrogen-enriched stars above $\sim$30~\Msun argues for rotationally 
induced mixing as the most likely explanation. However, none of the considered models 
can match the observed trends correctly, especially in the high mass regime.} 
{We confirm mass discrepancy for objects in the low mass O-star regime. We conclude that 
the rotationally induced mixing of helium to the stellar surface is too strong in some of 
the models. We also suggest that present inadequacies of the models to represent the N 
enrichment in more massive stars with relatively slow rotation might be related to 
problematic efficiencies of rotational mixing. We are left with a picture in which invoking 
binarity and magnetic fields is required to achieve a more complete agreement of the observed 
surface properties of a population of massive main- sequence stars with corresponding 
evolutionary models.}

\keywords{stars: massive -- stars: early type -- stars: fundamental parameters 
--  stars: mass loss -- stars: evolution }

\titlerunning{Galactic O stars}
\authorrunning{N. Markova et al.}

\maketitle

%________________________________________________________________
 %

\section{Introduction}\label{intro}

Over the past two decades, a growing body of theoretical and observational 
evidence has been assembled indicating that rotation is as important a 
factor for massive star evolution as mass loss. Consequently, several  
grids of evolutionary models for massive single stars, accounting for 
rotation in addition to mass loss, have been computed and made available to 
the international community (see e.g. \citealt{ekstroem}, \citealt{brott11} 
and \citealt{chieffi13}). While the models make detailed predictions of 
the surface properties of massive stars as a function of mass, initial chemical 
composition, and initial rotational rate, it is not in advance clear if and to 
what extent the physical processes included in the evolutionary calculations  
are  comprehensive  and  adequately describe the real nature of the stars. 

To address this important issue, several large surveys of massive OB stars in 
the Galaxy and the Magellanic Clouds  (e.g. the MiMeS survey, \citep{wade16}; 
the VLT FLAMES survey of massive stars, \citep{evans06}; and the VLT FLAMES-
Tarantula survey \citep{evans11}) have been initiated with the primary goal 
to collect sufficiently accurate and complete datasets of physical parameters 
to be used for testing and verifying the assumptions  made in the models.  
The outcome of these (but also other smaller) surveys gave rise to new 
challenging questions regarding various aspects of the physics included
in the models, in particular  mass loss, gravity darkening, critical velocities, 
instabilities, and rotational mixing (for a detailed review on this issue, 
see \citet{mm14}).

Since mixing affects all of the outputs of the models, this issue has been 
most extensively debated. Observations have clearly revealed the existence 
of N-rich slow rotators and N-poor fast rotators (see e.g. \citealt{hunter08, 
hunter09}; \citealt {morel08}; \citealt{gonzalez12a, gonzalez12b}; \citealt{nieva14}), 
which is a pattern that is not expected from the current evolutionary calculations 
for single stars. Nevertheless, theoreticians have argued that  ``since the
N-enrichment resulting from internal mixing is a multivariate function of rotation, 
mass, age, binarity, metallicity, and magnetic fields'' \citep{maeder14}, it is 
rather premature to question the concept of rotational mixing based on evidence 
accounting for the effect of rotation alone. Additionally, these theoreticians 
point out that an overall agreement between model predictions and observations 
can be achieved when all important parameters have been considered. 

Another long-standing problem refers to the systematic overestimate of evolutionary 
masses compared to spectroscopically derived masses, known as the mass discrepancy 
(see \citealt{herrero92}). While continuous improvements in model atmospheres and 
model evolutionary calculations have reduced the size of the discrepancy (e.g. 
\citealt{repo}), however without eliminating it completely \citep{mokiem07, hohle10, 
MP15}, there are studies  which argue that, at least for O stars in the Milky Way 
(MW), the mass problem has been solved (e.g. \citealt{WV10}, \citealt{massey13},). 

The situation regarding the lack of consistency between observed and predicted 
properties of massive hot stars became even more complicated when \citet{martins14}  
and \citet{keszthelyi16} have demonstrated that as the evolutionary calculations 
rely on various prescriptions to describe the physical processes driving the evolution, 
and as these prescriptions vary from code to code,  the outcome of a comparison 
between model predictions and observations can be significantly different when 
employing different model grids.

In this study we investigate the correspondence between evolutionary model 
predictions and observations for a sample of 53 O stars in the MW, trying 
to put constraints on various parameters or processes that might influence 
the outcome of the comparison. For 30 of these, own determinations of the 
photospheric and wind parameters were obtained using original spectral 
observations and applying the non-LTE line blanketed code FASTWIND (Sect.
~\ref{mod_analysis}); for the rest, literature data have been used instead. 
The complete database is analysed in terms of important physical relations
(Sect.~\ref{mod_res}) and confronted with model predictions from \citet{ekstroem} 
and \citet{brott11} with particular emphasis on stellar mass, nitrogen enrichment 
and rotational rate (Sects.~\ref{evol_mass}, ~\ref{mass_discrepancy}, and 
~\ref{rot_mix}, respectively). A general discussion of the main results is 
presented in Sect.~\ref{discussion} while Sect.~\ref{summary} summarises the
main conclusions.

\section{Stellar sample and main observational properties}
\label{obs}

The original sample, underlying this study, is described and used in 
\citet{markova14} (from now on Paper II) except for $\zeta$ Pup, which 
was discarded because of the low quality of the available spectrum. The 
sample comprises 30 O-type stars in the MW covering all luminosity classes 
(LC) and  spectral types (ST) from O3 to O9.7. Twenty-three of these are 
members of cluster and associations and the rest are objects from the field. 
The spectroscopic observations were collected with the FEROS spectrograph 
\citep{kaufer99} mounted on the ESO/MPG 2.2\,m telescope at La Silla. Each 
spectrum covers a wavelength range from about 350 to about 920\,nm and has 
a  resolving power of $R$=48\,000.
\begin{table*}
\begin{center}
\caption[]{Basic parameters of our targets (sorted by membership and spectral 
type). Cluster and association members are listed in the upper part, field stars 
in the lower. Uncertain membership is denoted by ':'. For objects with more than 
one entry see text.}
\label{sample}
\tabcolsep2.5mm
\begin{tabular}{lllllccll}
\hline 
\hline
\multicolumn{1}{l}{Object ID}
&\multicolumn{1}{c}{ST}
&\multicolumn{1}{l}{Membership}
&\multicolumn{1}{l}{d}
&\multicolumn{1}{l}{$R$}
&\multicolumn{1}{c}{$V$}
&\multicolumn{1}{r}{$B-V$}
&\multicolumn{1}{l}{Sp. status}
&\multicolumn{1}{c}{\MV}\\
%&\multicolumn{1}{l}{Comments}\\
\multicolumn{1}{l}{}
&\multicolumn{1}{c}{}
&\multicolumn{1}{l}{}
&\multicolumn{1}{l}{(kpc)}
&\multicolumn{1}{l}{}
&\multicolumn{1}{c}{(mag)}
&\multicolumn{1}{c}{(mag)}
&\multicolumn{1}{l}{}
&\multicolumn{1}{c}{(mag)}
\\
\hline    
HD~64568  &O3 V((f*))z&Pup OB1/NGC 2467 &2.51 H78  &3.10   &9.40  &0.074   &SB1? &$-$3.79\\  
HD~64568a &           &                 &5.52 KH00 &3.10   &      &        &     &{\bf $-$5.50}\\ 
HD~63005 &O6.5~IV((f))&Pup OB1          &2.51 H78  &3.10   &9.13  &-0.028  & S   &$-$3.74\\
HD~63005a &           &                 &5.69 KH00 &3.10   &      &        &     &{\bf $-$5.45}\\
HD~46223  &O4~V((f))  &Mon OB2/NGC 2244 &1.51 H78  &3.10   &7.27  &0.218   & S  &{\bf $-$5.26}\\
HD~46202  &O9.2~V     &Mon OB2/NGC 2244 &1.51 H78  &3.10   &8.18  &0.177   & S  &{\bf $-$4.22}\\
HD~93843  &O5~III(fc) &Car OB1          &2.6 H78   &3.10   &7.32  &-0.030  &SB1?&$-$5.63\\
HD~93843a &            &                &3.68 KS10 &3.10   &      &        &    &{\bf $-$6.38}\\
HD~91572  &O6.5~V((f))z&Car OB1         &2.6 H78   &3.10   &8.22  &0.036   &SB1 &$-$4.93\\
HD~91572a &            &                &2.77 KS10 &3.10   &      &        &    &{\bf $-$5.06}\\  
HD~91824  &O7 V((f)z   &Car OB1         &2.6 H78   &3.10   &8.17  &-0.055  &SB1 &$-$4.93\\
HD~91824a &            &                &3.1 KS10  &3.10   &      &        &    &{\bf $-$5.08}\\   
HD~94370  &O7(n)fp     &Car OB1 G87     &2.6 H78   &3.10   &7.94  &0.077   &SB2?&$-$5.25\\ 
HD~94370a &            &                &2.6 KS10  &3.10   &      &        &    &{\bf$-$5.25}\\ 
HD~94370b &            &field, GOSV3    &          &       &      &        &    &$-$5.60\\ 
HD~92504  &O8.5~V(n)   &Car OB1:        &2.6 H78   &3.10   &8.42  &-0.053  & S  &{\bf $-$4.45}\\ 
HD~94963  &O7 II(f)    &Car OB2 G87     &2.0 H78   &3.10   &7.16  &-0.087  &SB2?& $-$4.94\\
HD~94963a &            &                &2.6 KS10  &3.10   &      &        &    &{\bf $-$5.53} \\  
HD~94963b &            &field, GOSV3    &          &       &      &        &    & $-$5.90\\  
HD~93204  &O5.5~V((fc))&Tr 16           &2.90 HSB12&4.40 HSB12&8.44&0.095  & S  &$-$5.65\\
HD~93204a &            &                &3.50 P01  &3.70 P01  &    &       &    &{\bf $-$5.78}\\                               
CPD$-$59\,2600&O6~V((f))&Tr 16          &2.90 HSB12&4.40 HSB12&8.83&0.210  &SB1 &$-$5.77\\
CPD$-$59\,2600a&       &                &2.20 P01  &4.17 P01  &    &       &    &{\bf $-$5.05}\\
CPD$-$58\,2620&O7~Vz   &Tr 14           &2.90 HSB12&4.40      &9.27&0.180  & S  & $-$5.20\\
CPD$-$58\,2620a&       &                &2.20 P01  &3.50 T03  &    &       &    &{\bf $ -$4.16}\\
HD~93222  &O7~V((f))z  &Cr 228          &2.60 F95  &3.20      &8.10&0.045  & S  &{\bf $-$5.11}\\ 
CD$-$47\,4551&O5~Ifc   &Vela OB1:       &1.75 R00  &3.70 R00  &8.39 &0.890 &WCB &{\bf $-$7.15}\\ 
HD~75211   &O8.5~II((f))&Vela OB1       &1.75 R00  &3.70 R00  &7.51  &0.397   &SB1 &{\bf $-$6.12}\\
HD~76968   &O9.2 Ib    &Vela OB1: R00   &1.75 R00  &3.70 R00  &7.08  &0.133   &SB1 &{\bf$-$5.66}\\
HD~76968a  &           &field:, GOSV3   &          &          &      &        &    &$-$6.20 \\ 
CD\,$-$44\,4865&O9.7~III&Vela OB1       &1.75 R00  &3.70 R00  &9.43  &0.691   & S  &{\bf $-$5.49} \\
HD~78344   &O9.7 Iab   &Vela OB1: R00   &1.75 R00  &3.70      &9.09  &0.890   & S  &{\bf $-$6.45}\\
HD~75222   &O9.7 Iab   &Vela OB1 R00    &1.75 R00  &3.70 R00  &7.42  &0.380   & S  &{\bf $-$6.23}\\
HD~75222a  &           &field, GOSV3    & ---      &          &      &        &    &$-$6.50 \\
HD~151804  &O8~Iaf     &Sco OB1/NGC6231 &1.91 H78  &3.10      &5.23  &0.066   &WCB &{\bf $-$7.24}\\
HD~152249  &OC9~Iab    &Sco OB1/NGC6231 &1.91 H78  &3.10      &6.46  &0.193   & S   &{\bf $-$6.41}\\
HD~152003 &O9.7 Iab Nwk &Sco OB1/NGC6231&1.91 H78  &3.10      &7.03  &0.374   &S   &{\bf $-$6.40}\\
\hline
HD~169582  &O6~Iaf    & field           &          &         &      &        &S     &$-$7.0 \\
CD\,$-$43\,4690&O6.5 III &field         &          &         &      &        &S     &$-$5.6 \\
HD~69464   &O7~Ib(f)  &field            &          &         &      &        &S     &$-$6.3 \\
HD~97848   &O8~V      &field            &          &         &      &        &S     &$-$4.4 \\
HD~302505  &O8.5~III  &field            &          &         &      &        &S     &$-$5.6 \\
HD~148546  &O9~Iab    &field G87        &          &         &      &        &S     &$-$6.5 \\
HD~69106   &O9.7~IIn  &field            &          &         &      &        &S     &$-$5.3  \\
\hline
\end{tabular}
\end{center}
\small
{\bf Notes}: Spectral types are taken from  \citet{sota14}) with 
complementary data from \citet{markova11} and \citet{markova14}; 
$V$ and $B-V$ magnitudes are from the {\it Galactic O-star Catalogue} 
\citep{maiz04}) with individual data from the {\it Hipparchus Main 
Catalogue}.  Absolute magnitudes,\MV, have been determined following the 
procedure outlined in Sect.~\ref{absmag}. Boldface numbers in Column 9 
are those used in the present study. S = single star; ? = suspected 
binarity;  WCB = wind colliding binary. Reference. GOSV3 = \citet{sota08};
H78 = \citet{hump78}; HSB12 = \citet{HSB12}; G87 = \citet{gies87}; 
KH00 = \citet{KH00}; KS10 = \citet{KS10}; P01 = \citet{patriarchi01}; 
R00 = \citet{reed00}; T03 = \citet{tapia03}.
\end{table*}

\normalsize The basic parameters of the sample are summarised in Table~\ref{sample}.  
We used the third version of the Galactic O-stars catalogue (GOSV3, \citealt{sota08}) 
as a primary source for cluster and association membership, but we consulted other 
sources (e.g. \citealt{hump78}, \citealt{gies87}, \citealt{reed00}) as well. Since 
for four sample stars (HD~94370, HD~94963, HD~75222, and HD~76968) the memberships 
listed in GOSV3 and in earlier sources disagree, an additional entry (here but 
also in Table~\ref{para} and Table~A1) is provided to account for both alternatives. 

\subsection{Absolute magnitudes, distances, and reddening}
\label{absmag}

In order to compute stellar radii (required for masses, luminosities, and mass-loss 
rates), absolute magnitudes (\MV) are needed that still pose a problem for Galactic 
objects. For the members of clusters and associations, \MV was computed using (i) 
photometry from the GOSV3 as a primary  and the {\it Hipparcos Main Catalogue} (I/239) 
as a secondary source, (ii) mean intrinsic colours,  $(B-V)_{0}$, of $-0.31$ (dwarfs 
and giants) and $-0.28$ (supergiants) from \citet{wegner94}, and (iii) distances and 
reddening as described below.  

In the absence of more accurate astrometric distances\footnote{At the time 
when this analysis was performed, GAIA measurements were still lacking  while 
$HIPPARCOS$-based distances are considered as no longer reliable in the distance 
range considered here (e.g. \citealt{Z99, Schroeder04}).}, photometric distances 
were consistently used in the present study. Particularly, for all but the 
members of Vela OB1 and the three young open clusters, Trumpler~14 (Tr 14), 
Trumpler~16 (Tr~16), and Cr 288, mean distance estimates from \citet{hump78} 
combined with a standard value of total to selective extinction, $R$=3.1, were 
initially considered to calculate \MV. The obtained estimates turned out to 
agree well with the values expected from corresponding absolute magnitude 
calibrations (typically within $\pm$0.25 mag; see e.g. \citealt{W73} and 
\citealt{MP06}), but few outliers with too low absolute values\footnote{In the 
following, we use absolute values of \MV, such that larger numbers refer 
to a larger visual brightness.} of \MV\ did also appear: one in Car~OB1 
(HD~93843), another in Car OB2 (HD~94963), and two in Pup OB1 (HD~64568 
and HD~63005). Since the reddening law in the direction towards 
these stellar aggregates is believed to be normal and a possible multiplicity 
cannot make a star appear underluminous compared to a single one of the 
same ST and LC, our results suggest that for these particular objects the 
adopted mean association distances might underestimate the actual distances. 
Luckily, for all but one (see below) of the considered stars located in 
the above clusters and associations, individual distance estimates are
available in the literature (from \citealt{KH00} and \citealt{KS10}). 
Since with these values the problem with the outliers could be
successfully solved, while the outcome for the rest did not change
significantly, these estimates have been consistently adopted for
(almost) all stars that are members of Car~OB1/OB2 and Pup OB1 (second
entry in Table~\ref{sample}). We instead used the mean distance estimate, 
as provided by \citet{hump78} for HD~92504 because of the lack of 
alternative data. 

\begin{figure}
{\includegraphics[width=8.5cm,height=5.5cm]{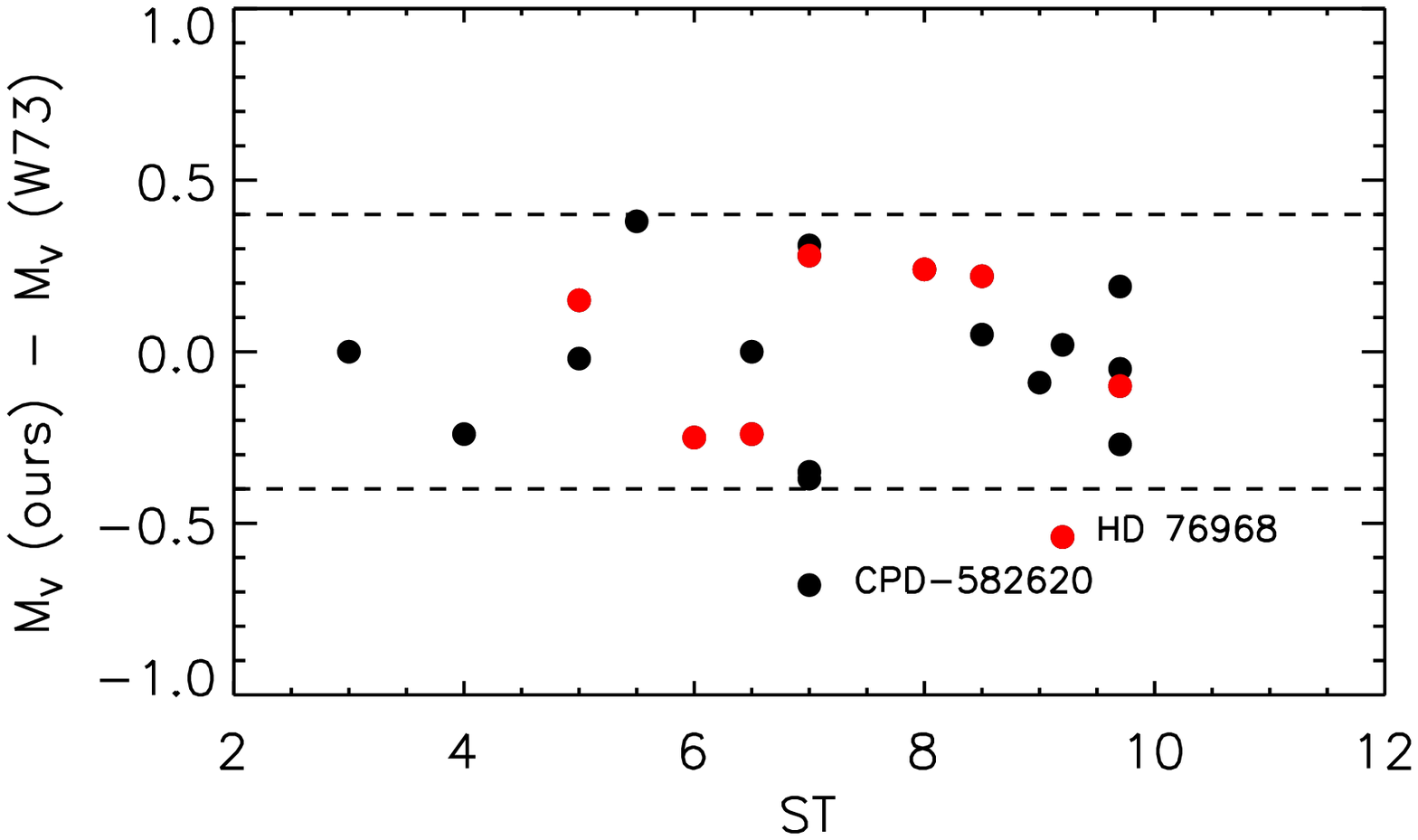}}
{\includegraphics[width=8.5cm,height=5.5cm]{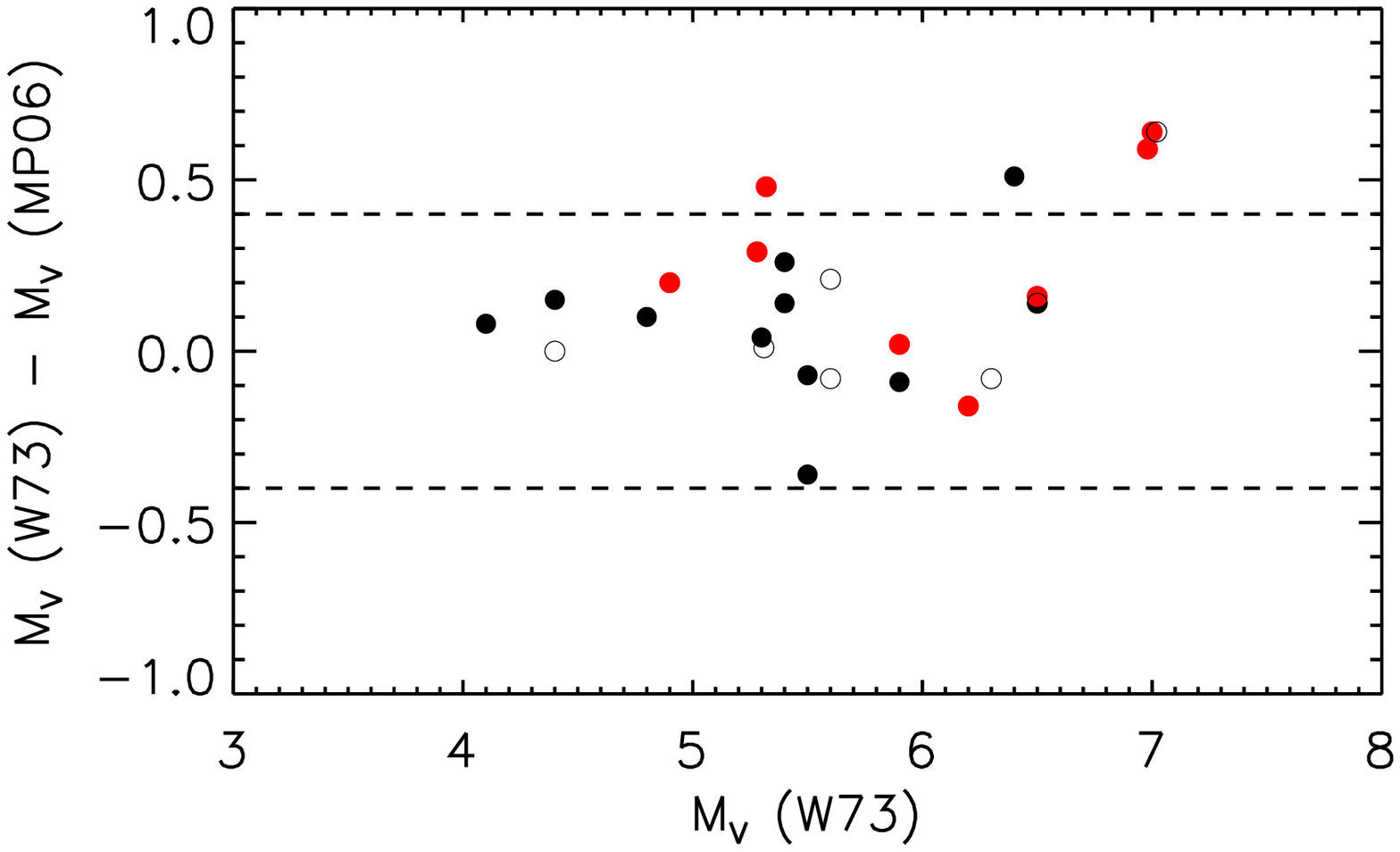}}
\caption{{\it Upper panel}: Comparison between our system of \MV--
estimates (boldfaced numbers in Column 9 of Table~\ref{sample}) and 
those as proposed by the calibrations from \citet{W73}.
{\it Lower panel}: Absolute magnitudes for the sample stars proposed by 
the calibrations of \citet{W73} vs. similar data inferred from the 
calibrations by \citet{MP06}. Filled and open circles denote 
cluster and association members and field stars, respectively. On 
each plot, absolute values of \MV\ were used, such that positive 
differences indicate objects that appear brighter in the first 
dataset compared to the second. Data points in red indicate 
confirmed SBs. For more information, see Sect.~\ref{sample}.
} 
\label{fig1}
\end{figure}

For Tr~14 and Tr~16, significantly different distance estimates (from
about 2.0 to about 4.0~kpc) can be found in the literature, depending
on the adopted reddening law, which is anomalous with $R \approx$
4--5 (see \citealt{SB08} and references therein). To account for this
problem, we proceeded twofold: first, for all stars which are
members of these clusters, we adopted the same value of R$_{V}$ =
4.4$\pm$0.2 and $d$ = 2.9$\pm$0.3~kpc, as derived by \citet{HSB12}
using 141 early-type stars with high proper motion membership
probability\footnote{These estimates are in reasonable agreement with
similar results from \citealt{vazquez96} and \citealt{tapia03}, but
disagree with \citet{CP01} who derive $R{_V}$ = 3.48$\pm$0.33 and $d$
= 3.98$\pm$0.5~kpc for Tr~16, and $R{_V}$ = 4.16$\pm$0.21 and $d$ =
2.5$\pm$0.3~kpc for Tr~14.} (first entry in Table~\ref{sample}).
Second, we considered individual $R$ and $d$ estimates (from
\citealt{patriarchi01}, second entry in Table~\ref{sample}) for each target. 
The former approach resulted in objects that are systematically
brighter (by 0.25 to 0.47~mag) than expected from the callibrations 
by Walborn\footnote{We chose this calibration as a reference because
unlike many others it separates between normal and bright giants as
well as between supergiants of class Ia, Ib, and Iab.}, while the 
second approach did not lead to any systematic trend; we note that
the SB1 system CPD$-$59\,2600 might be allowed to appear overluminous 
compared to a single star of the same ST, but the somewhat larger 
(absolute) \MV\ for the other two targets is difficult to explain. 
Thus, we accepted the latter approach for our final solution for 
those stars that are members of Tr~14 and Tr~16. Indeed, the (absolute) 
\MV\ derived for CPD$-$58\,2620 appears as too low now (see 
Fig.~\ref{fig1}), but this might still be due to a very young
nature, as suggested by its morphological OVz designation
\citep{walborn09}.  

Regarding the young cluster Cr~228,  all distance estimates, available 
in the literature, cluster around 2.6~kpc, and they were all derived 
assuming $R_{V}$ = 3.2 (see \citealt{F95} and references therein). Thus, 
we used these values to calculate the \MV\ for HD~93222.  

Finally, for Vel OB1 we adopted $d$ = 1.75~kpc and $R_{V}$ = 3.7, as 
derived by \citet{reed00} based on a variable extinction analysis of 
70 stars. For all (save one) stars that are members of this association, 
the computed \MV\ values agree  well with the callibrations by Walborn 
(generally within $\pm$0.20~mag, see Fig.\ref{fig1}). Regarding the 
outlier HD~76968, we suggest that its too low (absolute)
\MV\ might be explained as an indication that this star is an object from 
the field rather than a member of the Vel OB1 associations.

To get insight into the accuracy and reliability of our system of 
absolute magnitudes (boldfaced values in Column 9 of Table~\ref{sample}), 
we proceeded as follows:

1) For the cluster and association members, we compared our \MV -- 
determinations with those proposed by the calibrations from 
\citet{W73} (upper panel of Fig.~\ref{fig1}), and found that an
agreement within $\pm$0.4~mag is obtained, except for the two 
outliers discussed above.

2) Since  the use of different calibrations can lead to significantly 
different results (see e.g. \citealt{walborn02}, \citealt{martins05a} 
and references therein), we confronted the \MV\ for the complete 
sample as resulting from the callibrations by Walborn to similar data 
inferred from  the theoretical calibrations by \citet{MP06}. As 
demonstrated in the lower panel of Fig.~\ref{fig1}, the two datasets 
agree within $\pm$0.4~mag and the agreement is even better  for the 
field stars; we do not find clear evidence of a systematic trend.   

\begin{figure}
{\includegraphics[width=8.5cm,height=5.5cm]{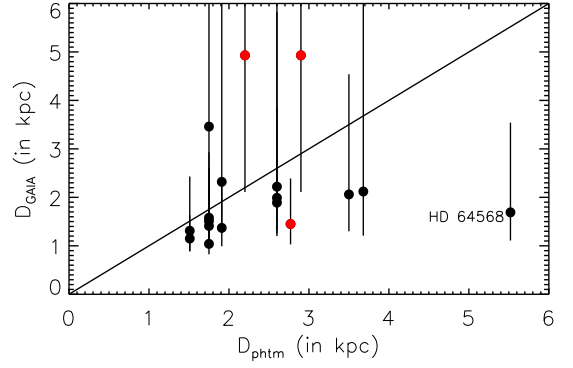}}
\caption{Photometric distances adopted in the present study vs. 
GAIA (DR1) distances. Confirmed SBs are denoted in red. Since 
not all targets have been observed by GAIA, the number of the 
data points is smaller than in the upper panel of Fig.~1.}
\label{fig2}
\end{figure}

3) To account for first results from the GAIA mission  (\citealt{GAIA16}, 
\citealt{arenou16}), we compared photometric distances adopted in the 
present study with estimates calculated using GAIA parallaxes (Fig.~\ref{fig2}). 
Given the (still) large error bars, which should become smaller within 
the next data releases, for all save two stars with measured  parallaxes 
(17 in total), no clear evidence of a significant discrepancy is found, 
although a trend for the objects with $d_{\rm phtm} >$~2pc to appear 
more distant than suggested by GAIA might be present. While this 
possibility needs to be confirmed by future analyses based on larger 
samples, we would like to point out that if the GAIA distance to 
HD~64568 were used to calculate \MV, the resulting luminosity would be 
in stark conflict with the \Teff\  as derived by our model atmosphere analysis.

Taken together, we conclude that the accuracy of our system of \MV\
for the cluster and association members is better than $\pm$0.4~mag 
and we do not find clear evidence of systematic uncertainties.  Given
corresponding results from the lower panel of Fig.~\ref{fig1}, the
same error has been consistently adopted for the field stars, whose 
\MV\ were inferred from the Walborn calibrations (for the sake of
consistency).

\subsection{Binarity }\label{binarity} 

As binarity can significantly influence the properties of massive 
stars (e.g. \citealt{langer12} and references therein), it is 
particularly important for studies like this to distinguish 
between single stars and members of close binary or multiple systems. 
While the objects in the sample have been initially selected as 
presumably single stars (see Paper~I), a re-consideration of their 
status, using newest data from  the Galactic O-star Spectroscopic 
Survey  Catalogue (GOSSSC III/274, \citealt{sota14}), has revealed 
that five of these objects (HD~91572, HD~91824, HD~75211, HD~76968, 
and CPD$-$59\,2600) have meanwhile been recognised as SB1 systems, 
two (CD~$-$47\,4551 and HD~151804) are considered as SBs with 
colliding winds, and four are  suspected to be either SB1 (HD~64568, 
HD~93843) or SB2  (HD~94963 and HD~94370). These results are taken 
into account in the following analysis.

\section{Modelling}\label{mod_analysis}

The model atmosphere analysis was performed by means of the latest version 
(V10.1) of the code FASTWIND (see \citealt{puls05} for the previous 
versions and \citealt {gonzalez12a} for the current one), following a 
three-step procedure. First, we used model grids with solar background 
abundances as derived by \citet{asplund05}\footnote{We are aware that 
an improved solar composition has been published by \citet{asplund09}, 
but since the C, N, O, Ne, and Fe abundances derived in the two studies 
agree within less than 0.05~dex, and since the main effects of different 
solar composition on the opacities is expected to appear in the post 
main-sequence evolution \citep{MP06}, the outcome of our analysis is not 
significantly influenced by the use of the older estimates.} to roughly 
constrain the parameters of the targets, namely effective temperature 
(\Teff), surface equatorial gravity (\logg), helium, and nitrogen content (with 
respect to particle number, \Yhe = N(He)/N(H) and [N] = 12 + log~N/H), 
and wind-strength parameter $Q$ = \Mdot/(\vinf*\Rstar)$^{1.5}$ (see 
\citealt{puls96}). The grids were calculated by Rivero Gonz{\'a}lez and 
additionally extended towards cooler temperatures and lower surface 
gravities by one of us (NM), such that they cover the complete parameter 
space appropriate for O stars in the MW. Second, such estimates were 
fine-tuned by calculating a grid of much higher resolution around the 
initial constraints with tailored values of \Teff, \logg, \Yhe, [N], and 
wind terminal velocity (\vinf) for each target, and different values of
the velocity field exponent $\beta$. Third, for each target, we derived 
the value of stellar radius (\Rstar)  and determined the final value of 
mass-loss rate, \Mdot. \newline \newline 
The {\it effective temperature} was constrained from the helium and nitrogen 
ionisation balance. For the cooler objects (ST later than O4), we 
relied on the former using the latter as a consistency check; for the 
hotter objects, we gave larger weight to the nitrogen balance as long as 
a reasonable fit to the He~lines could be maintained. The typical 
uncertainty of our \Teff\ determinations (obtained from the best line fit, 
estimated by eye) ranges from $\pm$1000 to 1500~K. \newline \newline 
The {\it surface gravity} was obtained from the best fit to the wings of 
\Hgama\ and \Hdelta. The error on these estimates is $\pm$0.1~dex for 
the objects with \vsini$<$120~\kms and $\pm$0.15~dex for those with
\vsini$\ge$120~\kms. For CD$-$47\,4551 and HD~151804 only, we adopted 
a somewhat larger uncertainty of $\pm$0.2~dex to account for the lower 
quality fit to the P~Cygni profiles of \Hgama\ and \Hdelta. \newline \newline
The {\it stellar radius} was determined from the de-reddened absolute magnitudes 
and the theoretical Eddington fluxes, following the procedure provided by
\citet{kudri80}. From \Teff\ and \Rstar, the stellar luminosity was calculated. 
The typical error, estimated following the philosophy outlined in 
\citet{markova04} and \citet{repo}, is  $\pm$0.08~dex in $\log\Rstare$ 
(corresponding to roughly 20\% in \Rstar), and $\pm$0.17~dex in 
$\log L$ at maximum. Since SBs did not show any peculiarity regarding 
their \MV\ (red data points in Fig.~\ref{fig1}), the same errors were adopted 
for these stars as well. \newline \newline 
{\it Wind terminal velocity.} For approximately half of the sample, we used individual 
estimates of \vinf\ as provided by \citet{howarth97} and \citet{prinja90}; for 
the other half, corresponding data from the calibrations by \citet{kudri00} 
were used instead (in Column~7 of Table~\ref{para}, numbers flagged by asterisks). 
For all targets, a typical uncertainty of $\pm$100~\kms\ was consistently
adopted.\newline \newline 
{\it Mass-loss rate and velocity field exponent $\beta$.} \Mdot\ was constrained 
from the best fit to \Ha\ and \heiid\ as a primary, and \niiia\ and \niva\ 
emission lines (when present) as a secondary diagnostics\footnote{As 
demonstrated by Rivero Gonz{\'a}lez et al.
(2011, 2012a), the N~III and N~IV emission lines can be used to
constrain the mass-loss rate, provided the nitrogen abundance is
derived independently from other lines.}. 
For objects with strong winds, $\beta$ is determined from the best fit
to \Ha\ in parallel to \Mdot; for those with weak winds (\Ha\ in
absorption), $\beta$=0.9 was initially adopted and (if necessary)
subsequently iterated along with \Mdot\ to improve the fit to \Ha. We
estimate an error of $\pm$0.1 in $\beta$ and of $\pm$0.13~dex (\Ha in
emission) and less than $\pm$0.23~dex (\Ha in absorption) in
\Mdot.\newline \newline 
{\it Clumping.} For all targets $unclumped$ wind models were used to 
constrain their wind properties. For a number of our objects, significantly 
less \heiid\ absorption (i.e. more wind emission) than observed was 
predicted from the best fit to \Ha. At least for objects with \Teff 
$\la$ 37,000~K, such discrepancy might be a direct indicator of a 
structured wind \citep{KUP06}, and we flagged the corresponding entries 
(Column 9, Table~\ref{para}) with ``a''. We note, however, that clumping 
might be present in all objects, even if there is no direct evidence in 
the optical spectra, and that all provided mass-loss rates might be upper 
limits to be reduced by the square root of the unknown effective clumping 
factor (presently estimated to lie in the range between 5 and 10; e.g. 
\citealt{puls08} and references therein). \newline \newline 
{\it Microturbulence.} All models were calculated assuming a depth-
independent microturbulent velocity of 10~\kms\ for the atmosphere structure, 
and of 15 (hotter) and 10~\kms\ (cooler objects) for the formal integral. 
No attempts were made to improve the quality of fit varying this 
parameter.\newline \newline
The {\it helium abundance} was determined by fine-tuning the fit to strategic 
helium lines, namely  He~II~ at $\lambda\lambda$ 4200, 4541, 6527, and 6683, 
and  He~I at $\lambda\lambda$4771, 4387, 4713,  and 6678. The accuracy of 
these estimates is $\pm$0.02, except for HD~151804 for which a somewhat 
larger error of 0.05 was adopted owing to the lower quality of the fit.\newline \newline 
The {\it nitrogen abundance} was constrained from the best fit to all
strategic N lines, giving larger weight to those that are not affected by 
stellar winds, and are furthermore strong enough to allow reliable estimates 
to be obtained
\footnote{The lines from the quarter system, N~III $\lambda\lambda$
4510 -4514 - 4518; the N~III triplet lines $\lambda\lambda$4634 - 4640 - 4641, 
and the N~III transitions at $\lambda\lambda$4003, 4195, and  4200; the latter 
three are used in the final abundance analysis.}
Although, for the majority of stars, an error of $\pm$0.1~dex in [N] was 
estimated, a more conservative error of 0.2~dex is consistently adopted 
to account for possible uncertainties caused by the fixed value of \vmic\ 
(see \citealt{gonzalez12a}). \newline \newline
{\it Projected rotational velocity and macroturbulence.}  The macroturbulent 
velocity, \vmac, was determined by a direct comparison between observed and 
synthetic profiles; the latter was broadened using a fixed value of \vsini,\ 
as derived in Paper II, and trial values of \vmac. We refrained from using 
\vmac\ values as published in Paper~II because these originate from the 
analysis of only one metal line, while macroturbulence is known to vary from 
one metal species to another. \newline \newline
{\it Radial velocities} (\Vr) were estimated from the measured positions of 
helium and metal (when present) absorption lines in the spectrum. 
\begin{table*}
\begin{center}
\caption[]{Final results of our sample of Galactic O stars, sorted according 
to \Teff. Cluster and association members are listed in the upper and field 
stars in the lower part. In addition to standard abbreviations, $g_c$ is the 
centrifugally corrected surface gravity, and V$_{t}$ the macroturbulence. 
Boldfaced numbers in Column 11 indicate \Vr\  measurements that are first 
(and thus unique) for the corresponding stars. By convention, [N] = 12 + log(N/H). 
Typical errors: less than 0.4~mag in \MV; less than 0.17~dex and 20\% in \logl\ 
and \Rstar, respectively; 
$\pm$0.2~dex in [N], and $\pm$0.02 in \Yhe. An uncertainty of $\pm$0.13~dex 
(\Ha in emission) and less than $\pm$0.23~dex (Ha in absorption) was estimated 
for log~\Mdot (unclumped).}
\label{para}
\tabcolsep0.95mm
\begin{tabular}{llllrcllclrlll}
\hline
\hline
\multicolumn{1}{l}{HD/CPD}
&\multicolumn{1}{c}{\MV}
&\multicolumn{1}{c}{\Teff}
&\multicolumn{1}{c}{\loggc}
&\multicolumn{1}{c}{\Rstar}
&\multicolumn{1}{c}{\logl}
&\multicolumn{1}{c}{\vinf}
&\multicolumn{1}{c}{$\beta$}
&\multicolumn{1}{c}{$\log$ \Mdot}
&\multicolumn{1}{c}{[\vsini,V$_{t}$]}
&\multicolumn{1}{c}{\Vr} 
&\multicolumn{1}{c}{\Mspec}
&\multicolumn{1}{c}{[N]} 
&\multicolumn{1}{c}{\Yhe}\\
\multicolumn{1}{l}{}
&\multicolumn{1}{c}{(mag)}
&\multicolumn{1}{c}{(kK)}
&\multicolumn{1}{c}{(cgs)}
&\multicolumn{1}{r}{(\Rsun)}
&\multicolumn{1}{c}{(\Lsun)}
&\multicolumn{1}{c}{(km/s)}
&\multicolumn{1}{c}{}
&\multicolumn{1}{c}{(unclm)}
&\multicolumn{1}{l}{(km/s)}
&\multicolumn{1}{c}{(km/s)}
&\multicolumn{1}{c}{(\Msun)}
&\multicolumn{1}{c}{} 
&\multicolumn{1}{c}{}
\\
\hline
HD~64568a        &$-$5.50 &48.0$\pm$1.5 &4.00$\pm$0.10 &11.5 &5.80 &3200$^{*}$&0.9 &-5.84&[55,96]&75$\pm$3 &48.5$\pm$17.9&8.18&0.10\\%77 
HD~46223        &$-$5.26 &43.5$\pm$1.5 &3.95$\pm$0.10 &10.9 &5.58 &2800 &0.8 &-6.10&[72,84]&40$\pm$4 &38.9$\pm$14.4&8.58&0.10\\
HD~93204a       &$-$5.78 &40.5$\pm$1.0 &3.91$\pm$0.10 &14.4 &5.70 &2890 &0.9 &-5.90&[105,105]&8$\pm$5 &60.9$\pm$22.5&7.78&0.10\\
CPD$-$59~2600a  &$-$5.05 &40.0$\pm$1.0 &4.01$\pm$0.10 &10.4 &5.40 &3065 &0.9 &-5.96&[120,90]&{\bf 7$\pm$2}&40.3$\pm$14.9&7.78&0.08\\
HD~93843a       &$-$6.38 &39.0$\pm$1.5 &3.66$\pm$0.10 &19.7 &5.91 &2730 &0.9 &-5.35&[90,40] &$-$5$\pm$4   &64.1$\pm$23.8&7.98&0.10\\
HD~91572a       &$-$5.06 &38.5$\pm$1.0 &3.90$\pm$0.10 &10.6 &5.35 &2410 &0.9 &-6.20&[49,73] &0$\pm$3      &32.7$\pm$12.1&8.37&0.10\\
HD~91824a       &$-$5.08 &39.0$\pm$1.0 &3.90$\pm$0.10 &10.6 &5.37 &2285 &0.9 &-6.82&[47,67] &$-$40$\pm$3  &32.7$\pm$12.1&8.48&0.10\\
HD~63005a &$-$5.45 &38.5$\pm$1.0 &3.75$\pm$0.10 &12.9 &5.52 &2120 &0.9 &-6.29&[63,87] &{\bf 59$\pm$3}&34.4$\pm$12.7&8.58&0.15\\ 
CPD$-$58~2620a&$-$4.16 &38.5$\pm$1.0&3.95$\pm$0.10  &7.0  &4.99 &2600. &0.9 &-7.00&[39,59]&{\bf $-$10$\pm$4}&16.0$\pm$5.9&7.98&0.10\\ 
HD~93222        &$-$5.11 &38.0$\pm$1.0 &3.90$\pm$0.10 &11.0 &5.36 &2700 &0.9 &-6.21&[52,90] &5$\pm$3      &35.2$\pm$13.0&7.98&0.10\\
CD$-$47\,4551  &$-$7.15 &38.0$\pm$1.5 &3.60$\pm$0.20 &28.8 &6.19 &2100 &0.9 &-4.95&[50,110]&12$\pm$2     &120.9$\pm$44.9&8.08&0.12\\  
HD~94963a&$-$5.53 &36.0$\pm$1.0 &3.51$\pm$0.10 &14.0 &5.47 &2300$^{*}$&1.0 &-5.82&[82,82] &6$\pm$3      &23.1$\pm$8.6&8.38&0.10 \\      
HD~94963b&$-$5.90 & & &16.6 &5.62 & & &-5.70& & &32.4$\pm$12.0&&\\
HD~94370a       &$-$5.25 &36.0$\pm$1.0 &3.73$\pm$0.15 &12.3 &5.36 &2600$^{*}$&0.9 &-5.80&[185,84]&0$\pm$2      &29.9$\pm$11.1&7.78&0.10\\
HD~94370b       &$-$5.60 &             &     &14.4 &5.50 &          &    &-5.70&        &             &40.5$\pm$15.1&&\\
HD~92504        &$-$4.45 &35.0$\pm$1.0 &3.87$\pm$0.15 &8.5  &4.99 &1900$^{*}$&0.9 &-7.13&[155,82]&$-$20$\pm$2  &19.7$\pm$7.3&7.78&0.10\\
HD~75211  &$-$6.12 &34.0$\pm$1.0 &3.52$\pm$0.15 &18.9 & 5.63 &2100$^{*}$&0.9 &-6.14&[145,58]&20$\pm$3     &43.3$\pm$16.1&8.58&0.13\\
HD~46202        &$-$4.22 &34.0$\pm$1.0 &4.00$\pm$0.10 &7.9  &4.88 &1200 &0.8 &-7.19&[15,34]  &35$\pm$3     &22.8$\pm$8.4&7.88&0.10\\ 
HD~152249 &$-$6.41 &31.5$\pm$1.0 &3.21$\pm$0.10 &20.9 &5.59 &2010 &1.0 &-5.56$^{a}$&[65,93]  &5$\pm$4      &25.7$\pm$9.5&7.88&0.10\\
HD~151804       &$-$7.24 &30.0$\pm$2.0 &3.11$\pm$0.20 &36.5 &5.99 &1445 &1.6 &-4.75$^{a}$&[67,75]  &20$\pm$3 &62.1$\pm$23.9&8.98&0.30\\
CD$-$44\,4865   &$-$5.49 &30.0$\pm$1.0 &3.46$\pm$0.10 &15.3 &5.26 &1600$^{*}$&0.9&-6.37&[60.79]  &46$\pm$4 &24.4$\pm$9.0&7.98&0.10\\
HD~152003 &$-$6.40 &30.5$\pm$1.0 &3.16$\pm$0.10 &24.1 &5.66 &1300 &1.3 &-5.42$^{a}$&[77,80]  &8$\pm$3  &30.7$\pm$11.4&7.78&0.10\\
HD~75222  &$-$6.23 &30.0$\pm$1.0 &3.16$\pm$0.10 &22.1 &5.56 &1840 &1.0 &-5.53$^{a}$&[67,80]  &58$\pm$2 &25.7$\pm$9.5&8.38&0.10\\
HD~75222a &$-$6.50 &             &     &25.0 &5.67 &     &    &-5.44&         &        &32.8$\pm$12.2&&\\
HD~78344  &$-$6.45 &30.0$\pm$1.0 &3.16$\pm$0.10 &25.2 &5.60 &1700$^{*}$&1.15&-5.30$^{a}$&[64,64]&5$\pm$1 &33.3$\pm$12.3&8.58&0.20\\
\hline                                  
HD~169582       &$-$7.00 &37.0$\pm$1.0 &3.50$\pm$0.10 &27.2 &6.10 &2100 &0.9 &-5.19&[73,105] &5$\pm$2      &86.1$\pm$32.1&8.98&0.20\\
CD~$-$43~4690&$-$5.60 &37.0$\pm$1.0 &3.61$\pm$0.10 &14.1 &5.53&2600$^{*}$ &0.9 &-5.91&[93,90]&30$\pm$4   &29.5$\pm$10.9&8.38&0.10\\
HD~97848        &$-$4.40 &36.5$\pm$1.0 &3.90$\pm$0.10 &8.2  &5.03 &2400$^{*}$ &0.9 &-6.72&[42,74]&$-$5$\pm$2 &19.6$\pm$7.2&8.38&0.10\\
HD~69464        &$-$6.30 &36.0$\pm$1.0 &3.51$\pm$0.10 &20.0 &5.78 &2300       &0.9 &-5.55&[83,92]&48$\pm$7   &46.9$\pm$17.3&8.28&0.10\\
HD~302505&$-$5.60 &34.0$\pm$1.0 &3.60$\pm$0.10 &14.9 &5.43 &2300$^{*}$ &0.9 &-6.26&[43,65]&{\bf 1$\pm$4} &32.4$\pm$12.0&8.18&0.10\\
HD~148546 &$-$6.50 &31.0$\pm$1.0 &3.22$\pm$0.10 &24.4 &5.70&1780  &0.9 &-5.25$^{a}$&[100,95]&$-$45$\pm$3  &35.7$\pm$13.2&8.98&0.20\\
HD~76968a &$-$6.20 &31.0$\pm$1.0 &3.25$\pm$0.10 &21.3 &5.58&1815  &1.0 &-5.61$^{a}$&[55,62] &$-$25$\pm$4  &29.8$\pm$11.0&8.18&0.10\\
HD~69106        &$-$5.30 &30.0$\pm$1.0 &3.55$\pm$0.15 &14.2 &5.09 &1340       &0.9 &-6.85&[310,105]&20$\pm$3    &21.8$\pm$8.1&8.00&0.10\\
\hline
\end{tabular}
\end{center}
\small
{\bf Notes}: "*" marks \vinf\, adopted from the calibrations by
\citet{kudri00};
"a" indicates a HeII~4686 mass-loss rate inconsistent with 
the wind emission implied by \Ha, directly pointing to the 
presence of wind inhomogeneities (see text).
\end{table*}
\normalsize
For the majority of stars, the spectral line shifts agree within the accuracy 
of individual estimates ($\pm$ 5~\kms), allowing  a mean value of \Vr\ to be  
obtained and used as an input parameter for the fitting procedure. To our 
knowledge, for four of the targets  
our \Vr\  estimates are first and thus unique (boldfaced numbers in Column~11
of Table~\ref{para}). For all but two stars a good correspondence (within 
10~\kms) between our \Vr\  determinations and those provided in the GCMRV 
(General Catalogue of Mean Radial velocities (III/213),  \citealt{barbier}) is 
established. Both of the only two outliers, HD~91\,824 and HD~69\,106, are 
known \Vr\  variables. 

The main physical parameters, derived as described above, and their corresponding 
errors, are listed in Table~\ref{para}.

\section{Results of the model atmosphere analysis}\label{mod_res}
\subsection{General comments}\label{gen_comments}

As our sample is relatively large and to avoid lengthy discussions, we mostly 
refrain from describing the objects one by one and focus on specific 
peculiarities and problems.

\begin{itemize}
\item[i)] For the two hottest stars in the sample, HD~64568 and HD~46223, 
we were not able to obtain good quality fits to \nv\ in parallel to the 
rest of lines in the spectrum: these lines appear stronger than predicted 
by the models and shifted to the red (by about 20~\kms), compared to the 
measured mean radial velocity. Since similar results have been reported  
by \citet{bouret12} using the CMFGEN code, the problem should not be 
related to a specific issue within the FASTWIND modelling but most likely 
indicates a more general problem (see also \citealt{gonzalez12b}). 
Consequently, a larger error of 1.5~kK in \Teff\ was adopted for these stars.
\item[ii)] At \Teff$<$36~kK, the models predict more N~III~4634-42 
absorption (or less emission) than actually observed. Since in 
the presence of an accelerating velocity field and under Galactic 
conditions, the key process, determining the N~III~triplet emission, 
is the overpopulation of the upper level due to the coupled 
N~III and O~III resonance lines \citep{gonzalez11}, and since in 
the current  version of FASTWIND this coupling is not accounted 
for, this shortcoming might explain our failure to reproduce  
the N~III 4634-42 lines strength correctly.  
\item[iii)] For \Teff\ between $\sim$39~kK and $\sim$ 44~kK, 
\niva\ is predicted to appear in emission or neutral whereas observed in 
absorption. Since lower values of \Mdot\ are precluded by the observed 
strength of \Ha, and given similar results from \citet{gonzalez12a}, 
we suggest the issue  might either imply a certain problem in the 
FASTWIND modelling regarding this particular transition or might indicate 
that clumping may play an important role. 
\item[iv)] In many cases, the \vmac\ needed to fit N lines is lower than 
the value required by the rest of lines in the spectrum. This might imply 
a depth dependent macroturbulence.
\item[v)] For a number of stars, \heiid\  is either broader compared to 
the rest of the spectral lines (HD~91572 and HD~91824 ) or displays more 
wind emission in the blue part of the profile than predicted by the best 
fit to \Ha\ (HD~46202, HD~93843, HD~94963, HD~75211, HD~69464, and HD~94370). 
Since all but two of these objects have been recognised or suspected as 
SB1/SB2 systems, binarity seems to be the most likely cause to explain these 
results.
\end{itemize}
In addition to these more general problems, there are  also other problems 
that only refer to individual stars. These are summarised and discussed in 
the following.
\begin{table*}
\caption[]{Comparison between fundamental parameters derived  and used in 
the present work and previous results}\label{common}
\tabcolsep.7mm
\begin{center}
\begin{tabular}{lllllllllll}
\hline
\hline
\multicolumn{1}{l}{Object}
&\multicolumn{1}{l}{ST}
&\multicolumn{1}{c}{\MV}
&\multicolumn{1}{c}{\Teff}
&\multicolumn{1}{c}{\loggc}
&\multicolumn{1}{c}{\Yhe}
&\multicolumn{1}{c}{[N]}
&\multicolumn{1}{c}{\logl}
&\multicolumn{1}{l}{\Rstar}
&\multicolumn{1}{l}{\Mspec}
&\multicolumn{1}{l}{Ref.}\\
\multicolumn{1}{l}{}
&\multicolumn{1}{c}{}
&\multicolumn{1}{c}{(mag)}
&\multicolumn{1}{c}{(kK)}
&\multicolumn{1}{c}{(cgs)}
&\multicolumn{1}{c}{}
&\multicolumn{1}{c}{}
&\multicolumn{1}{c}{}
&\multicolumn{1}{l}{(\Rsun)}
&\multicolumn{1}{l}{(\Msun)}
&\multicolumn{1}{l}{}\\
\hline
HD~46223 &O4 V((f)) &$-$5.26$\pm$0.4 &43.5$\pm$1.5 &3.95$\pm$0.1&0.10$\pm$0.02&8.58$\pm$0.2&5.58$\pm$0.17 &10.9$\pm$2.0&38.9$\pm$14.4 &This work \\
         &          &$-$5.22 &43.0$\pm$1.0 &4.0$\pm$0.1&0.10&8.85&5.60          & ----         &48.3$\pm$19.3 &M12/M15\\
HD~46202 &O9.2~V    &$-$4.22$\pm$0.4 &34.0$\pm$1.0 &4.0$\pm$0.1&0.10$\pm$0.02&7.88$\pm$0.2&4.88$\pm$0.17 &7.9$\pm$4.5   &22.8$\pm$8.4 &This work\\ 
         &          &$-$4.19 &33.5$\pm$1.0 &4.1$\pm$0.1&0.10&8.00&4.85$\pm$0.12 & ----         &29.0$\pm$12.4 &M12/M15\\
HD~93204 &O5.5 V((fc))&$-$5.78$\pm$0.4  &40.5$\pm$1.0 &3.9$\pm$0.1&0.10$\pm$0.02&7.78$\pm$0.2&5.70$\pm$0.17 &14.4$\pm$2.6  &60.9$\pm$22.5&This work\\
         &          & -----  &40.0$\pm$2.0 &4.0$\pm$0.1&0.10&----&5.51$^{+0.25}_{-0.20}$ &11.9$^{+4.23}_{-3.14}$ &52$^{+47}_{-25}$&M05\\
HD~94963 &O7~II(f)  &$-$5.53 &36.0$\pm$1.0 &3.51$\pm$0.10&0.10&8.38$\pm$0.2&5.47$\pm$0.17&14.0&23.1/32.4&This study\\
         &          & -----  &35.0$<\pm$2.0&3.51$<\pm$0.2&0.10    &8.67$^{+0.13}_{-0.19}$&-----&-----&-----&M17\\ 
HD~151804&O8 Iaf    &$-$7.24$\pm$0.4 &30.0$\pm$2.0 &3.1$\pm$0.2  &0.30$\pm$0.05&8.98$\pm$0.25&5.99$\pm$0.17&36.5$\pm$7.5 &62.1$\pm$23.9 &This work\\
         &          & -----  &30.0$\pm$1.0 &3.0$\pm$0.15&--- &----&5.68$^{a}$  & -----      & -----        &M15\\  
         &          & -----  &29.0$\pm$0.5   &3.0        &0.29$\pm$0.01&----&5.90        &35.4   & -----   &CE09\\
HD~152249 &OC9Iab    &$-$6.41$\pm$0.4 &31.5$\pm$1.0 &3.21$\pm$0.1 &0.10$\pm$0.02 &7.88$\pm$0.2&5.59$\pm$0.17 &20.9$\pm$4.3        &25.7$\pm$9.5 &This work\\    
         &          & -----  &31.0$\pm$1.0 &3.25$\pm$0.15&----&8.11&5.61$^{a}$ & -----      & -----       &M15\\
\hline
\end{tabular} 
\end{center}
\small
{\bf Notes}: 'a' indicates data adopted from the calibrations of 
\citet{martins05a}. Reference. M05~=~\citet{martins05b}; M12~=~\citet{martins12b}; 
M15~=~\citet{martins15b}; M17~=~\citet{martins17};  CE09~=~\citet{ce09}.
\end{table*}
\normalsize
\paragraph{\bf CPD $-$59\,2600:} The only peculiarity revealed throughout our 
analysis is that \Yhe=0.08 is needed to reproduce the strategic He lines. As 
this star is a SB1 system \citep{sota14}, one might argue that due to dilution 
of the global spectrum by the secondary, the He lines might appear weaker than 
normal (see e.g. \citealt{carolina14}). Within this hypothesis, however, the 
N lines  should also appear weaker than expected for a single star of the same 
ST and LC,  a possibility that was not  confirmed by our analysis  that indicates 
a nitrogen content equal to the baseline solar abundance for this star (see 
Table~\ref{para}). Unless the dilution effects (reduced strength of N lines) have 
been, by chance, completely compensated by enriched N content due to binary evolution, 
our results would imply  that CPD$-$59\,2600 might be  a helium-deficient star. 
We note that another presumably helium deficient O star (HD~15570) has been 
recently observed and analysed by \citet{bouret12}.
\newline
\newline
{\bf CD~$-$43\,4690:} Our analysis revealed that the He~I lines are significantly 
broader and indicate a \Vr\ by \~25~\kms\ larger than observed for the rest of 
absorption lines in the optical spectrum. Both of these results might be easily 
accounted for assuming that  CD~$-$43\,4690 might be a SB.  
\newline
\newline
{\bf HD~94370:} This star has been classified as O6.5~III \citep{W73}, O7.5~III(f) 
\citep{M88}, 7.5~IInn (Paper II), and  O7(n)fp \citep{sota14}. While \Teff = 36~kK, 
as derived by us,  agrees  well with the value expected for a single giant of 
O7 subtype, \loggc = 3.73~dex is more appropriate for a subgiant rather than a giant 
(see Fig~\ref{fig4}). Additionally, \Ha\ displays an emission component that is 
not consistent with the strength of the absorption trough, and  \heiid\ also appears 
peculiar  (see item v) above). All this supports a binary nature of HD~94370, as 
suggested  by \citet{sota14} based on \Vr\ measurements.
\newline
\newline
{\bf HD~169582:} A FASTWIND model with  \Teff = 37~kK, \logg = 3.5, \Yhe = 0.2, and 
[N] = 8.98 provides acceptable fits to all strategic lines, except for the N~V doublet, 
which appears too strong in absorption. Additionally, the position of various absorption 
lines are not consistent: some of these indicate \Vr = 10~\kms, others indicate \Vr\ 
of about zero or even $-$9~\kms. These results strongly suggest that HD~169582 might 
be a SB, but no evidence for the presence of a companion has been reported so far in 
the  literature.
\newline
\newline
{\bf CP$-$47\,4551 and HD~151804}: Despite our efforts we were able to obtain a good 
fit quality to all strategic lines with one set of parameters for either of these 
stars. Given that both stars are likely SBs  with colliding winds (see e.g. \citealt{sota14} 
and references therein) and that the former also possesses a magnetic field 
\citep{hubrig11}, this result is easy to understand. Consequently, larger error bars 
on the derived parameters were adopted for these stars, to account for their specific 
nature.
\newline
\newline
\noindent
From what has been outlined, it should be clear that the model atmosphere analysis of 
the sample did not pose serious problems. The noted difficulties refer either to (a) 
specific line transition(s) in a specific temperature regime, or invoke particular 
objects, generally SB1/SB2. In all these cases, the impact on the derived parameters 
can be easily accounted for in the error bars or even neglected.

\subsection{Comparison to previous results}\label{comp_res}

In this subsection we compare results from the quantitative analysis 
performed in this work with similar results obtained by other 
investigators for individual stars in common (Sect.~\ref{comp_individual}) 
and for other Galactic O stars with  similar but not identical properties 
(Sects.~\ref{teff_logg}, \ref{Mdot}, and \ref{Nabn}). In the latter case, 
the comparison is indirect through  several functional relationships, and 
involves additional, (presumably) single objects analysed in terms of main 
photospheric and wind parameters, (e.g. \logl, \Teff, \logg, \Rstar, \Mdot, 
\Mspec, He and N content, and \vsini\ accounting for macroturbulence) using 
methodologies similar to ours. Since all these data were derived by means 
of the code CMFGEN \citep{HM98} in combination with UV and optical spectroscopy, 
these data should be (to a large extent) internally consistent. 

\subsubsection{Comparison of individual objects}\label{comp_individual}

Table~\ref{common} lists fundamental stellar and wind parameters derived in 
the present and previous studies for five stars in common.  Obviously, 
almost perfect agreement between our determinations and those from the 
cited works is established.
\begin{figure}
{\includegraphics[width=8.5cm,height=5.5cm]{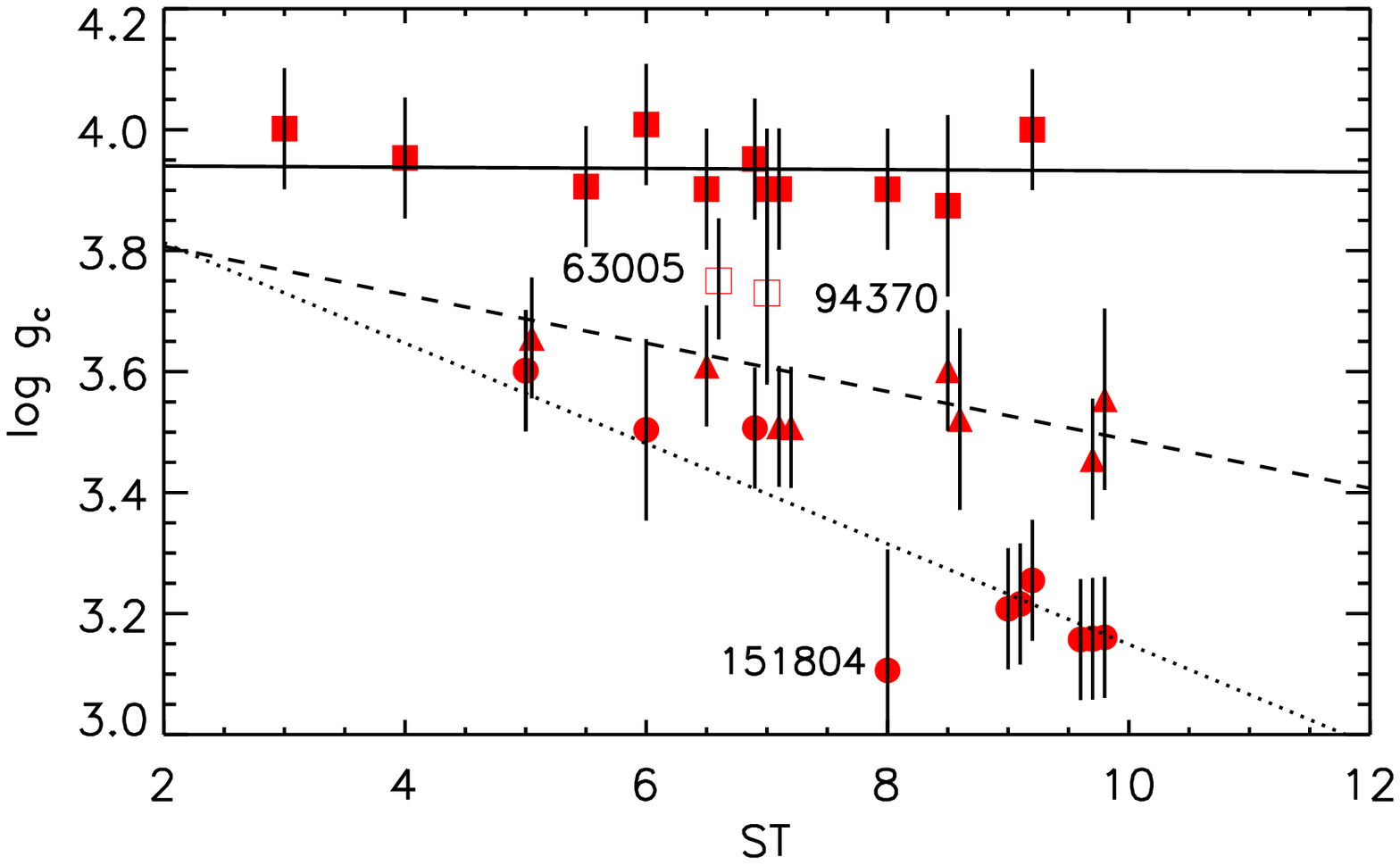}}
{\includegraphics[width=8.5cm,height=5.5cm]{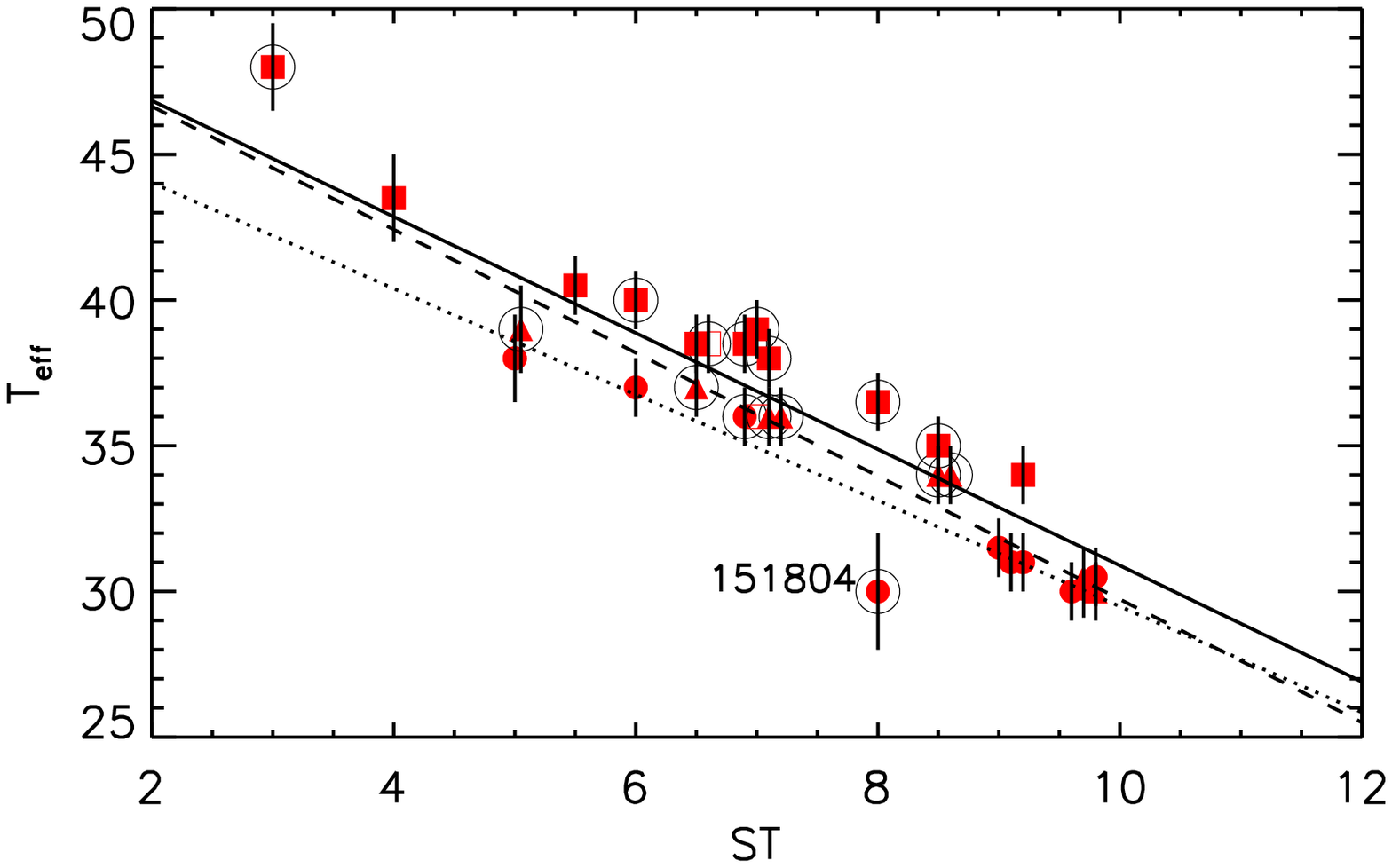}}
\caption{Surface gravities corrected for centrifugal acceleration 
(upper panel) and effective temperatures (lower panel) for the stars 
listed in Table~\ref{para}, as a function of spectral type. LC I 
objects are denoted by circles, LC~II/III objects by triangles, and 
LC~IV and V objects by open and filled squares, respectively. The 
calibrations from \citet{martins05a} are overplotted:\ the solid line 
indicates dwarfs, the dashed line indicates giants, and the dotted 
line indicates supergiants. Fast rotators are additionally denoted 
by large circles. For clarity, some objects were shifted slightly horizontally.} 
\label{fig3}
\end{figure}

\subsubsection{Temperature  and gravity scales}\label{teff_logg}

Fig.~\ref{fig3} shows the \Teff\ and \loggc\ distributions of the  stars 
listed in Table~\ref{para} as a function of ST. The calibrations from 
\citet{martins05a}, based on quantitative spectroscopy  of O stars in the 
MW, are overplotted. From the upper panel, one can see that within each 
of the three luminosity subgroups, the agreement between our \loggc\  
determinations and the values proposed by the calibrations is reasonably 
good (within the error bars), and there are only three real outliers: 
HD~63005, which is actually not a dwarf but a subgiant; HD~151804, which 
is likely a SB2 with a very strong wind (the strongest one in the sample); 
and HD~94370, which is listed in the GOSV3 catalogue without any luminosity 
class designation, but according to earlier sources has been classified 
as an LC III/IV object \citep{W73, M88}. 

From the lower panel of Fig.~\ref{fig3}, on the other hand, we find that 
our \Teff\  values for the giants and supergiants are consistent with the 
values proposed by the corresponding calibrations,  and that HD~151804 is 
the only real outlier (see above), whereas a small offset towards  higher 
temperatures may be present for the sample dwarfs of intermediate and 
late ST.  

While our finding about systematically higher \Teff\ for O7--O9.7 dwarfs 
is consistent with similar findings from \citet{S14}, we have not been 
able to confirm the trend of decreasing \loggc\ towards earlier subtypes, 
as demonstrated by their Fig.~1. Since our dwarf subsample is relatively 
small, and in particular smaller than that studied by these authors (11 
against 33 in the latter case), we complemented our subsample with \Teff\ 
and \loggc\ estimates for 27 (presumably single) O-type dwarfs analysed  
by  \citet{martins12a, martins12b}, Martins et al. (2015a,b), and 
\citet{marcolino} to improve the statistics. We note that we did not employ 
any data from \citet{repo} and \citet{martins05b}, since these were used  
by \citet{martins05a} to construct our comparison calibrations.
\begin{figure}
{\includegraphics[width=8.5cm,height=5.5cm]{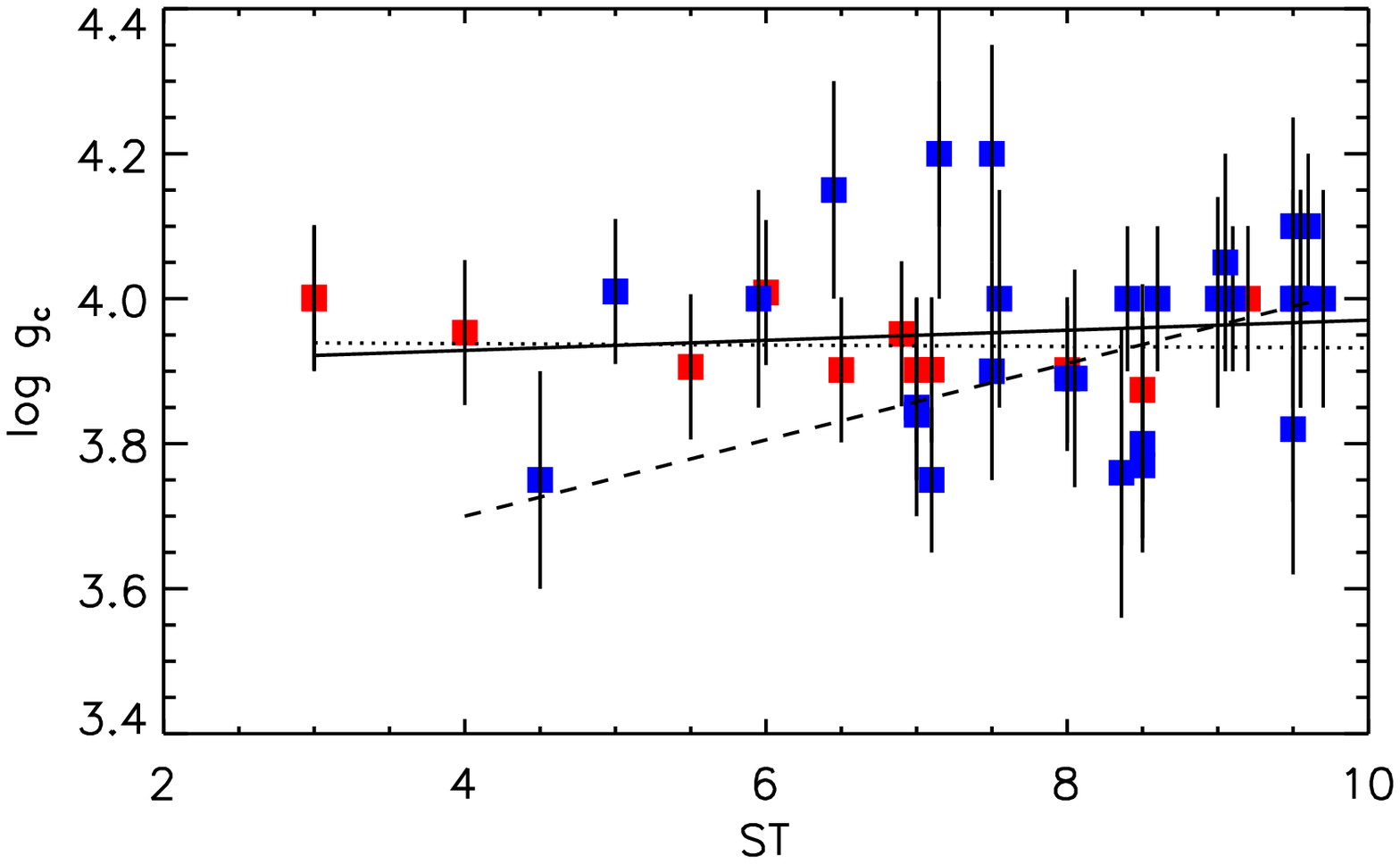}}
{\includegraphics[width=8.5cm,height=5.5cm]{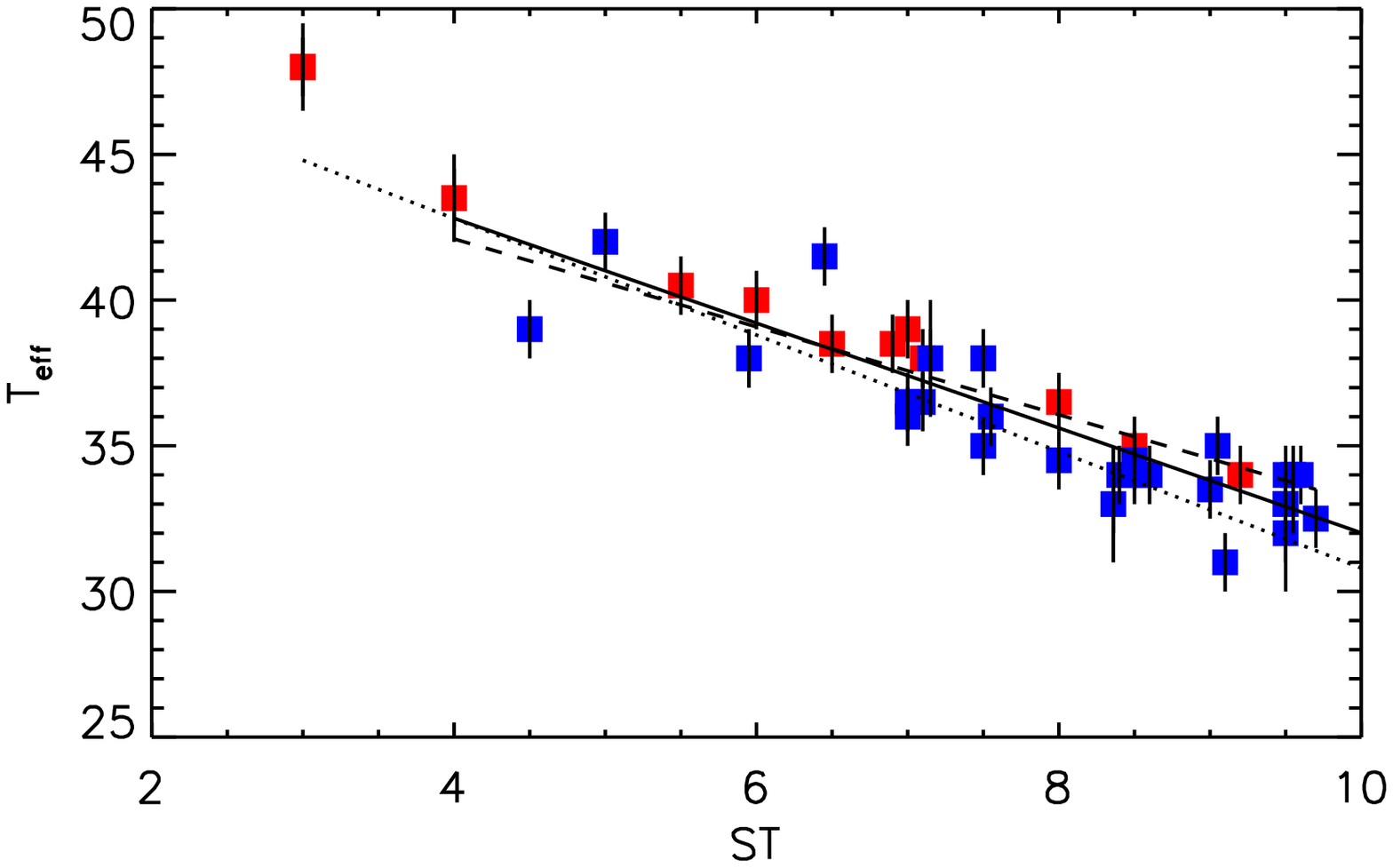}}
\caption{ST--\loggc\ and ST--\Teff\ relations for the extended dwarf 
sample (see text). The FASTWIND targets (present study) are highlighted 
in red, and objects analysed by means of CMFGEN in blue. The solid line 
provides the least-squares fit for the complete sample. Dotted and dashed 
lines denote callibrations by Martin and the regression obtained by \citet{S14}, 
respectively.} 
\label{fig4}
\end{figure}

The  distribution of the extended dwarf sample in the ST--\loggc\ and 
the ST--\Teff\ planes is illustrated in Fig.~\ref{fig4}. Despite the 
sizable scatter at a given ST, the displayed data confirm our findings 
from Fig.~\ref{fig3}. Firstly, surface equatorial gravities for Galactic 
O-type dwarfs appear to be generally consistent with the callibration by 
Martins  without any systematic trend. Secondly, current effective 
temperatures  for O7--~O9.7 dwarfs may be higher than those proposed by 
the calibration by about 1 to 2~kK. To get additional insight into the 
former result, we investigated the \loggc--~log\Teff\ distribution of 
the O-type dwarfs studied by \citet{martins15b} (see their Fig.~3), and 
found that also these data do not provide any clear evidence for a 
systematically lower \loggc\ towards  hotter \Teff. Consequently, we 
suggest that misclassification and/or underestimated \logg  values might 
both contribute to explain why the \loggc\ pattern obtained by \citet{S14} 
is so different from that revealed by the data shown in the upper panel 
of Fig.~\ref{fig4}.

Another point is that \citet{massey13} warned about a specific discordance 
between FASTWIND and CMFGEN \logg  determinations, finding that the former 
are systematically lower than the latter by about 0.12~dex. While this finding 
is not confirmed by our analysis (see upper panel of Fig.~\ref{fig4}),  a 
theoretical explanation in terms of differences in the quasi-hydrostatic 
treatment in various model atmosphere codes has been recently proposed by 
\citet{sander15}. 

Finally, \citet{gonzalez12b} and \citet{carolina14} recently reported about 
a possible non-uniform ST--\Teff\ relation for LMC O-type dwarfs and found 
that the slope is steeper for the hotter (ST<O4) than for the cooler stars 
(ST>O4). Since the temperature derived for our only dwarf of O3 subtype is 
significantly higher than proposed by the linear fit to the rest of the 
extended dwarf sample, this result might imply that a similar non-uniform 
ST--\Teff\ relation could also apply for O-dwarfs in the MW.

\subsubsection{Mass loss and wind clumping}\label{Mdot}

\begin{figure}
{\includegraphics[width=8.5cm,height=5.5cm]{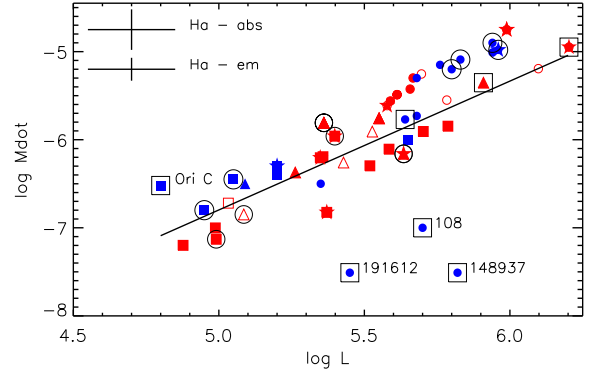}}
\caption{Unclumped mass-loss rates for FASTWIND (red) and CMFGEN 
(blue) targets as a function of \logl. The  LC~I, II/III, IV/V objects 
are denoted by circles, triangles, and squares, respectively; open and 
filled symbols represent field stars and cluster members, respectively. 
Star symbols denote SBs as indicated in \citet{sota14}. Magnetic stars 
and fast rotators (\vsini$>$110~\kms) are additionally highlighted by 
large squares and circles, respectively. 
The solid line represents a least-squares fit to the predictions 
from \citet{vink00}. For further explanation, see Sect.~\ref{Mdot}.}
\label{fig5}
\end{figure}

Fig.~\ref{fig5} shows the mass-loss rate, as derived in the present study 
employing unclumped wind models, as a function of \logl. Similar data are 
overplotted in blue for 23 O-type stars obtained by means of the CMFGEN code 
using \Ha  (Martins et al, 2012a,b; 2015a) or UV resonance  + \Ha\  lines 
\citep{bouret12} as wind diagnostics\footnote{For the CMFGEN targets with 
clumped winds, the corresponding unclumped \Mdot\ were calculated using the 
maximum clumping factor in the outer wind, as derived in the corresponding 
studies.}.
From these data it is evident that for all but the outliers marked with their 
IDs (see below), the spectroscopically derived (unclumped) mass-loss rate 
increases with increasing luminosity, where the FASTWIND and the CMFGEN 
targets participate in a similar way. While this trend is qualitatively 
consistent with theoretical expectations, a comparison to the predictions 
by \citet{vink00} -- computed using the best-fit parameters derived for each 
target (own and adopted) -- reveals a serve discordance (by up to a factor 
of 3) for all but two of the more luminous supergiants, and a reasonable
agreement (within the error bars) for the rest of the sample stars. Closer 
inspection furthermore shows that all supergiants with \Mdot(unclumped) 
significantly larger than the predictions by Vink display evidence of 
structured winds, in terms of the aforementioned discordance of HeII\,4686 
and \Ha\,  (FASTWIND targets, flagged in Column 9 of  Table~\ref{para} with 
``a''), or in terms of direct fitting of UV and optical spectral lines 
with clumped wind models (CMFGEN targets).

This situation closely resembles the results presented by \citet{repo}. 
Already then, it was argued that the discrepancy should be due to the 
neglect of clumping in the mass-loss diagnostics, and in this situation 
this also seems to be the most likely explanation. Discrepancies in 
\Mdot by a factor of three correspond to clumping-factors of the order 
of ten, which is a typical number derived from various clumping diagnostics 
(as summarised e.g. by \citealt{puls08}, \citealt{PSM15} and \citealt{martinez17}). 
Whether there is an additional discrepancy between our data and the simulations 
by Vink et al. cannot be decided though, since this would require a detailed, 
multiwavelength mass-loss analysis accounting for micro- and macro-clumping. 

Overall, however, we note that most objects follow the predicted trend, and 
there are only few real outliers, denoted by their ID in Fig.~\ref{fig5}. 
All of these outliers are slowly rotating magnetic stars,  which among other 
peculiarities have demonstrated rotationally modulated stellar and wind 
properties \citep{martins12a}. As even weak magnetic fields have  the potential 
to channel the wind material towards the magnetic equator \citep{udDoula02}, 
the hypothesis of an oblique magnetic rotator has been suggested as a possible 
explanation for their peculiar behaviour. For HD~191612 this possibility was  
confirmed by 2D \citep{jon12}) and 3D \citep{naze16} magneto-hydrodynamical 
simulations  within the so-called  dynamical magnetosphere model. A similar 
explanation may also  apply to $\theta^{1}$ Ori~C and HD~148937, as suggested by 
\citet{jon12}. Since the density structure of such a dynamical magnetosphere is 
very different from a spherically symmetric wind, realistic mass-loss determinations 
have to account for corresponding models and a multi-D radiative transfer in both 
the optical and  UV. Studies accounting for these requirements  
are in progress, and we will have to check how the mass-loss rates discussed here 
translate to the actual quantities. 

Conversely, given the results shown in Fig.~\ref{fig5}, one might speculate that 
the spectra analysed by \citet{martins12a} were taken when the wind-confined  
disc-like structure was viewed either face-on (i.e. maximum wind emission  -- 
$\theta^{1}$ Ori~C) or edge-on (minimum wind emission -- HD~108, HD~148937 and 
HD~191612).  At least for HD~108 and HD~191612 this turned out to be the case 
(see \citealt{martins12a}.)

\subsubsection{Nitrogen abundances}\label{Nabn}

\begin{figure}
{\includegraphics[width=8.5cm,height=5.5cm]{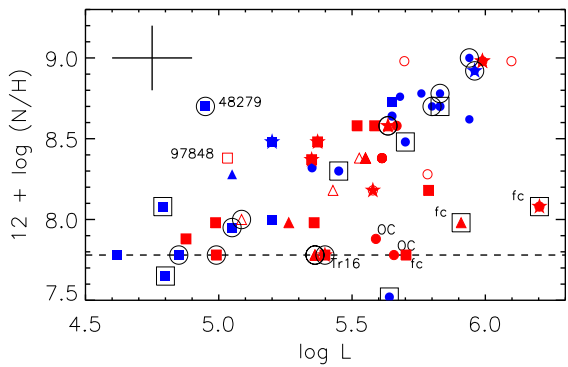}}
\caption{Surface N abundance vs. \logl. Same symbols and colours as in 
Fig.~\ref{fig5}. The members of the OC and Ofc categories are 
explicitly denoted. For more information, see Sect.~\ref{Nabn} }
\label{fig6}
\end{figure}

Fig.~\ref{fig6} shows the run of N abundance for the sample stars (own and adopted) 
as a function of \logl. Despite the sizable scatter at a given \logl, a well-defined 
trend of more enrichment in more luminous (and thus more massive) stars can be 
observed within each of the three LC subgroups. This finding is qualitatively 
consistent with evolutionary calculations for a coeval stellar population, which 
predict that because of rotational mixing and mass loss, more massive stars should 
be more chemically enriched than their less massive counterparts. 

On the other hand, given that fast rotators and SBs do not demonstrate any specific 
pattern, one might argue that this contrasts with theoretical expectations about 
the role of rotation and binarity regarding the surface chemical enrichment of 
massive stars. Such interpretation, however, would be rather premature, since  
other physical agents different from stellar mass, binarity, and rotation --- such 
as age or binarity history (see e.g. \citealt{m09}, \citealt{deMink09}) --- can 
also contribute. 

Apart from the stars determining the main trends in Fig.~\ref{fig6},  
there are also those which deviate by more than 1$\sigma$. In the 
dwarf subsample (objects denoted by squares), the most outstanding 
outliers are HD~97848 and HD~48279, which show extreme N enrichment,  
and CPD$-$59\,2600 (\logl=5.40~dex, SB1) and HD~93204 (\logl=5.71), 
which are basically unenriched. Several reasons may play a role in 
determining the N pattern  of these objects. Particularly, for the 
field star HD~97848 an underestimated luminosity, due to uncertain 
distance, might be responsible or contribute to explain its relatively 
large N enrichment. For CPD$-$59\,2600 and HD~93204, the apparent 
youth of their host cluster Tr~16  might play a role; this possibility 
seems to be additionally supported by the subsolar He abundance 
derived for CPD$-$59\,2600 (see Sect.\ref{gen_comments}). Regarding 
HD~48279, a former binarity with mass transfer and/or tidal 
interactions between the two components is a possibility 
\citep{martins12b}.

Concerning the most outstanding outliers among the more evolved 
objects (giants and supergiants denoted by triangles and filled 
dots, respectively), there are two important features to be noted. 
First, these objects all indicate a N enrichment significantly 
lower than that derived for other stars of the same LC and nearly 
same \logl\ and \Teff, and, second, all of these outliers have been 
recognised as morphologically peculiar objects with very weak nitrogen 
lines. These outliers are the following:\newline\newline 
a) HD~152249 (\logl=5.59) is a member of the OC category, defined 
by C~III~4650 absorption much stronger than that of N~III~4634-40-42.\newline 
b) HD~152003 (\logl=5.66) and $\zeta$~Ori~A (\logl=5.64) are denoted
as Nwk objects, meaning its N lines are too weak.\newline 
c) HD~93843 and CD$-$47\,4551 (\logl=5.94/6.19) are Ofc stars
characterised by C~III 4647-50-52 emission of similar strength as
that of N~III 4634-40-42 (see \citealt{walborn10}).\newline 
d) HD~94370 is an Onfp star demonstrating a reverse P Cygni profile
in \heiid and a variable C~III~4647-50-52 emission equal or larger
than the N~III~4634-40-42 emission.\newline\newline 
Interestingly, also the dwarf star HD~93204, for which a peculiarly
low N enrichment has been derived (see above), is a member of the 
 Ofc category. While the physical nature of the Ofc, OC, and 
Onfp stars is still unclear (see e.g. Walborn et al., 2010b, 2011), 
our results clearly indicate that they  follow their own, specific 
N enrichment pattern that runs in parallel to the main trend, but 
at significantly lower values.  Since three of the six more massive 
and evolved outliers are found to possess weak magnetic fields (see 
\citealt{hubrig11} and \citealt{meynet11}), one might speculate whether 
this specific property is  responsible for or, at least contributes 
to, their peculiarly low N abundance.

From the results outlined in this section, it should have become 
clear that the physical relations determined from the FASTWIND 
targets are fully consistent with those displayed by the CMFGEN targets. 
Thus far, it seems justified that we complement our original sample 
of 30 O stars in the MW with 23 such stars selected from the studies 
by \citet{martins12a, martins12b}, \citet{martins15a}, and \citet{bouret12}, 
to improve the statistics and completeness of the database. Accordingly, 
the total number of stars underlying the following analysis rises to 53 
and comprises 20 supergiants, 8 normal/bright giants, 22 dwarfs, and 3 
objects without LC designation. We did not incorporate all external dwarfs 
as used in Sect.~\ref{teff_logg} because some of these objects have not 
been analysed in terms of wind properties \citep{martins15b} or chemical 
enrichment, in particularly N abundances \citep{marcolino}.

\section{Evolutionary masses}\label{evol_mass}
\subsection{Potential uncertainties}\label{gen_uncert}

Evolutionary masses (\Mevol) can be estimated by comparing the 
derived location of a given star in the Hertzsprung-Russell 
diagram (classical or spectroscopic, see below), or the Kiel
diagram\footnote{Like the spectroscopic HRD, this diagram depends 
only on the distance-independent quantities \Teff\ and \loggc.} 
(KD), with evolutionary tracks calculated from a set of 
pre-selected values of initial masses (\Minit) and initial rotation 
velocities (\vinit).  The accuracy of these estimates depends 
(i) on the uncertainties in \Teff, \loggc\ , and \logl\ derived 
from quantitative spectroscopy, (ii) on the interpolation 
procedure between different tracks, and (iii) on the tracks 
themselves (see e.g. \citealt{martins14}).
\newline
\newline
{\it Observational uncertainties.} For Galactic objects, the 
main source of errors on \logl\ are uncertain distances. While 
for most sample stars that are members of cluster and associations, 
the adopted photometric distances agree well (within the error 
bars) with the estimates inferred from current GAIA parallaxes, 
for some of them a tendency to appear more distant than determined 
from the GAIA measurements seems to emerge (see Fig.~\ref{fig2}). 
To put additional constraints on this issue, we proceed twofold. 
Firstly, we distinguish clearly between \Mevol\ obtained for cluster and
association members and for the field stars with filled and open 
symbols, respectively. Second, in parallel to the classical Hertzsprung 
Russell Diagram (HRD), we also consider the so-called spectroscopic 
HRD (sHRD), which does not require knowledge of stellar distances 
(see Paper~II and \citealt{LK14}). 
\begin{figure*}
{\includegraphics[width=8.5cm,height=5.5cm]{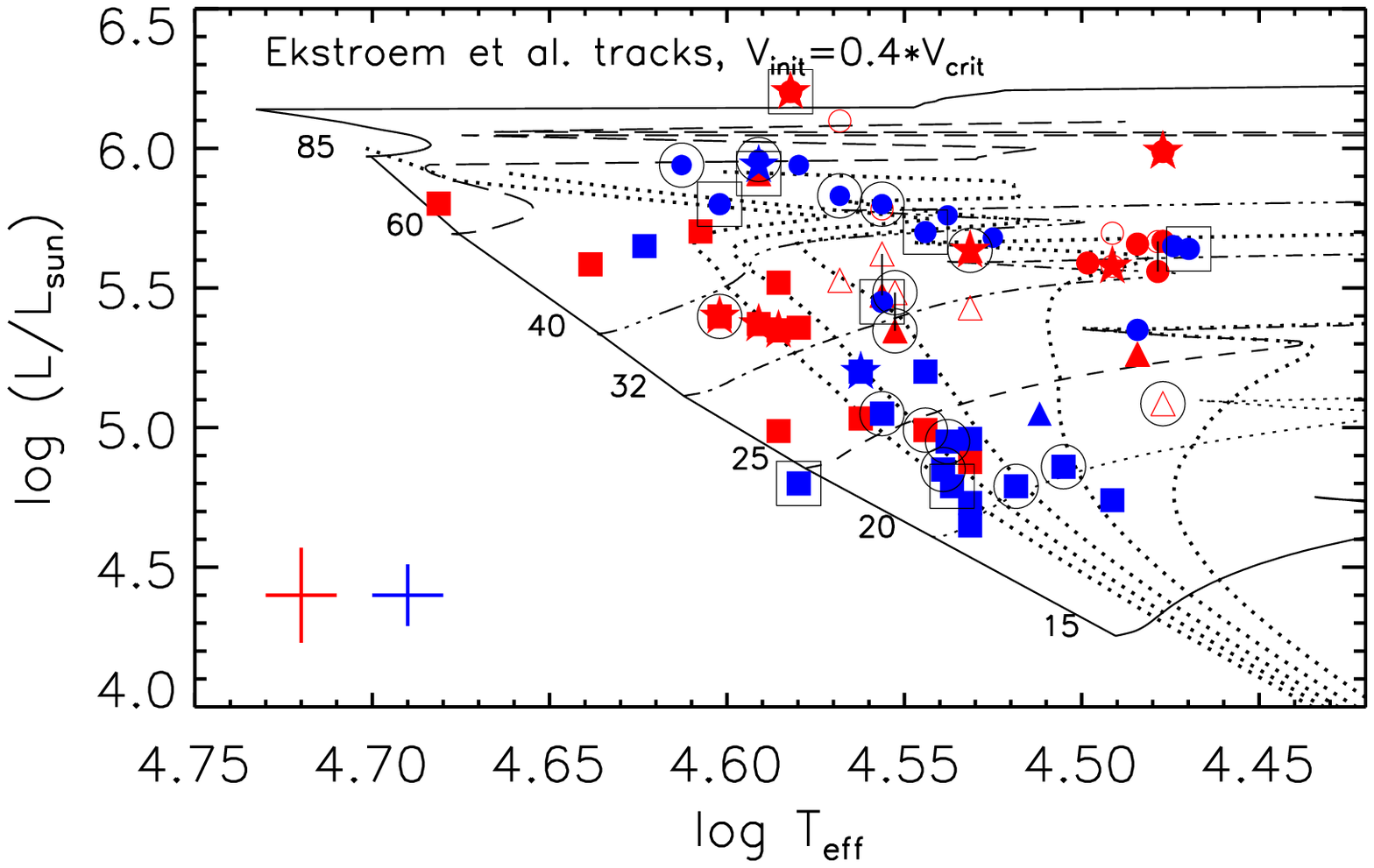}}
{\includegraphics[width=8.5cm,height=5.5cm]{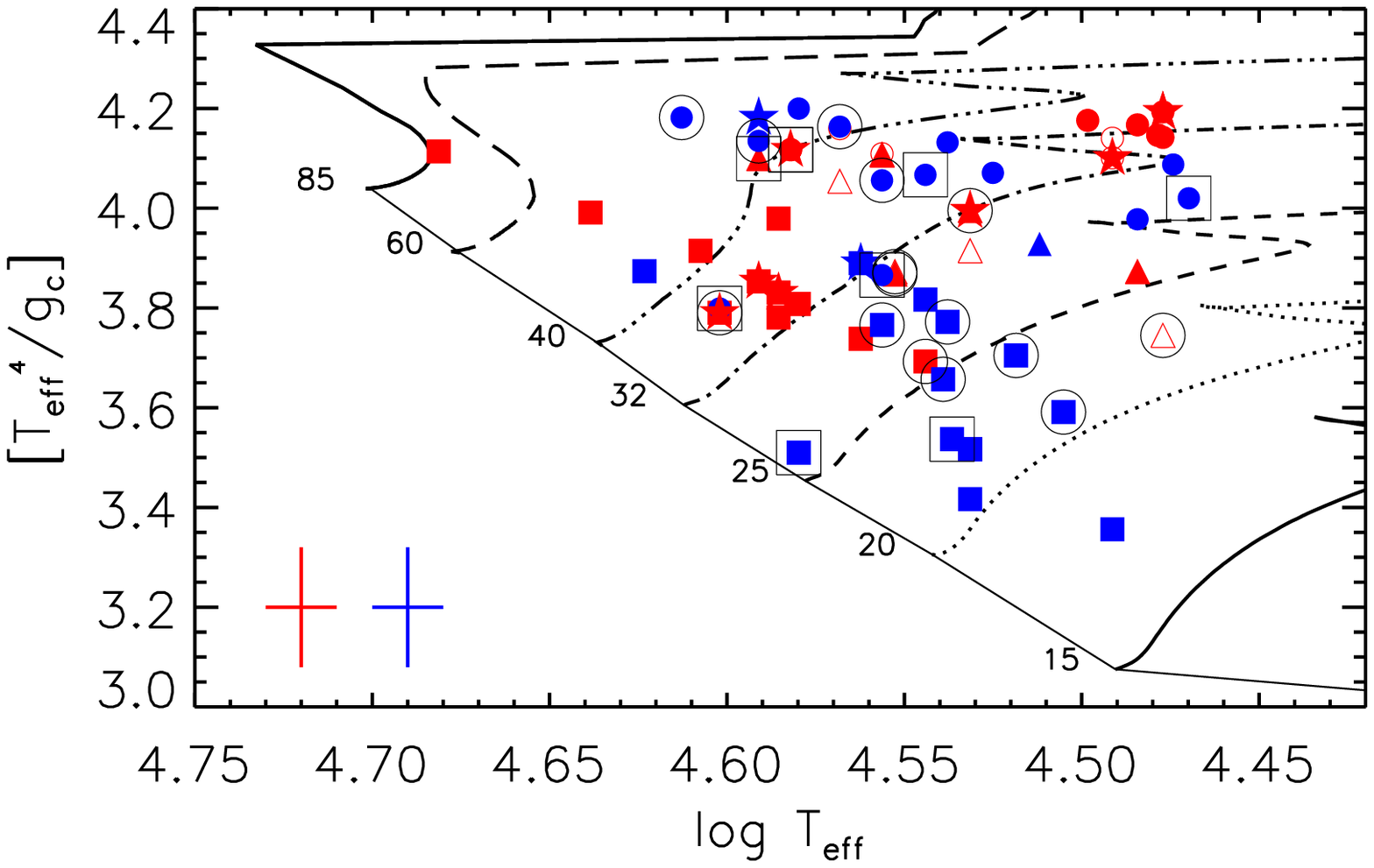}}

{\includegraphics[width=8.5cm,height=5.5cm]{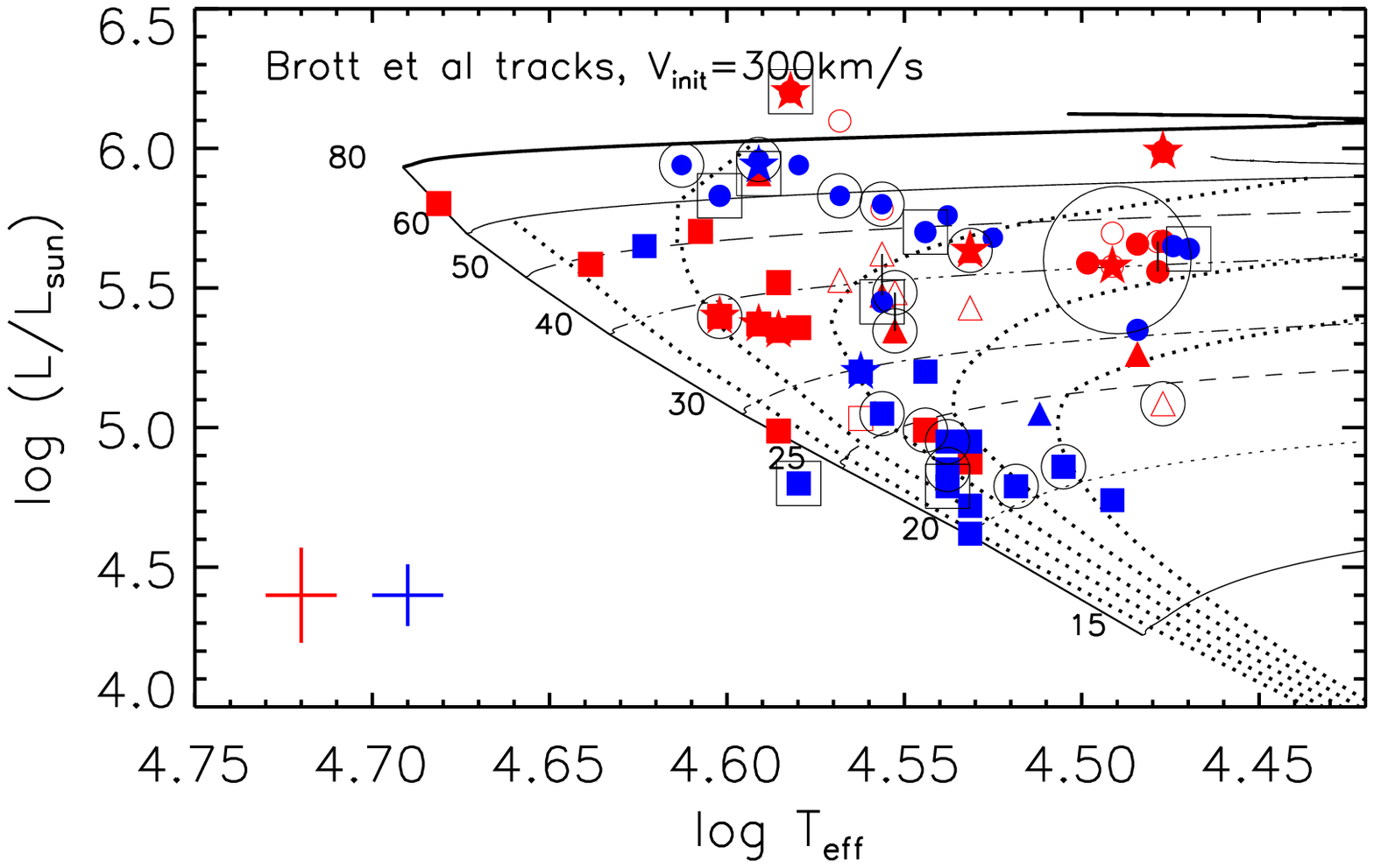}}
{\includegraphics[width=8.5cm,height=5.5cm]{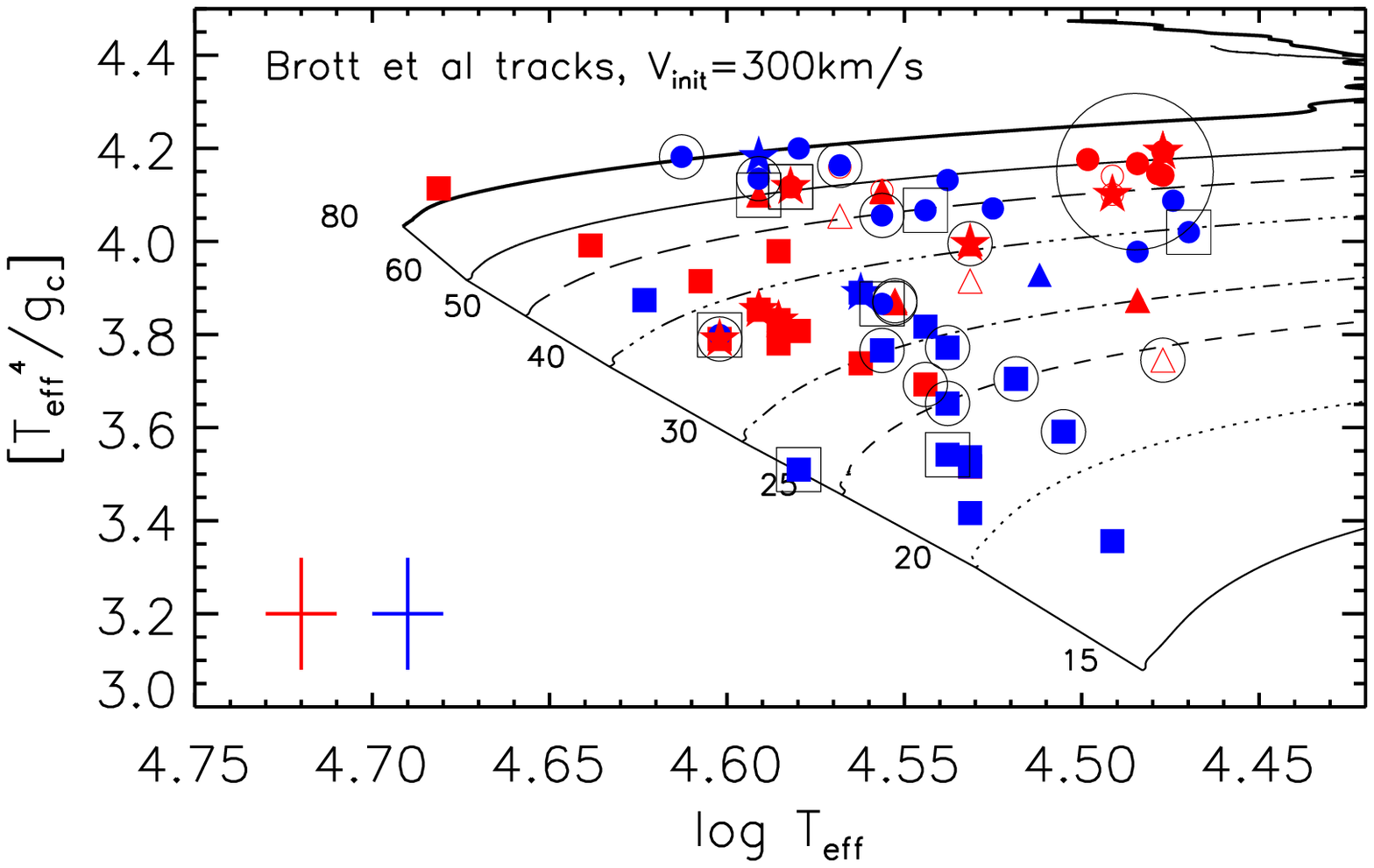}}
\caption{Classical (left) and spectroscopic (right) HR diagrams,  
built using the Ekstroem et al. and the Brott et al. tracks with 
rotation. Overplotted are corresponding isochrones for 1, 2, 3, 
4, and 5~Myr (for the HRD only). Same symbols and colours as in 
Figs.~\ref{fig5} and \ref{fig6}. On the left plots, vertical 
lines connect the two luminosity solutions for each star with 
two entries in Table~\ref{para} (clearly seen when zoomed in). 
For more explanation, see Sect.~\ref{hrd_shrd}.} 
\label{fig7}
\end{figure*}

Another source of observational uncertainties refers to various 
physical assumptions and approximations as implemented in model 
atmosphere calculations (see e.g. Sect.~\ref{teff_logg}). By using 
CMFGEN and FASTWIND data in parallel, we hope to reveal and constrain 
such effects (if present).
\newline
\newline
{\it Differences caused by input physics in evolutionary calculations.} 
To address this issue, two sets of widely used evolutionary tracks for 
solar metallicity have been considered: one from \citet{ekstroem} and 
the other from \citet{brott11}. (We refer to these as the current 
Geneva and Bonn tracks/models, respectively). Since the former were computed 
for \vinit~=~0.4~\vcrit (corresponding to velocities from 270 to 370~\kms 
for the mass range between 14 and 85~\Msun), while the latter cover a wide 
range between zero to 600~\kms, we employed (when not stated otherwise) 
the Bonn tracks with \vinit$\approx$300~\kms in the present analysis for 
the sake of consistency. \newline \newline
{\it Uncertainties caused by the approach used to interpolate between 
the available tracks.} To determine the evolutionary masses of the 
sample stars (own and adopted), we used a self-developed IDL-routine 
that interpolates between available tracks and isochrones in the 
corresponding diagram.  For the Bonn tracks with \vinit$\approx$300~\kms, 
the reliability of our estimates was checked by comparing to similar data
derived by means of the BONNSAI tool \citep{schneider14}\footnote{The 
BONNSAI masses were inferred using \logl, \Teff, \logg, and \vsini\ as 
observables, and adopting a \citet{salpeter55} initial mass function, a 
Gaussian distribution of \vinit\, with $\mu$=300~\kms and $\sigma$=0.1 
\kms (to be as close as possible to the grid considered by us), a random 
orientation of rotational axes, and a flat age distribution as independent 
priors.}.  Since the two datasets are found to agree within 3 to 4~\Msun, 
which is generally lower than the typical error of our \Mevol determinations 
accumulated from uncertainties in the observationally derived \Teff\ and 
\logl, we conclude that the contribution of our IDL routine to the total 
error budget is rather low, and can be neglected therefore.

\subsection{Evolutionary masses from classical and spectroscopic HR diagrams}\label{hrd_shrd}

Since initial mass, \Minit, fixes the track to which empirical 
stellar properties have to be compared and because of the predicted 
dependence of these
properties on stellar mass, it is especially important to know to
which degree the choice of a particular model grid and diagram might
influence the outcome of a comparison between model predictions and 
observations. In this and the next sections, we elaborate on this
issue in more detail.

Fig.~\ref{fig7} display the classical and spectroscopic HR
diagrams for the sample built using the current Geneva and Bonn
tracks with rotation. From these plots, one can see that the sample
covers an area between 4.46 and 4.69~dex in log\Teff, and between 4.6
and $\sim$6.2~dex in \logl with a deficit of very luminous stars with
hottest and coolest temperatures. The corresponding limits in units of
$ \left [ \Teffe^{4}/g \right ] = \log \left ( \Teffe^{4}/g \right ) -
\log \left ( \Teffe^{4}/g \right )_\odot$ are 3.35 and 4.2~dex,
corresponding roughly to an electron scattering Eddington factor
($\Gamma_{e}$) of 0.1 and 0.4, respectively. Additionally, we also see
that the sample comprises very young objects, located on (and even
before) the zero age main sequense (ZAMS), as well as more evolved 
ones of $\sim$5~Myr age. While the majority of stars are in the 
main-sequence (MS) phase,  there are also others that appear either 
as core hydrogen burning objects close to the end on the MS (in the 
Bonn grids) or as post-MS objects (in the current Geneva grids). 
This result reflects differences in the position of the terminal 
age MS between the selected grids  (for more information, see 
\citealt{castro14}). Another interesting feature to note is that 
according to the Geneva isochrones, our sample dwarfs appear 
systematically younger (by $\sim$1~Myr) than proposed by the Bonn 
isochrons.

We derived two mass estimates, \Mevol(HRD) and \Mevol(sHRD),  for each 
target (own or adopted) based on the diagrams shown in Fig.~\ref{fig7} 
and applying our interpolation routine. The typical uncertainties on 
these estimates, determined by inserting the limits of \logl, \Teff, and 
\loggc, ranges from $\sim$13 to $\sim$25\% for the low and the high mass 
end, respectively. For the FASTWIND targets, the obtained masses are 
listed in Table~A.1, together with their corresponding error. Regarding 
these data, three important features are noteworthy. First, as a consequence 
of their loci at the limits of the area covered by the tracks, for several 
stars it was not possible to derive error bars. A maximum error of 
$\sim$25\% was consistently adopted for these objects (numbers denoted in 
italics). Second, for the two most luminous stars in the sample 
(HD~169582 and CD$-$47\,4551), which are located above the 80\Msun Bonn 
track, the derived \Mevol(HRD) are upper limits. Third,  within each 
of the two grids, the mass estimates derived for the stars with two 
luminosity solutions (HD~94963, HD~94370, and HD~75222) are practically 
identical (within the error). Thus, a mean instead of two individual 
estimates for \Mevol(HRD) is considered to simplify the following
analyses.

\begin{figure}
{\includegraphics[width=8.5cm,height=5.6cm]{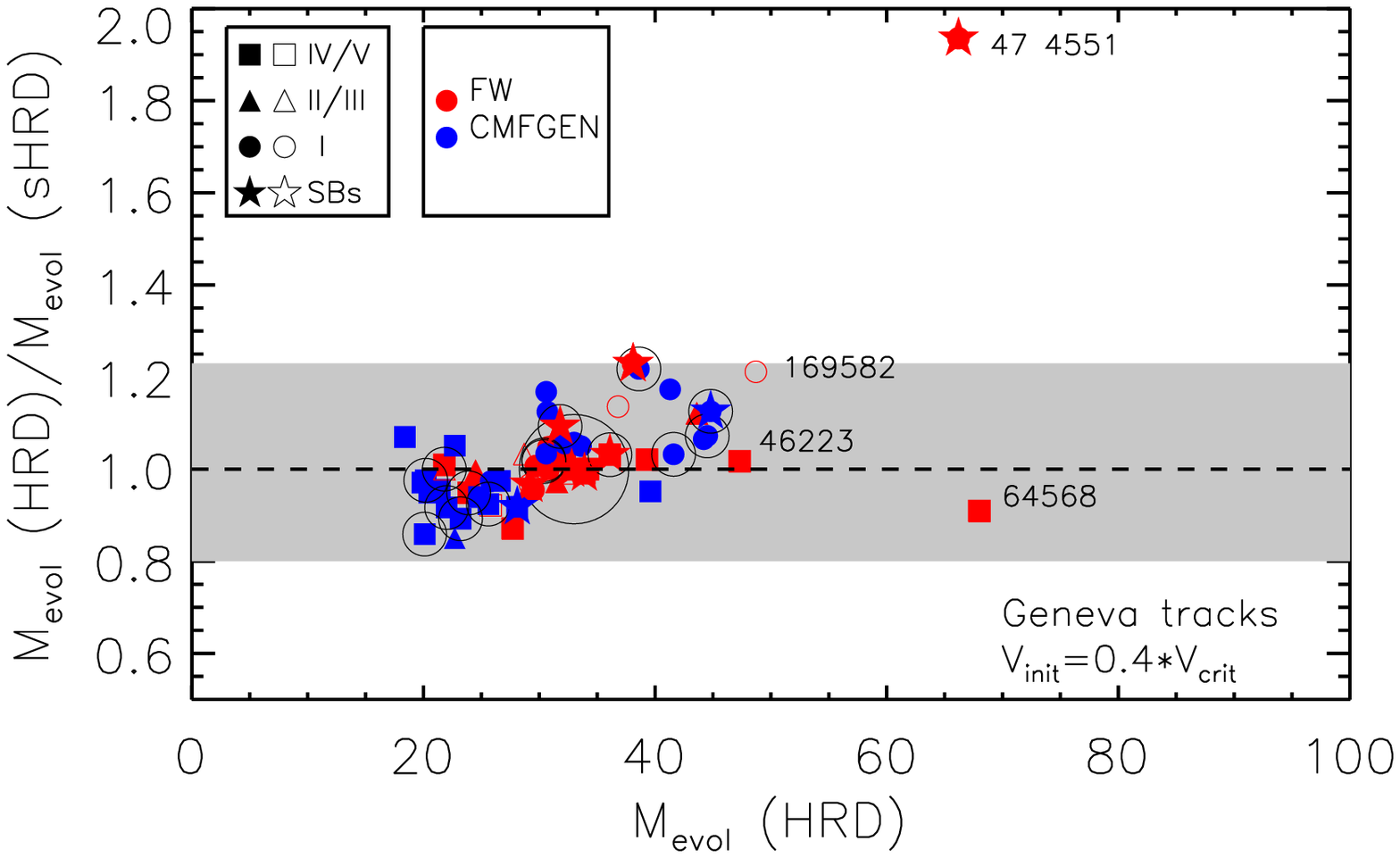}}

{\includegraphics[width=8.5cm,height=5.6cm]{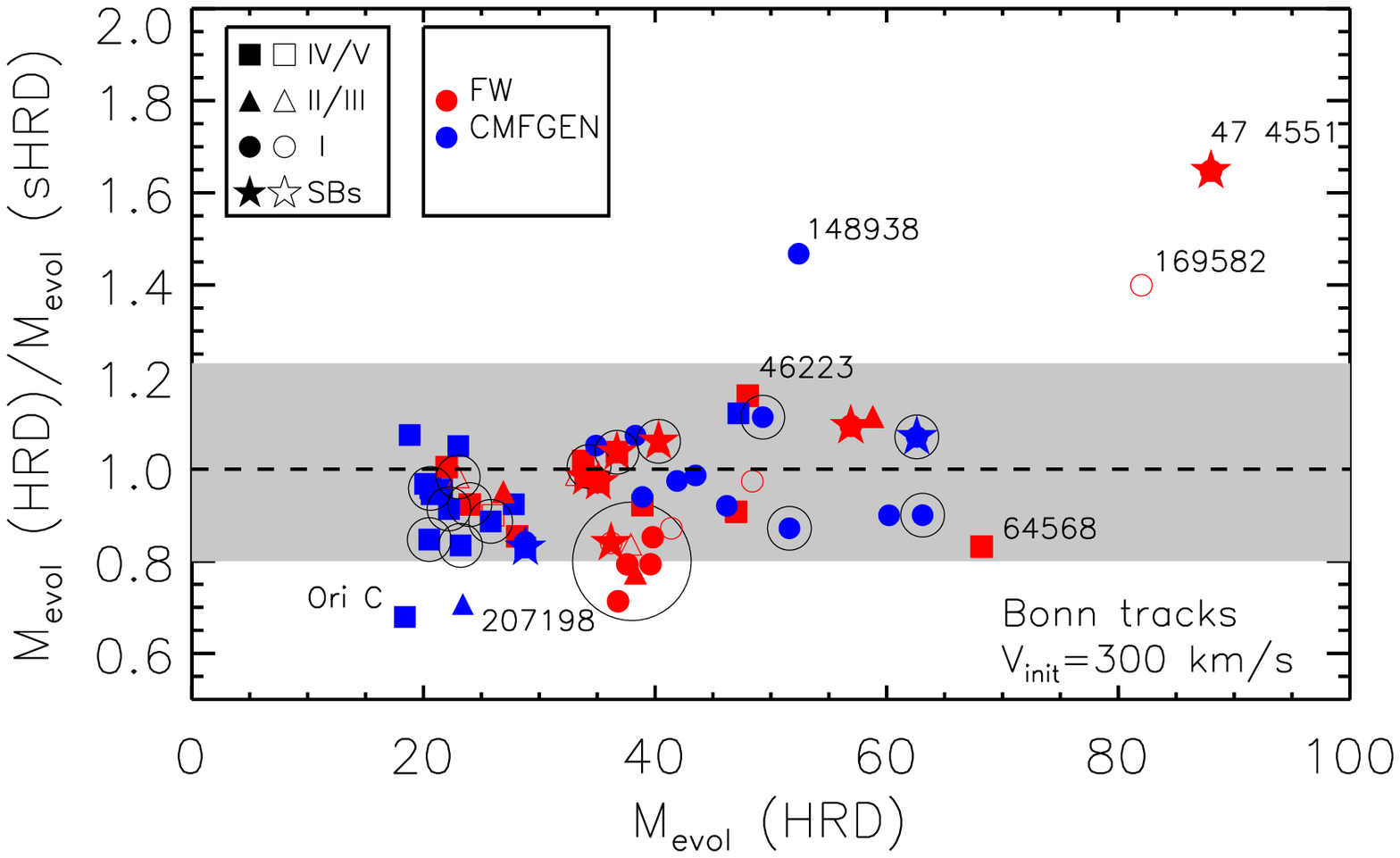}}
\caption{Ratio of evolutionary masses from classical 
(\Mevol(HRD)) and spectroscopic (\Mevol(sHRD)) HR diagrams 
built using the current Geneva and the Bonn tracks with rotation. 
The open symbols indicate field stars and the filled symbols 
indicate cluster and association members. Fast rotators are 
additionally highlighted by large circles. The shaded area 
corresponds to the typical 1$\sigma$ uncertainty at the individual 
mass ratios centred at unity.
For more explanation, see Sect.~\ref{hrd_shrd}.
} 
\label{fig8}
\end{figure}

Careful inspection of Fig.~\ref{fig7} reveals that within a given
model grid, the distribution of the stars in the HRD and the sHRD is
not identical; some data points are moving upwards or downwards
compared to the rest of sample.  There are two most notable examples. 
First, the more massive supergiants  from the CMFGEN subsample, 
which on the classical HRD are distributed between the 40~\Msun 
and the 60~\Msun\ Geneva tracks, while on the corresponding sHRD 
they appear as objects with \Minit\ between 32\Msun and 40\Msun. 
Second, the group of the coolest supergiants from the FASTWIND 
and the CMFGEN subsamples (highlighted by a large circle in Figs.~\ref{fig7} 
to \ref{fig14}), which in comparison to the Bonn tracks appear more 
massive on the sHRD than on the HRD by about 10 (and more) solar 
masses. These results imply that, apart from the used model grid, 
stellar masses might also depend on the kind of diagram used.

To get insight into this important issue, we investigated  the 
ratio of  \Mevol(HRD) to \Mevol(sHRD)  for the sample stars, 
using each of the two model grids.  From the upper panel of 
Fig.~\ref{fig8}, one can see that in spite of the generally good 
agreement between the two mass estimates inferred from  the current 
Geneva tracks with \vinit=0.4\vcrit\ (within $1\sigma$ 
uncertainty at the individual \Mevol(HRD)/\Mevol(sHRD) ratio), 
a systematic component is present. Over the mass range 
probed by the sample, the mass ratio increases gradually from 
values lower than unity (for \Mevol(HRD)$<$30\Msun) via such 
around unity (30\Msun$<$\Mevol(HRD)$<$35\Msun) to values higher 
than unity (for \Mevol(HRD)$>$35\Msun), where in the latter case 
only the giants and supergiants seem to be involved.
  
Analogous results, based on the Bonn tracks with \vinit$\approx$300~\kms,  
indicate good agreement  without any systematic trend for the  
majority of stars with \Mevol(HRD)$>$35\Msun\ and a disagreement 
for those with \Mevol(HRD)$<$30\Msun\ and the aforementioned group 
of the coolest supergiants from the FASTWIND and CMFGEN subsamples 
(data points enclosed by a large circle); in the latter two 
cases the objects tend to appear less massive in the HRD compared 
to the sHRD  typically by about 10\% and 20\%, respectively. 
Regarding the most outstanding outliers indicated by their ID in 
Fig~\ref{fig8},  some of these are from the CMFGEN sample (the 
magnetic star Ori C, HD~207198, and HD~148937); others are from 
the FASTWIND sample (HD~169582 and CD$-$47\,4551).  

Since systematic, distance-, and temperature-dependent errors in 
our results for \logl\ appear unlikely (see Sect.~6.2), an 
evolution different from that of normal single stars and/or 
inadequate physical ingredients implemented in evolutionary model 
calculations appear to be the only alternatives to explain the 
above results.

In a recent study, \citet{LK14} pointed out that close binary
evolution or homogeneous evolution caused by rapid rotation can 
make an object appear overluminous in the sHRD compared to the 
HRD. Close inspection of the data shown in Fig.~\ref{fig8} accounting
for the \vsini\, and the spectroscopic status of the objects however 
indicates that 
none of the fast rotators or the stars recognised or suspected 
to be SB systems show 
\Mevol(sHRD) significantly larger than \Mevol(HRD). This result 
suggests that close binary or homogeneous evolution are not 
likely to play a decisive role in determining the discordance between 
\Mevol(HRD) and \Mevol(sHRD) for any target in the sample.
On the 
other hand, it may well be that the peculiarly high \Mevol(HRD) 
compared to \Mevol(sHRD) (i.e. the reverse situation) derived for 
CD$-$474551 and HD~148937 might be due to present or former binarity, 
respectively (see Appendix~B).

Overall, the main implication of the above results is that the 
employment  of any of the two considered grids to study the 
properties of our sample might lead to  inconsistent (even 
discrepant) results, depending on the used diagram, i.e. 
classical versus spectroscopic HRD (or the KD).  This 
result is consistent with similar findings from \citet{carolina17} 
who report about "a non-negligible number" of O stars in 
the LMC that  appear more massive (by more than 20\%) in the 
KD compared to the classical HRD.

\subsection{Evolutionary masses inferred from Geneva 
and Bonn tracks in parallel}\label{Geneva_Bonn} 

\begin{figure}
{\includegraphics[width=8.5cm,height=5.6cm]{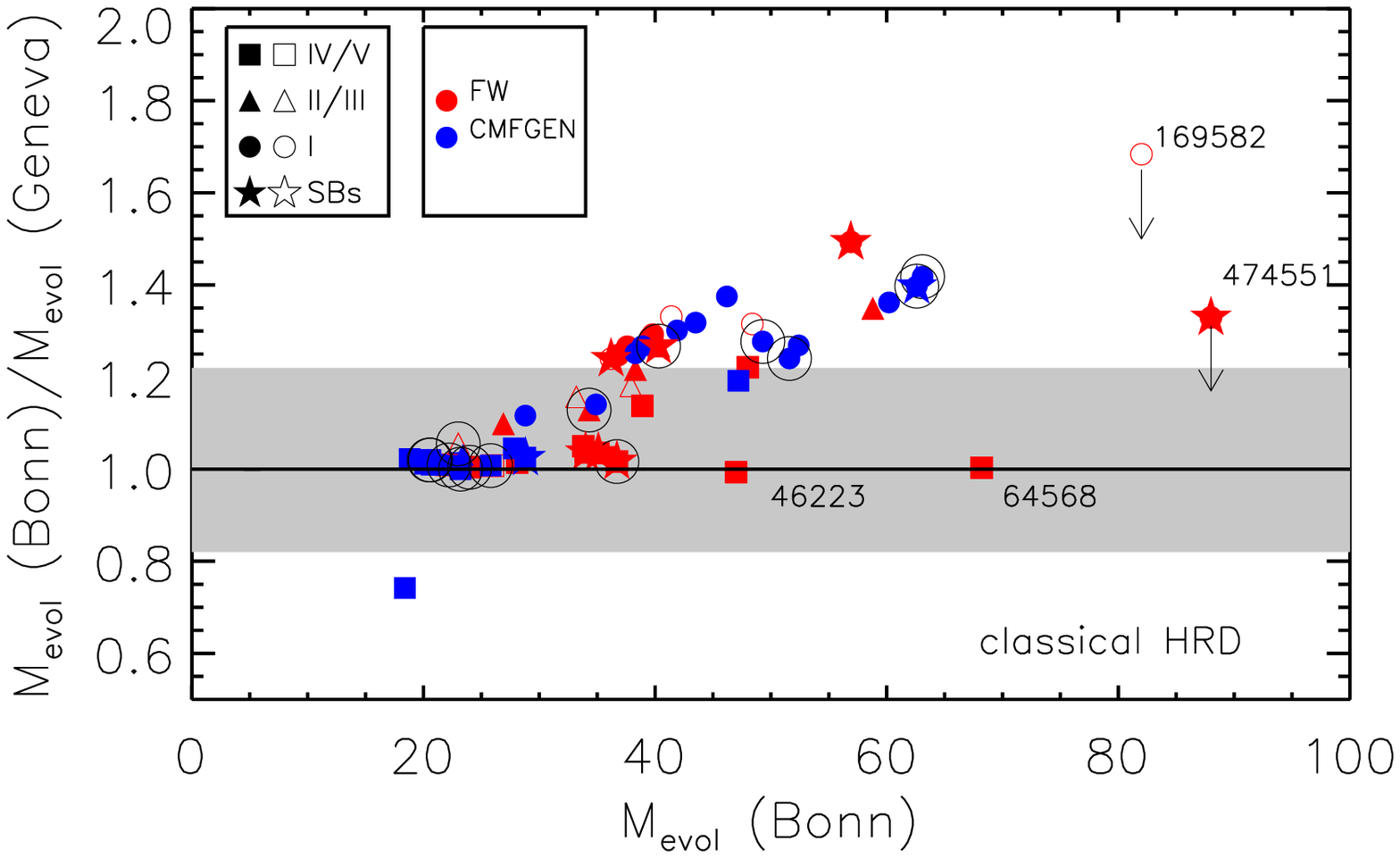}}
{\includegraphics[width=8.5cm,height=5.6cm]{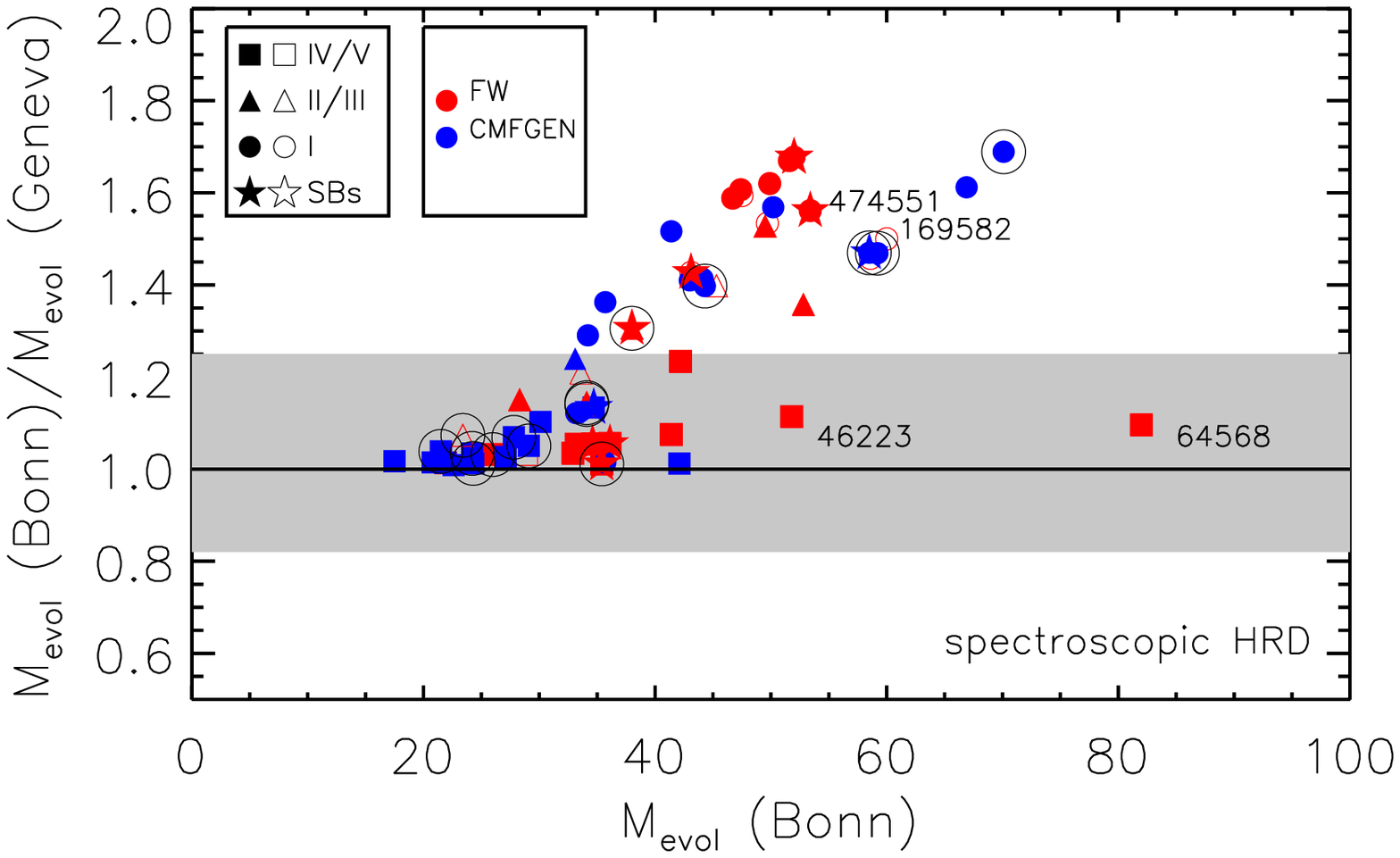}}
\caption{Ratio of evolutionary masses inferred from 
the Bonn and Geneva tracks with rotation using classical 
and spectroscopic HR diagrams.
Same symbols and colour coding as in Fig.~\ref{fig8}.
} 
\label{fig9}
\end{figure}

Recently, \citet{martins15a} noted that owing to differences in
luminosity of the 40~\Msun Geneva and Bonn tracks, the corresponding
mass estimates might differ by up to 25\% beyond the MS. A direct
comparison between evolutionary masses derived for the sample stars
using the same type of HR diagram but different model grids, however,
reveals that similar and even larger differences can appear during 
the MS phase as well. Particularly, our results (see Fig.~\ref{fig9}) 
indicate that  for masses above $\sim$30~\Msun, the use of the 
current Geneva tracks  with \vinit=0.4\vcrit results in stellar 
masses that are systematically lower than those inferred from the 
Bonn tracks for  \vinit$\approx$300~\kms.  The discrepancy is 
more pronounced towards higher masses and later evolutionary stages 
(dwarfs are practically unaffected), and is also stronger for 
\Mevol(sHRD) compared to \Mevol(HRD):  for the highest mass probed 
by the sample (excluding the objects denoted by their ID), the 
deviation reaches about 50\% and 70\% for the HRD 
and sHRD, respectively. 

Several physical ingredients and processes may contribute to explain 
the discrepant evolutionary masses given by the two model grids.
(For a detailed comparison between the input physics and 
its implementation in the Ekstr{\"o}m et al.  and the Brott et al. 
model grids, see the work 
by \citealt{martins14}.)

\subsubsection{Rotation.} 

\begin{figure}
{\includegraphics[width=8.5cm,height=5.6cm]{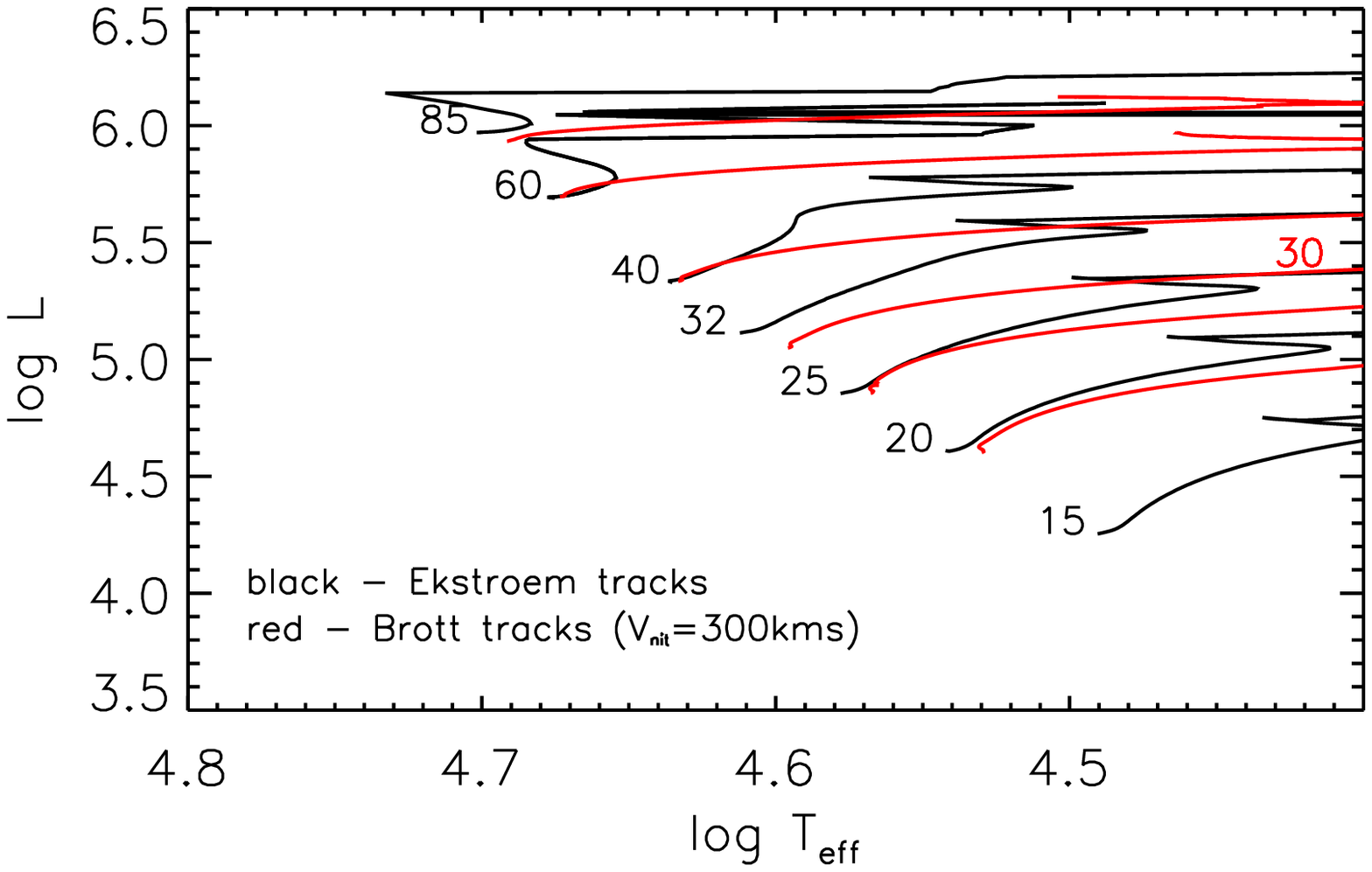}}

{\includegraphics[width=8.5cm,height=5.6cm]{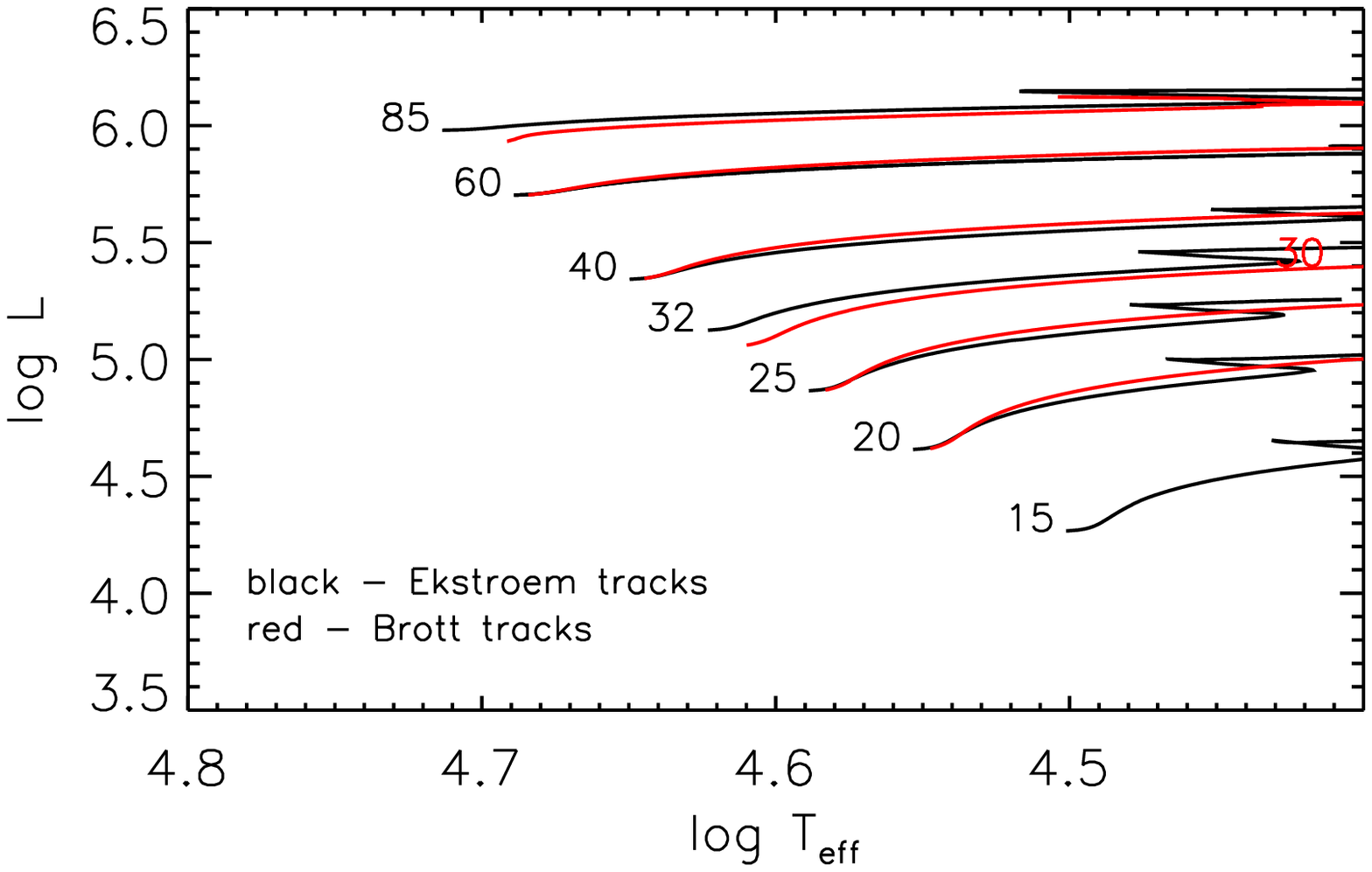}}
\caption{Comparison between evolutionary tracks computed by 
Ekstroem et al. and Brott et al., for Galactic single massive
stars with rotation (upper panel) and without (lower panel). 
Since the current Geneva grids do not include a 30~\Msun\ 
track, the 32~\Msun\ track is shown instead.} 
\label{fig10}
\end{figure}

Some insight into the results outlined above can be
obtained by comparing the rotating and the non-rotating Geneva and
Bonn tracks within the classical HRD, as shown in Fig.~\ref{fig10}.
From these data, one can see that while the non-rotating tracks from the 
two grids are
practically indistinguishable, large discrepancies appear when the
rotating tracks are considered.  The differences in \logl\ are small 
(smaller than $\sim$0.1~dex) for \Minit$\la$30~\Msun, and large for 
\Minit\ beyond this value, where soon after the ZAMS the Geneva
tracks appear systematically more luminous than the Bonn tracks. The
disagreement increases towards higher masses and cooler temperatures,
reaching $\sim$0.25~dex for \Minit = 60~\Msun. Since for a given 
star, the use of more luminous tracks would result in a lower current 
mass estimate than proposed by the less luminous tracks, and since the 
luminosity pattern demonstrated by the rotating Geneva and Bonn 
sequences is qualitatively consistent with the picture shown in the 
upper panel of Fig~\ref{fig9}, we conclude that rotation plays a 
decisive role in inducing the discordant masses given by the two 
model grids.

Because the current Geneva and the Bonn models for \vinit$\approx$300~\kms\ 
rotate at similar velocities on the ZAMS (see Sect.~\ref{gen_uncert}) 
for the same \Minit, differences in the implementation of rotation 
and related issues seem to be most likely responsible for the established 
mass disagreement. 

One such difference refers to the treatment of the effect of mean 
molecular weight barriers: in the Bonn models, these have been 
fully taken into account, whereas in the Geneva models horizontal 
turbulence is thought to limit their effects. A direct consequence 
of these alternative approaches is that at the same \Minit\ and 
almost same \vinit, the former models show very little mixing of 
helium into the radiative envelope during the MS phase, whereas 
substantial mixing of helium occurs in the latter, especially in 
the high mass regime.  
Since luminosity  strongly increases with the average mean molecular 
weight, $\mu$,\footnote{An analytic expression of the form  
L$\propto$M$^{\alpha}\mu^{\beta}$  (where $\alpha$ and $\beta$ 
are positive exponents, decreasing with stellar mass) was found 
to apply for massive stars in the MS phase by \citet{kw90}.} 
(i.e. with increasing average helium mass fraction) as a consequence 
of the mass-luminosity relation, this can explain why the Geneva 
models appear more luminous in the high mass regime; thus leading 
to significantly lower current masses.

A further difference refers to the internal angular momentum 
transport. The Bonn models treat the angular momentum transport 
as for a  diffusive process,  and they also account for internal 
magnetic fields, which is an approach that is more efficient 
than purely hydrodynamic transport mechanisms. The Geneva models, 
on the other hand, include angular momentum advection from the 
meridional circulations, which may transport angular momentum 
from the envelope inwards.  Direct consequences of this 
implementation are that angular velocity differences are small 
in the Bonn models and larger in the Geneva models, and that soon 
after the ZAMS the former rotate faster (at the surface) than 
the latter (see Sect.~\ref {obs_mod_vrot}). Since faster rotation 
is associated with more luminous tracks, one might expect that 
such differences in the internal angular momentum transport might 
contribute to the detected mass discordance as well.

While such expectation is legitimate, our results suggest that the
process is most likely dominated by the different treatment of the
$\mu$ barrier, rather than by differences in the treatment of the
internal angular momentum transport. 

\subsubsection{Mass loss.} 

In recent studies, \citet{MP15} and \citet{keszthelyi16} 
have pointed out that the mass-loss rates resulting from the current Geneva 
models with \vinit=0.4\vcrit\ and the Bonn models with \vinit$\approx$300~\kms  
can differ significantly, although both grids use the same 
mass-loss prescriptions from \citet{vink00}. Within our work, we found 
that for \Minit\ from 25 to 60~\Msun\ and soon after the ZAMS, the 
rotating Geneva models experience a mass-loss rate by about 0.01 
to 0.5~dex larger than displayed by the Bonn models at the same \Minit\ 
and \Teff. Such higher mass loss (because of higher $L$ in the 
rotating Geneva models, see above) is the main (or one of the major) 
contributors to the mass discordance between the two grids compared 
here. Since mass-loss effects accumulate with time and are larger for 
more massive stars, the mass discordance should be largest for evolved 
massive objects, which is nicely confirmed in Fig.~\ref{fig9}.

Mass loss also depends on metallicity \citep{vink01}, and it might be 
speculated that the Geneva models (with a metal mass fraction, 
$Z$ = 0.014) lose more mass than the Bonn models ($Z$ = 0.0088). This, 
however, is not true, since mass loss in the latter models has 
been calibrated to the (solar) iron abundance (see also 
\citealt{keszthelyi16}). 
\begin{figure*}
{\includegraphics[width=8.5cm,height=5.6cm]{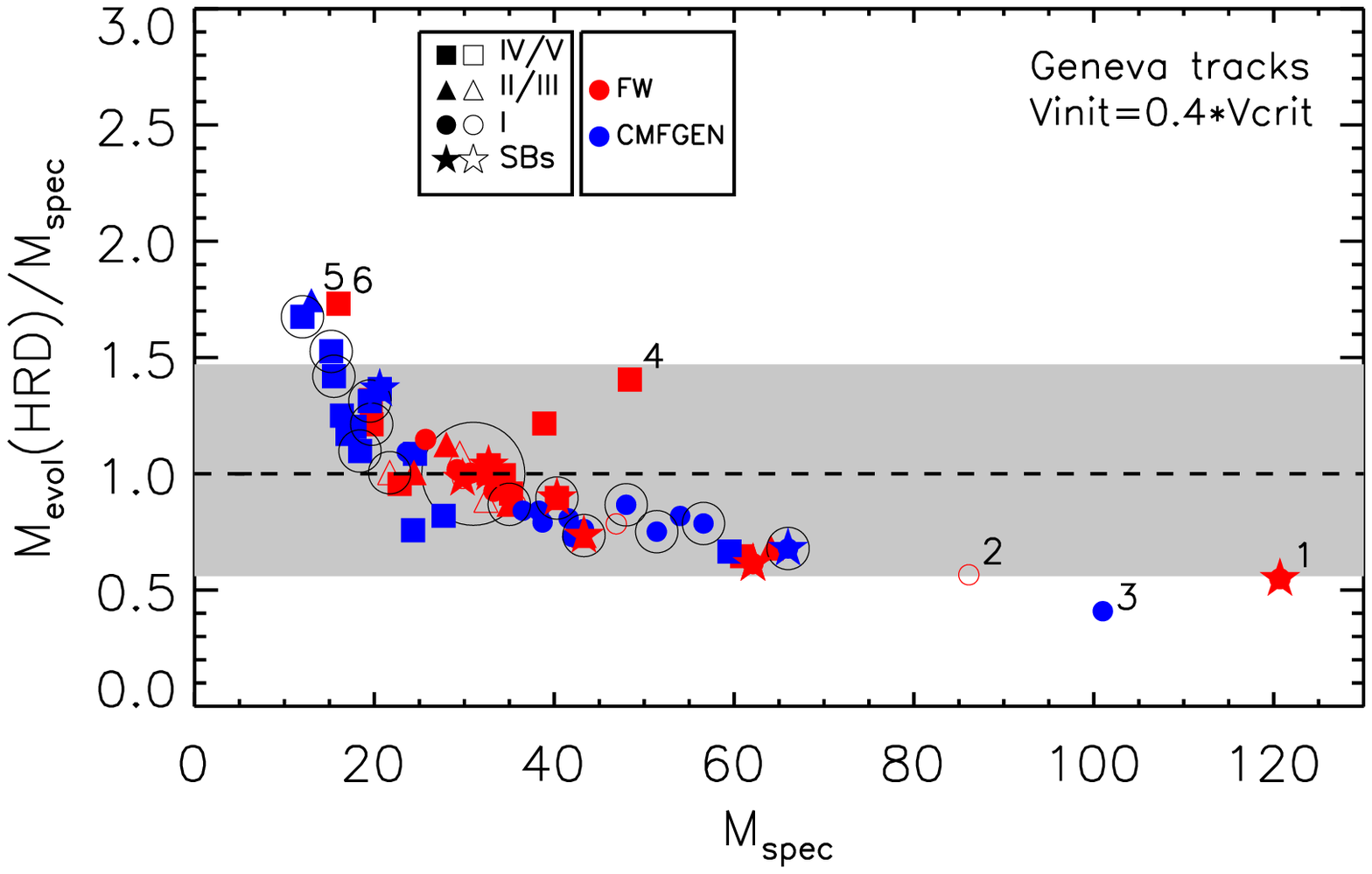}}
{\includegraphics[width=8.5cm,height=5.6cm]{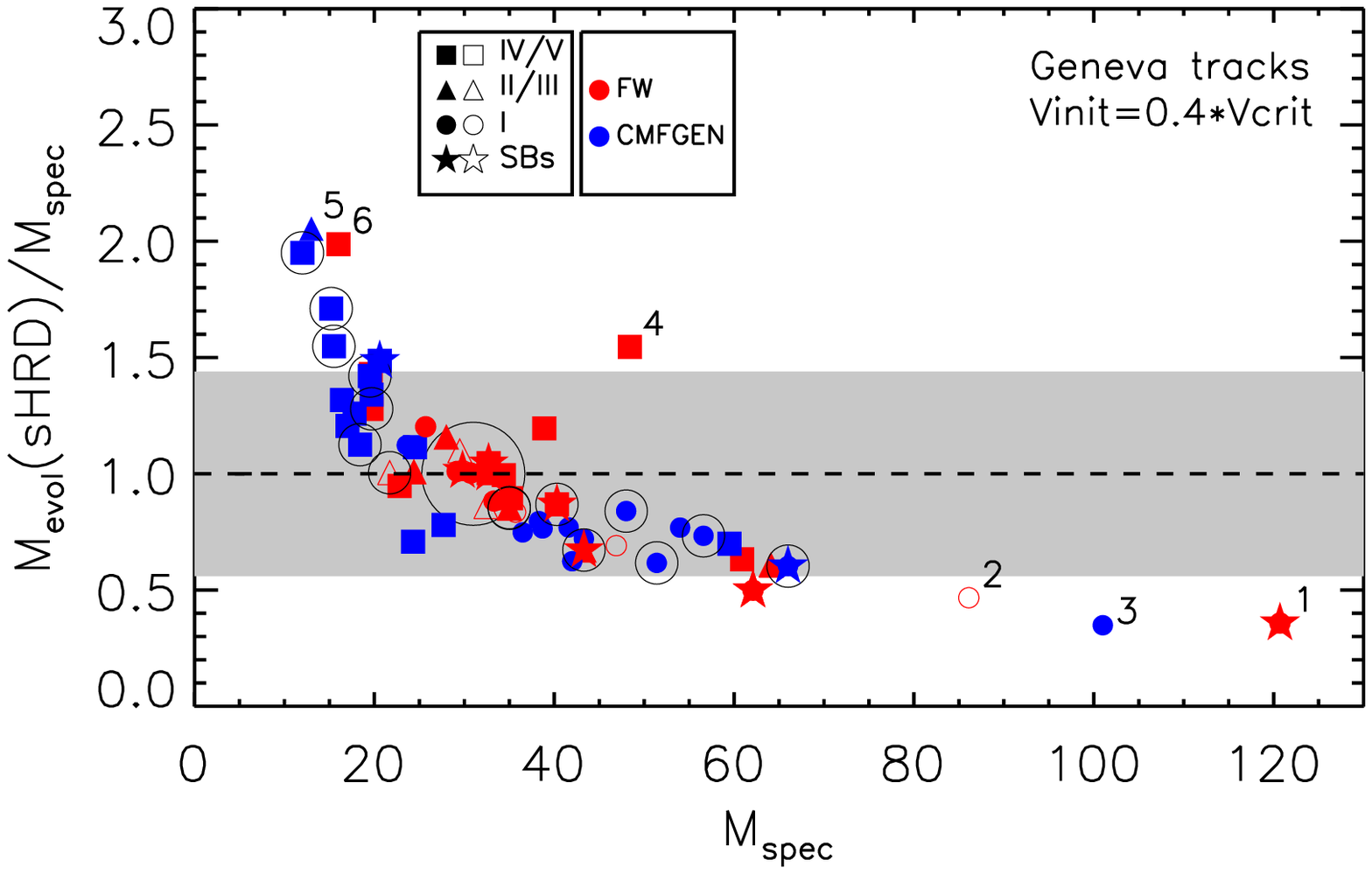}}

{\includegraphics[width=8.5cm,height=5.6cm]{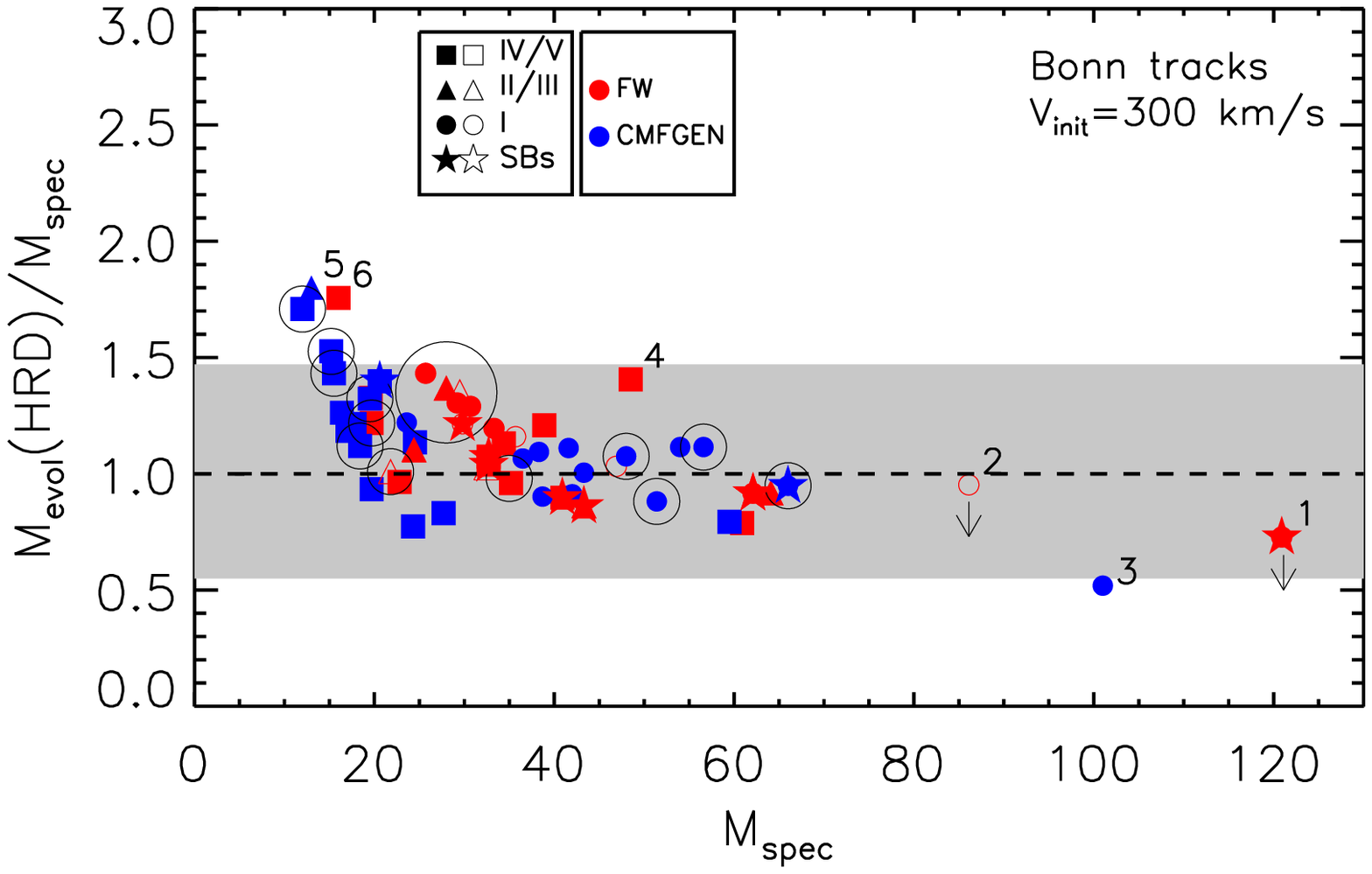}}
{\includegraphics[width=8.5cm,height=5.6cm]{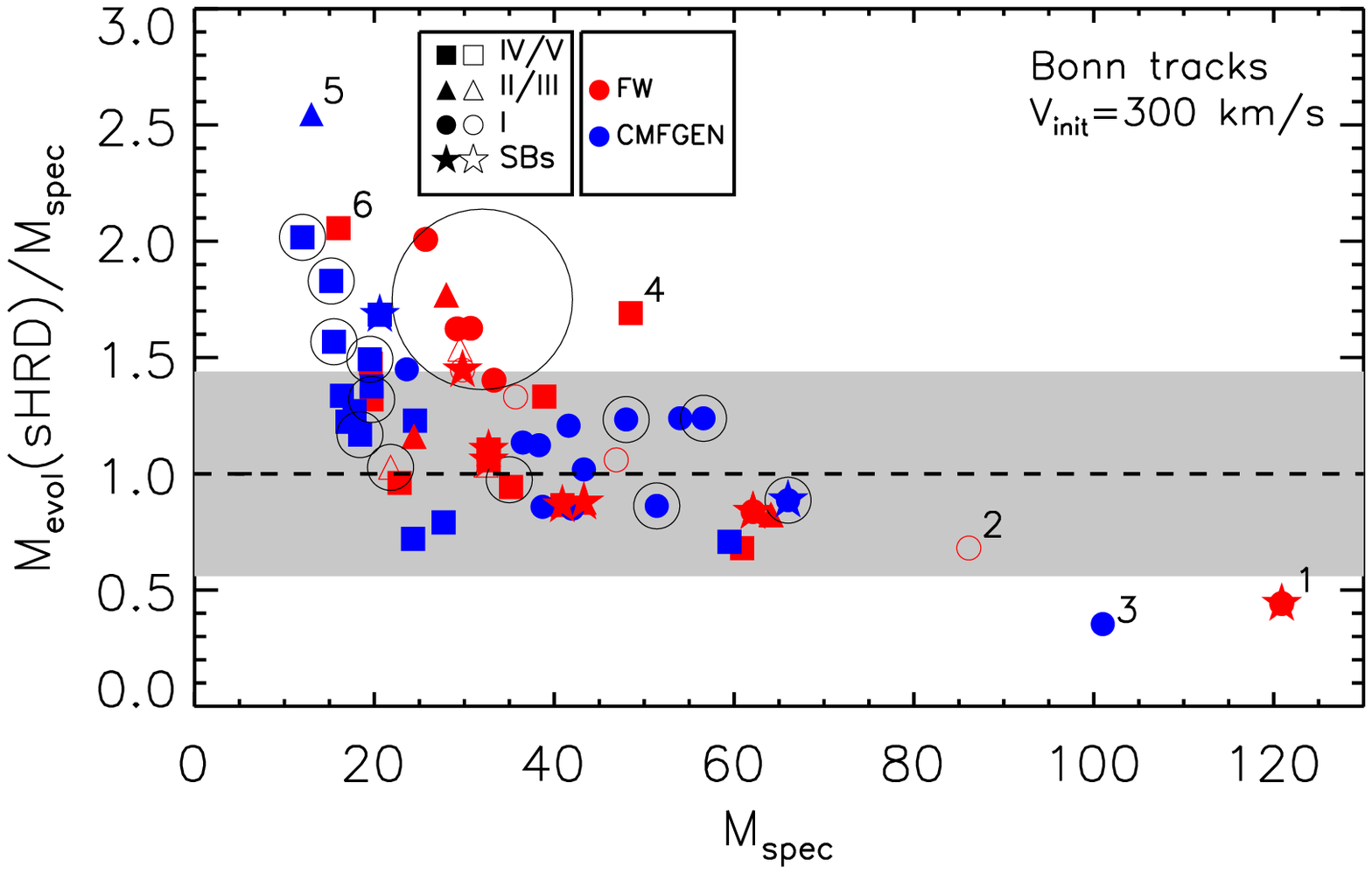}}
\caption{Ratio of  evolutionary and spectroscopic masses 
vs. spectroscopic masses for the sample stars. The former  
is inferred from the Geneva  and Bonn tracks with rotation.  
The dashed line is the one to one relation.  
Same symbols and colour coding as in Fig.~\ref{fig8}.
The most outstanding outliers common to
both grids are denoted  by numbers: No. 1 -- 
CP$-$47\,4551; No. 2 -- HD~169582; No.~3 -- HD~148937; No.~4 -- 
HD~64568; No.~5 -- HD~207198; No.~6 -- CPD$-$58\,2620.}
\label{fig11}
\end{figure*}

\subsubsection{Convection, semi-convection, and overshooting.}  

Generally, 
the extension of the convective regions can be determined using 
either the Schwarzschild (Geneva models) or Ledoux (Bonn 
models) criterion for convection. The major consequence of 
these two approaches is that in the former case a more extended 
convective region would be obtained, and the different region  
is semi-convective in the latter case\footnote{Semi-convection 
occurs when the Schwarzschild criterion for 
convection is fulfilled but the Ledoux criterion not.}. 

The convective core can be additionally enlarged using the so-called 
convective $overshooting$. While Ekstr{\"o}m et al. applied an 
overshoot parameter of 0.1  based on the observed width of the MS 
in their low mass models (\Minit between 1.35 and 9~\Msun), Brott 
et al. used a significantly larger overshoot parameter of 0.3. This 
overshoot parameter was calibrated to adjust the evolution of \vrot\ 
in their 16~\Msun\ model to reproduce the sudden drop in \vsini\ 
at \logg=3.2 observed for massive stars in the LMC. Since larger 
overshooting (Bonn grids) should translate into  a bigger star and 
thus lower gravity and higher luminosity  (for the same \Minit, \Teff, 
and \vinit), this might also lead to a mass discordance between the 
estimates derived from the two model grids. While the non-rotating Bonn 
sequences indeed appear slightly overluminous compared to the current 
Geneva sequences for the same \Minit\ (see Fig.~\ref{fig10}),  the 
differences are small and, in particular smaller (by about a factor 
of 5 to 6) than the typical uncertainty on the derived \logl, and thus 
insignificant within the context of evolutionary mass determinations. 
Based on a set of models computed by means of the MESA code, 
\citet{martins14} came to a similar conclusion.

Summarising, we conclude that while there are other reasons that may 
contribute, such as the specific treatment of convection, semi-convection, 
and overshooting, the problem with the discordant masses inferred from 
the rotating Geneva and the Bonn tracks with \vinit$\approx$300~\kms\ 
is most likely dominated by differences in the treatment of the mean 
molecular weight barriers; this results in models of substantially 
different luminosities and mass-loss rates.

\section{Spectroscopic versus evolutionary masses -- Mass discrepancy}
\label{mass_discrepancy}

\subsection{General comments}\label{mas_dis_general} 

Fig.~\ref{fig11} shows  the ratio  between  the evolutionary 
(\Mevol(HRD) and \Mevol(sHRD)) and  the spectroscopic (\Mspec) 
masses for the sample stars, where the former are inferred from the 
current Geneva and Bonn tracks with rotation. For the CMFGEN 
targets, \Mspec\ was adopted from the corresponding studies; 
for the FASTWIND targets, \Mspec\ was derived using 
corresponding data for \Rstar\ and \loggc\ as listed in 
Table~\ref{para}\footnote{For the normal and slow rotators  
(\vsini$\le$110~\kms, see Paper II), the centrifugal correction 
is  small, i.e. smaller than 0.01~dex; for the fast rotators, this correction can 
be significant, reaching values of up to $\sim$0.15~dex.}. 
The relative error on our \Mspec estimates, accumulated by uncertainties 
in \loggc\ and \Rstar, is $\approx$37\% at maximum\footnote{We note that this 
error does not account for systematic uncertainties in the derived 
\Teff\ and \logg-values, which may appear as a consequence of 
specific methods and approximations used  in the alternative 
atmosphere codes.}.

Several important features become apparent from this figure: 
\begin{itemize}
\item[i)] Despite the generally good agreement between \Mevol\  
and \Mspec\ (within the 1$\sigma$ error bars), suggestive 
evidence for the presence of a mass discrepancy is found, in 
terms of systematic trends and individual targets whose mass ratio 
deviates from unity by about 1$\sigma$ and more (objects marked 
with their IDs). 
\item[ii)] The discrepancy  depends on the model grid used and 
is mass dependent. For the Geneva tracks, a trend towards a
positive (\Mevol$>$\Mspec), negative  (\Mevol$<$\Mspec), and 
neutral (\Mevol$\approx$\Mspec) mass discordance is observed 
for \Mspec $\la$25~\Msun, $\ga$35~\Msun, and between these 
two mass regimes, respectively. For the Bonn tracks, a neutral
mass discordance  is present for \Mspec$\ga$35\Msun 
, whereas  a trend towards a positive mass discordance 
emerges  for  \Mspec smaller than $\sim$35~\Msun.
\item[iii)] Within a given model grid, the mass discordance is 
qualitatively similar but stronger for the sHRD compared to the 
HRD. 
\item[iv)] Fast rotators (\vsini$>$110~\kms) and SBs do not 
demonstrate any peculiarity but appear to follow the trends 
determined by the rest of the sample stars.
\item[v)]  The FASTWIND  and the CMFGEN  targets behave similarly.  
This result suggests that differences in \logg\ derived by 
means of the two codes (if present) are not likely to play 
a decisive role in determining the agreement between \Mevol\ 
and \Mspec. 
\end{itemize}

Additional  insight into the properties of the established mass 
discordance can be obtained if one considers the HR diagram (the 
classical and the spectroscopic one), and the \Mevol(HRD)/\Mspec 
ratio as a function of \logl\ for the sample stars accounting 
for the results outlined in item ii) above (as in Figs. 12, 
13, and 14).  In practice, to construct these figures we divided 
the stars into several \Mspec\ bins, depending on their specific 
mass discordence and used  different colours to represent these 
stars: green for the stars showing a trend towards a positive mass 
discordance (\Mspec <$\sim$25~\Msun\ and <$\sim$35\Msun\ for the 
Geneva and Bonn tracks, respectively), purple for those demonstrating  
a trend towards a negative mass discordance (\Mspec$\ga$35~\Msun 
for the Geneva tracks and \Mspec >80\Msun for the Bonn tracks), and 
magenta for the objects indicating  \Mevol/\Mspec around unity 
without any systematic trend (~25\Msun<\Mspec<~35\Msun for the Geneva 
tracks and \Mspec >35\Mspec for the Bonn tracks except for the three 
most massive targets.)

From these  data we find that in comparison to the Geneva models  
with \vinit=0.4\vcrit, all but one (outlier No.4) of our stars with  
\logl$\ga$5.65 and \Minit$\ge$40~\Msun appear over massive by 
$\sim$20\% to $\sim$50\% and by $\sim$20\% to $\sim$70\% if the 
HRD or sHRD were used to derive their \Mevol. For the objects with 
\logl$<$5.3 and \Minit$<$32~\Msun, on the other hand, a mass discordance 
with \Mspec\ smaller than \Mevol\  by  about 20\% and 29\% for the 
HRD and the sHRD, respectively, is observed. Concerning the stars  in 
the  intermediate mass and luminosity regime, they all demonstrate 
spectroscopic masses that are consistent with the evolutionary masses 
within less than 20\%  independent of the used diagram.

Analogous findings  for the Bonn grid with \vinit of $\sim$300\kms\ 
indicate \Mspec$<$\Mevol by typically 24\% (for the HRD) and 40\% 
(for the sHRD) for  the objects  in the low mass and luminosity regime 
(\logl$<$5.3 and \Minit$\la$30~\Msun),  and \Mevol$\approx$\Mspec for 
the rest of stars in the sample, except for the cooler stars with 
\Minit$\approx$40\Msun (data points enclosed by a large circle). 
These  tend to appear under massive, by about 30\% and 65\% 
(for the HRD and the sHRD, respectively) compared to the models.

\subsection{Possible origin}\label{obs_models}

Discrepant evolutionary and spectroscopic masses can be 
interpreted in terms of large uncertainties in observed stellar 
properties, particularly stellar luminosity and surface equatorial 
gravity, or in terms of inadequate physical ingredients implemented 
or adopted in evolutionary calculations. 

\begin{figure*}
{\includegraphics[width=8.5cm,height=5.6cm]{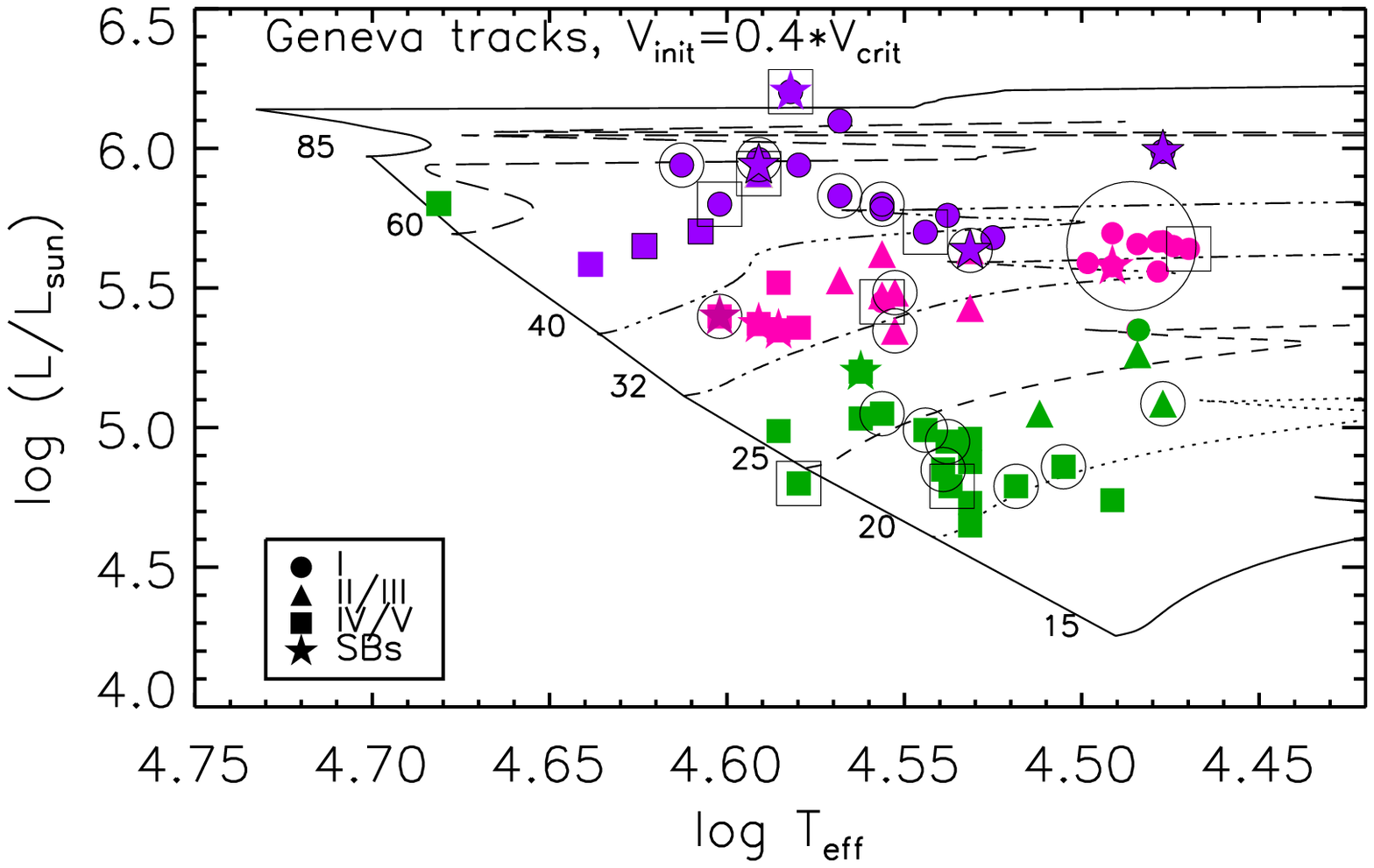}}
{\includegraphics[width=8.5cm,height=5.6cm]{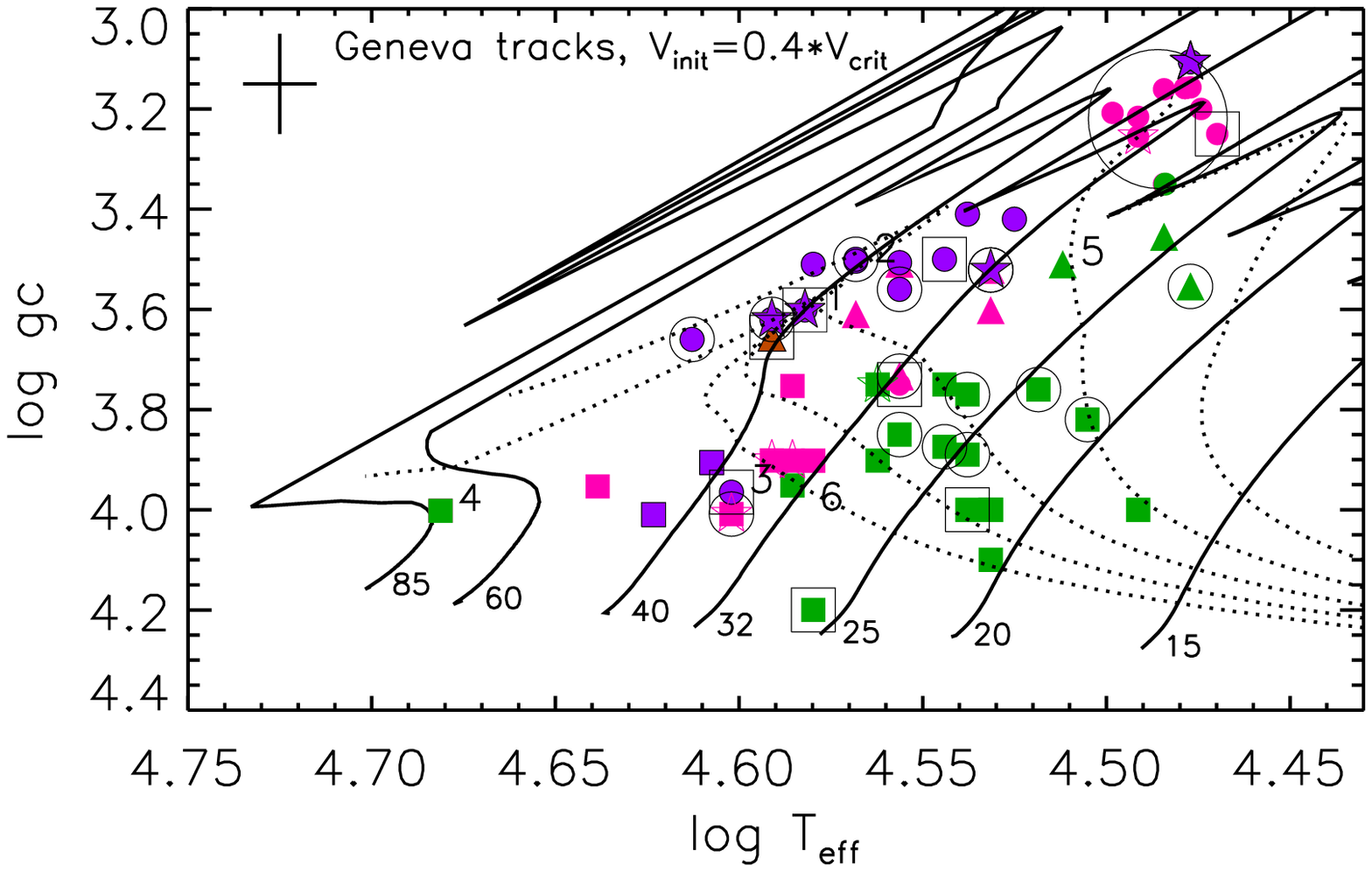}}

{\includegraphics[width=8.5cm,height=5.6cm]{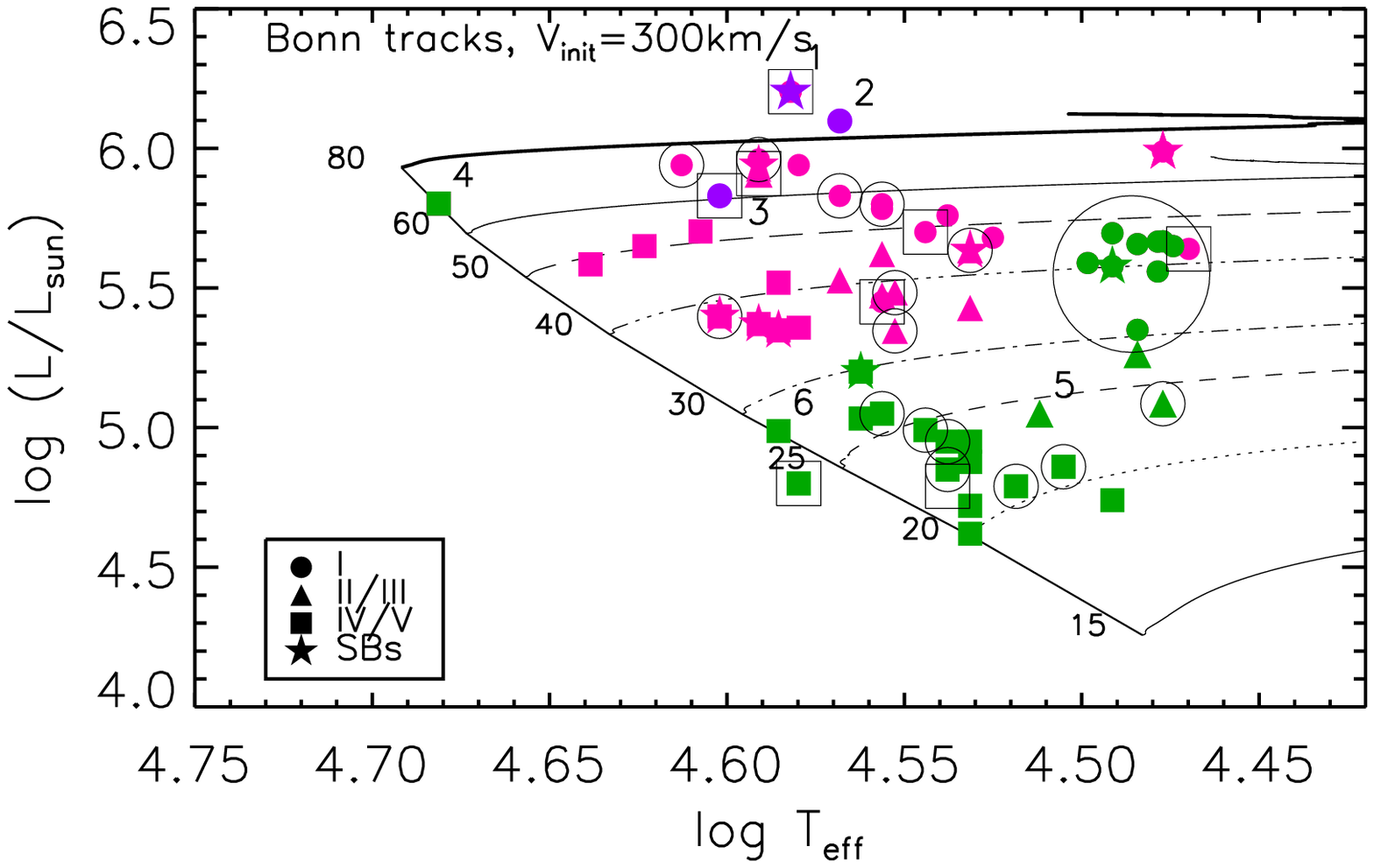}}
{\includegraphics[width=8.5cm,height=5.6cm]{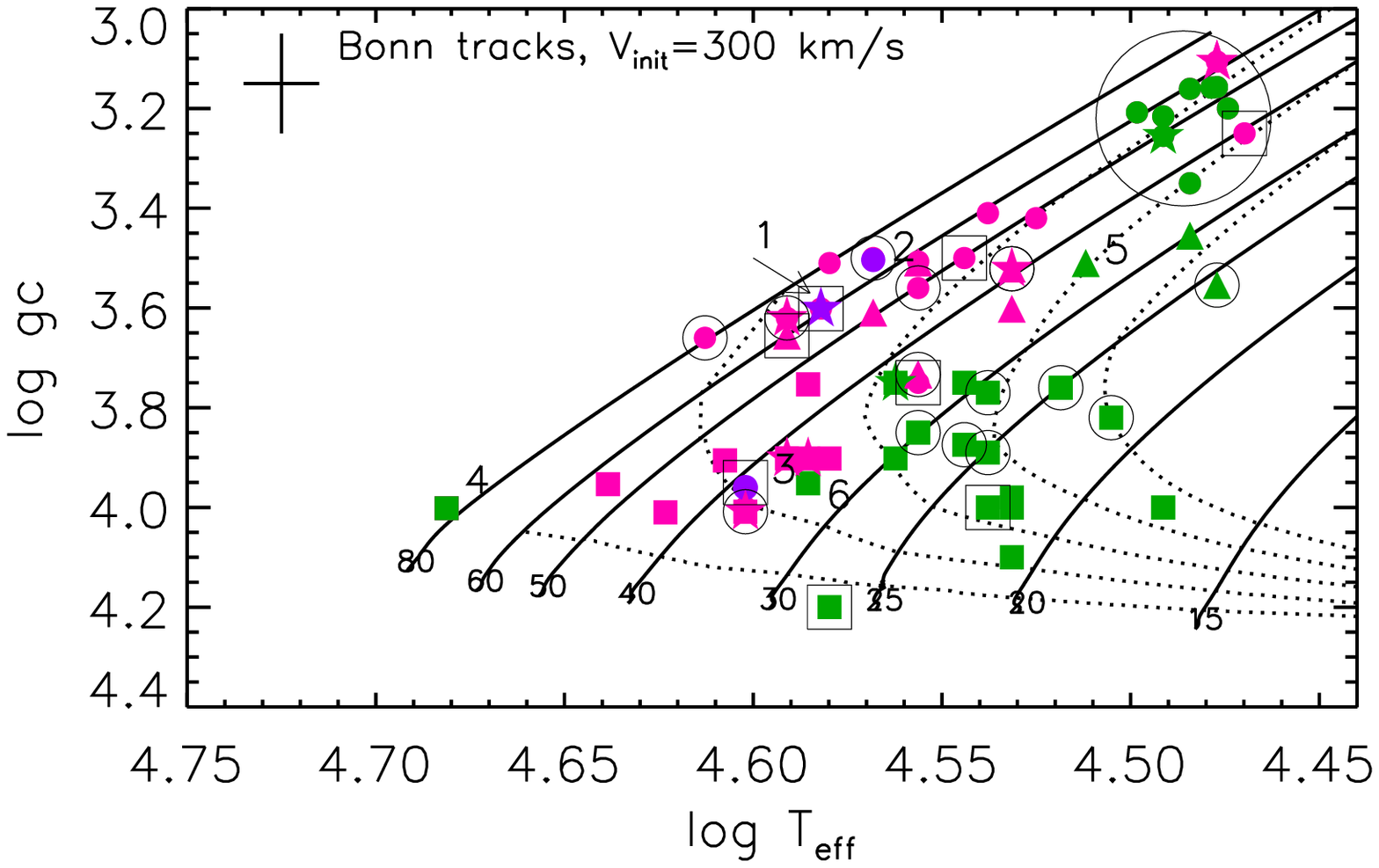}}
\caption{Classical HR (left) and  KDs (right) for sample stars, 
plotted against current Geneva and Bonn tracks with rotation. 
Symbol coding as indicated in the left plots; fast rotators and 
magnetic stars are additionally highlighted by large circles and 
squares, respectively. Objects showing a trend towards a positive 
(\Mevol>\Mspec) or  a negative (\Mevol<\Mspec) mass discordance 
are denoted in green and purple, respectively; those indicating  
\Mevol/\Mspec around unity without any systematic trend  in magenta.
Dotted lines represent  corresponding isochrones 
at 1 to 5~Myrs. Numbers from 1 to 6 denote the outliers as 
discussed in Sect.~\ref{mas_dis_general} and Appendix B. 
For more information see Fig.~\ref{fig11} and Sect.~\ref{obs_mod_logg}.
} 
\label{fig12}
\end{figure*}

Regarding the objects whose mass ratio  deviates most from 
unity (indicated by numbers from 1 to 6 in Figs.~\ref{fig11} to 
\ref{fig14}), independent of the model grid used, we suggest  
that large uncertainties in the observational parameters rather 
than inadequate model predictions might be responsible for 
their peculiarity (see Appendix A.2.).

\subsubsection{Stellar luminosity}

While systematic, distance- and temperature-dependent errors in 
our results for \logl\, appear unlikely\footnote{These data 
originate from two different model atmosphere codes that use 
different assumptions and approximations, and different 
methods and approaches to determine surface luminosities.} there 
are, at least, three lines of reasoning which suggest that the
established mass problem (in this but also in Sects. \ref{hrd_shrd}) 
should not be related to large errors in \logl\ caused by uncertain 
distances and/or reddening. Indeed, if errors in \logl\ were 
responsible for the mass discordance, then 1) field stars should 
behave differently from members of cluster and associations; 
2) the discrepancy should be qualitatively different if \Mevol(sHRD) 
instead of \Mevol(HRD) were considered\footnote{\Mevol(sHRD) 
is almost independent of \Rstar, while \Mevol(HRD) is directly 
related to it.}, and 3)  the adopted distance should systematically 
and significantly deviate from the values inferred from the GAIA 
parallaxes, that is at least for the stars that are members 
of clusters and associations. Since none of these items are confirmed by our results, 
it seems rather unlikely that such errors could lead to the mass 
patterns visible in  Figs.~\ref{fig8} and \ref{fig11}, although 
large uncertainties in \logl\ for individual objects cannot be 
excluded, of course.

\subsubsection{Surface gravities}
\label{obs_mod_logg}

In Fig.~\ref{fig12}, the classical HRD and the KD of the sample 
is plotted against the current Geneva tracks 
with rotation and the Bonn tracks for \vinit$\approx$300~\kms.  
To guide the eye and facilitate the comparison, different 
colours have been used to denote the objects that  experience 
different kinds of a mass discordence (see Sect.~\ref{mas_dis_general}). 

In an observational context, the data shown on the right reveal 
that sample dwarfs and subgiants (squares) with \loggc\ ranging 
from $\sim$4.1 to $\sim$3.75~dex are clearly separated from the 
normal giants and supergiants (circles), and only one object 
indicates \loggc=4.2~dex. For \Minit(KD)\ between 20 and 40~\Msun, 
this range is independent on stellar mass; for \Minit(KD)$\ge$40~\Msun, 
no constraint can be provided because of the limited number of very 
massive stars close to the ZAMS. Analogous findings for normal and bright 
giants indicate a lower limit of $\sim$3.4~dex for 
20~\Msun$<$\Minit(KD)$<$25~\Msun, 
and of $\sim$3.6~dex for 30~\Msun$<$\Minit(KD)$<$40~\Msun. For the 
supergiants, the corresponding values range from $\sim$3.6 for the
highest masses probed, to $\sim$3.2 for the most evolved objects with
masses of $\sim$32~\Msun (Geneva tracks) and $\sim$40~\Msun (Bonn
tracks). These findings are quantitatively consistent with similar
results from previous studies (e.g. \citealt{martins05a}, \citealt{MP06}, 
and \citealt{martins15b}), and thus also confirm the reliability of 
our \loggc\ determinations for the FASTWIND targets.

Accounting for the loci of the targets in the classical HRD, there 
appears to be  a rough  correspondence between observed and 
predicted \loggc\  values for the objects in the low mass and luminosity 
regime (data points in green): on both diagrams, these 
targets are generally located between the 20~\Msun\ and 32~\Msun\ 
Geneva tracks and between the 20~\Msun and 30~\Msun\ Bonn tracks. 
Closer inspection of the data however revealed that in the KD many 
of these targets 
tend to appear more massive and more evolved, and thus with \loggc\ 
lower than expected for the measured \Minit(HRD) by about 0.05 
to 0.15 ... 0.20~dex.

Regarding the objects in the high mass and luminosity regime, and 
in comparison to the Geneva tracks, some of these (magenta data points) 
show consistent \logg\, values, while others (purple data points) 
disagree, indicating gravities by about 0.2~dex larger than predicted. 
Compared to the Bonn tracks, however, we find a good correspondence for 
the majority of stars in this mass regime  (magenta data points) except 
for the objects  enclosed by a large circle, which display \loggc\  
values lower than predicted for their \Minit(HRD) by about 0.10 to 
0.15~dex, and the three mass outliers CPD$-$47\,4551 (No.~1), HD~169582 
(No.~2), and HD~148937 (No.3), which appear less massive in the KD compared 
to the HRD (data points in purple).  

One way to interpret the disagreement between observed and predicted 
\loggc\  in the low mass regime  is to assume that the former  
might have been systematically underestimated by the model atmosphere
analysis. Since the FASTWIND and CMFGEN targets are equally 
involved, our results would then imply that owing to some common 
deficiency both codes should provide gravities lower than the real 
gravities. A potential candidate to play this role is microturbulence. 

Although the origin of microturbulence is still debated for massive
stars (wind velocity fields: \citealt{kudri92}; subsurface convection
zones: \citealt{cantiello09}, \citealt{grassitelli15a}, 2015b;
combination of pulsations: \citealt{townsend07}; or the consequence of
a still rather simplified treatment of the complex stellar physics
involved in model atmosphere calculations), and $if$ this parameter is
not a fiction but a real physical quantity, it should result in a
turbulence pressure term in the hydrodynamic and quasi-hydrostatic
description leading to increased surface gravities in spectroscopic
analyses. Since neither CMFGEN nor FASTWIND accounts for this
possibility, the equatorial surface gravities derived by means of
these codes, at least at actual analyses, might be somewhat 
underestimated \footnote{For  a star 
with \Teff=40~kK, a microturbulence of 15 to 20~\kms would increase 
the value of \logg\ by $\sim$0.1 to 0.15~dex if accounted for 
as a pressure term, which would be roughly enough to reconcile 
model predictions and observations even in the more prominent cases.}.
\begin{figure*}
{\includegraphics[width=8.5cm,height=5.6cm]{fig12a.ps}}
{\includegraphics[width=8.5cm,height=5.6cm]{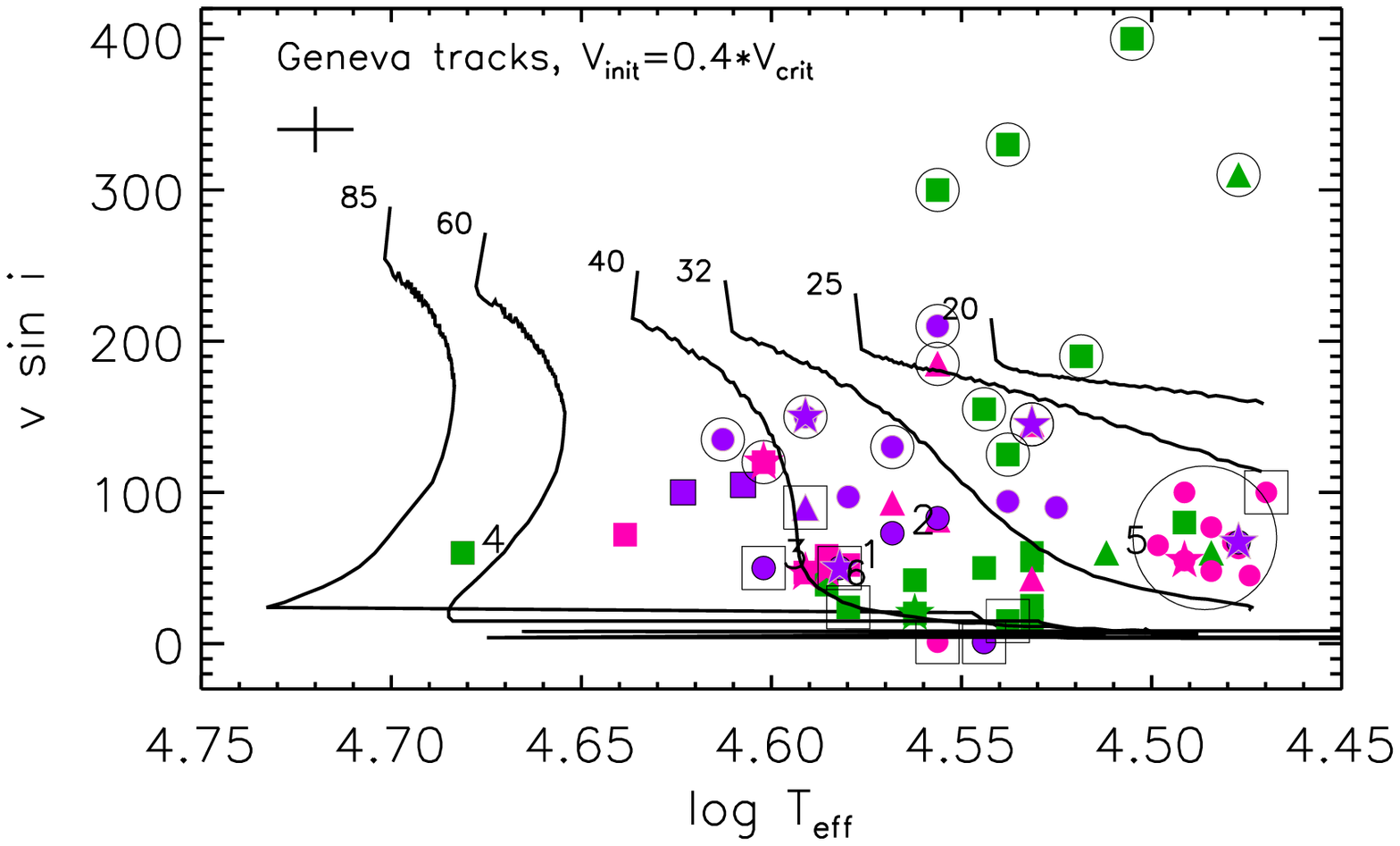}}

{\includegraphics[width=8.5cm,height=5.6cm]{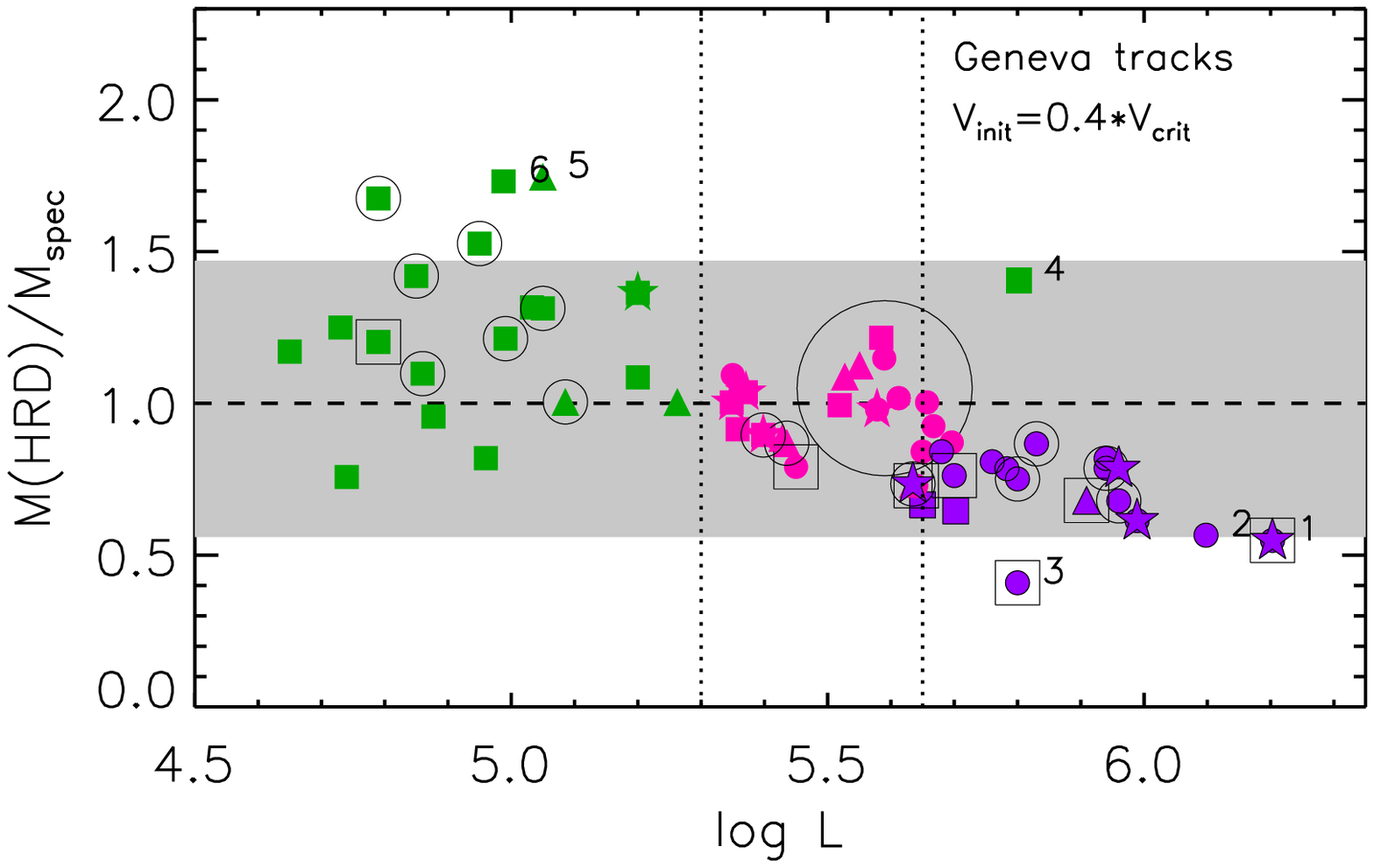}}
{\includegraphics[width=8.5cm,height=5.6cm]{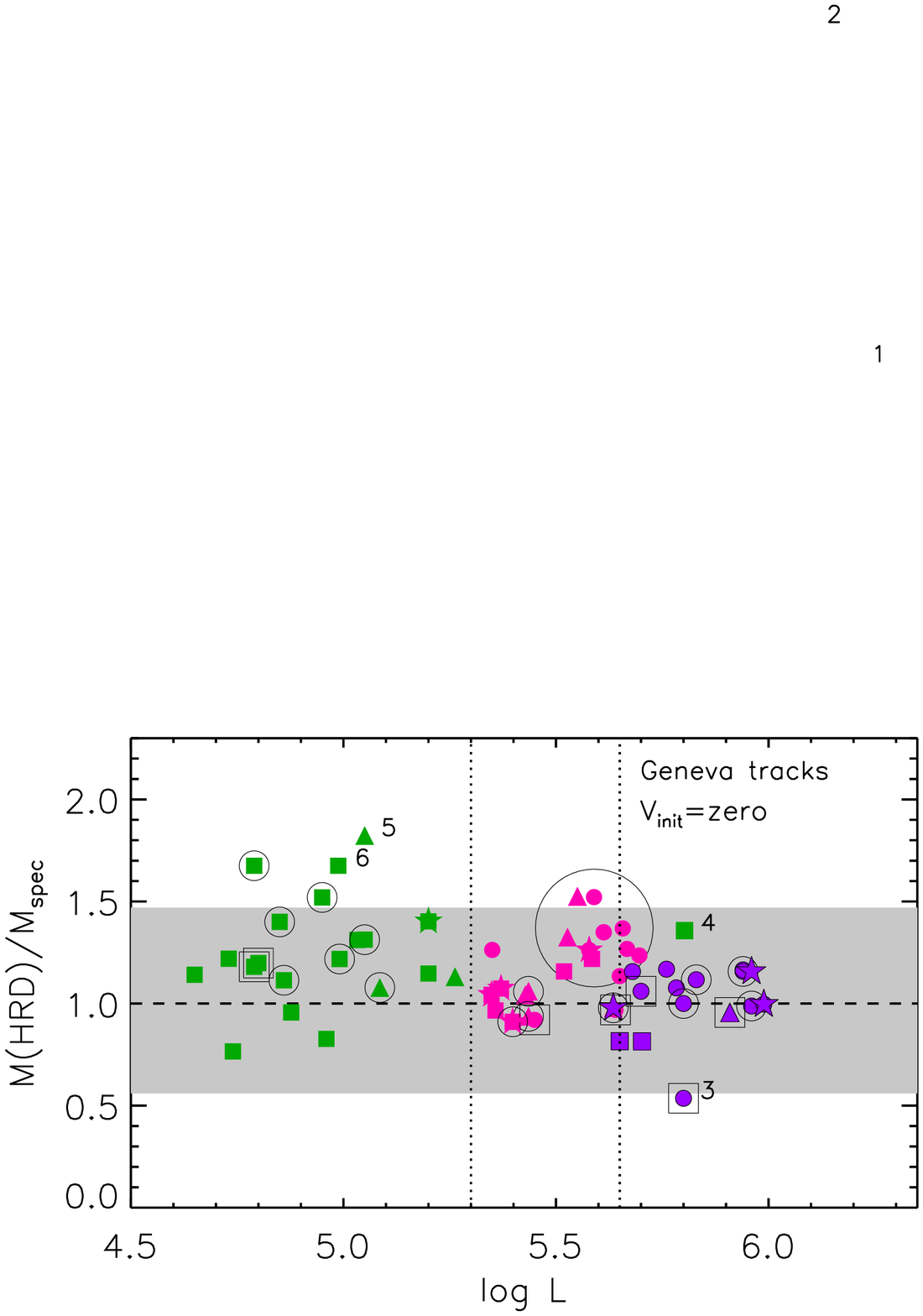}}
\caption{{\it Upper panels}: Classical HRD (left) and the 
\vsini  distribution as a function of log~\Teff (right) for 
the sample stars plotted against the current Geneva grids. 
The observed \vsini values account for the effects of 
macroturbulence; the model \vrot\ was scaled by $\pi$/4 
to account for (average) projection.  
{\it Lower panels}: Ratio of evolutionary and spectroscopic 
mass as a function of \logl, after using the rotating and
non-rotating Geneva tracks to calculate \Mevol(HRD). The shaded 
area corresponds to the typical 1$\sigma$ uncertainty in the 
individual mass ratios centred at unity. 
Vertical dashed lines indicate the luminosity bins as discussed 
in Sect.~\ref{mas_dis_general}.  Numbers from 1 to 6 denote the 
outliers as discussed in Sect.~\ref{mas_dis_general} 
and Appendix B. Symbols and colour coding  as in Fig.~\ref{fig12}.
} 
\label{fig13}
\end{figure*}

There might be, at least, one potential problem  with this hypothesis. 
If underestimated surface gravities caused by the neglect of 
microturbulent pressure in stellar atmosphere codes were responsible 
for the \loggc\ problem in the low mass regime, similar, and even 
larger discrepancies should be present in the high mass regime where 
a stronger influence of \vmic\ is expected  (see e.g. \citealt{cantiello09}, 
\citealt{massey13}, \citealt{markova14}, and references therein). Such 
a discrepancy, however has not been revealed by our analysis.

A possible way to overcome this difficulty is to assume that in 
the high mass  regime (\Minit$\ge$40~\Msun), the effect of 
neglected microturbulent pressure on \loggc\ might have been 
hidden by some other process(es), which also contribute(s), but 
in the opposite direction. Since in this mass regime, the evolution 
of surface gravity  is governed by rotation and mass loss, our 
results would then imply that the effects of faster rotation and 
eventually higher \Mdot\ should either compensate (for the Bonn 
tracks) or overcompensate (for the Geneva tracks) the effects of 
neglected microturbulent pressure. This would then lead to the 
present situation, where for the Bonn tracks 
\loggc(model)$\approx$\loggc(obs), while for the Geneva
tracks \loggc(model)$<$\loggc(obs).

\subsubsection{Inadequate rotational rate}\label{obs_mod_vrot}

Recently, \citet{martins14} have warned that the current Geneva 
models should be used with a good recognition that they are relevant 
for fast rotating objects only. With this caveat in mind, in the upper 
panels of Fig.~\ref{fig13} we plotted the distribution of the sample 
in the classical HRD and the \vsini--\Teff diagram using the same 
symbols and colours as in Fig.~\ref{fig12}.  The current Geneva  
predictions for \vinit=0.4\vcrit\ are overplotted. 

From the upper right panel one can see that our sample consists 
of stars with \vsini\, ranging from close to zero up to 
$\sim$400~\kms. About 68\% of these stars have \vsini$<$100~\kms:
23\%  rotate at \vsini\, between 100~\kms to 200~\kms and the 
rest are fast rotators with \vsini$>$200~\kms. Analogous results 
for 116 O stars in the MW studied by  \citet{SH14} indicate that 
63\%, 15\%,  and 22\% of the total sample are distributed in 
the same velocity bins as considered for our sample\footnote{We 
compare to \citet{SH14} because, similar to us, these authors 
consider macroturbulent broadening  in addition to the rotational 
broadening  while studies, such as \citet{penny96}, \citet{penny04}, 
and \citet{wolff06} do not account for the effects of macroturbulence.}. 
From this comparison it follows that our sample may lack objects with 
very fast rotation but appears to be representative for O-type  stars with
\vsini\, lower than 200~\kms.   
\begin{figure*}
{\includegraphics[width=8.5cm,height=5.6cm]{fig12c.ps}}
{\includegraphics[width=8.5cm,height=5.6cm]{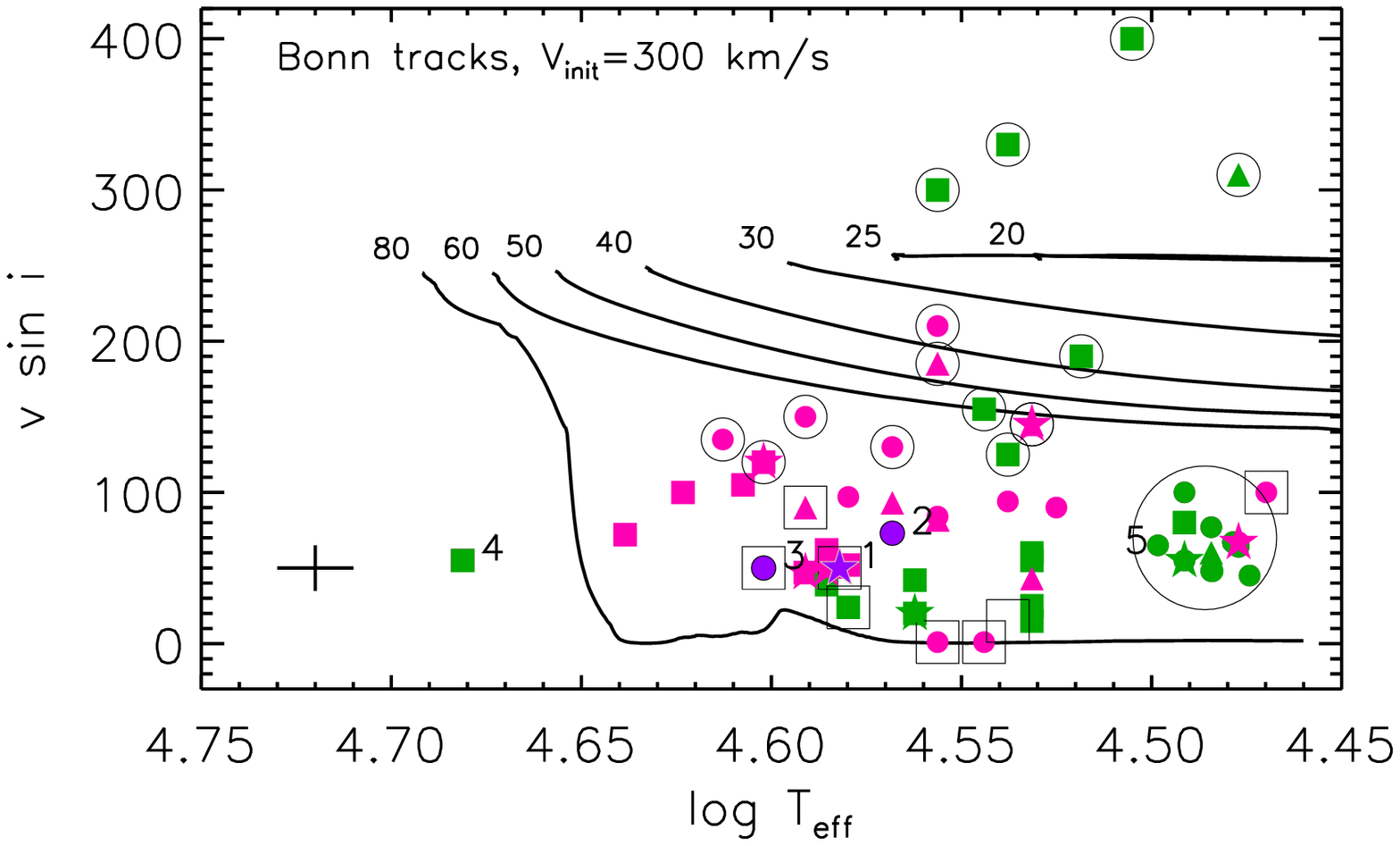}}

{\includegraphics[width=8.5cm,height=5.6cm]{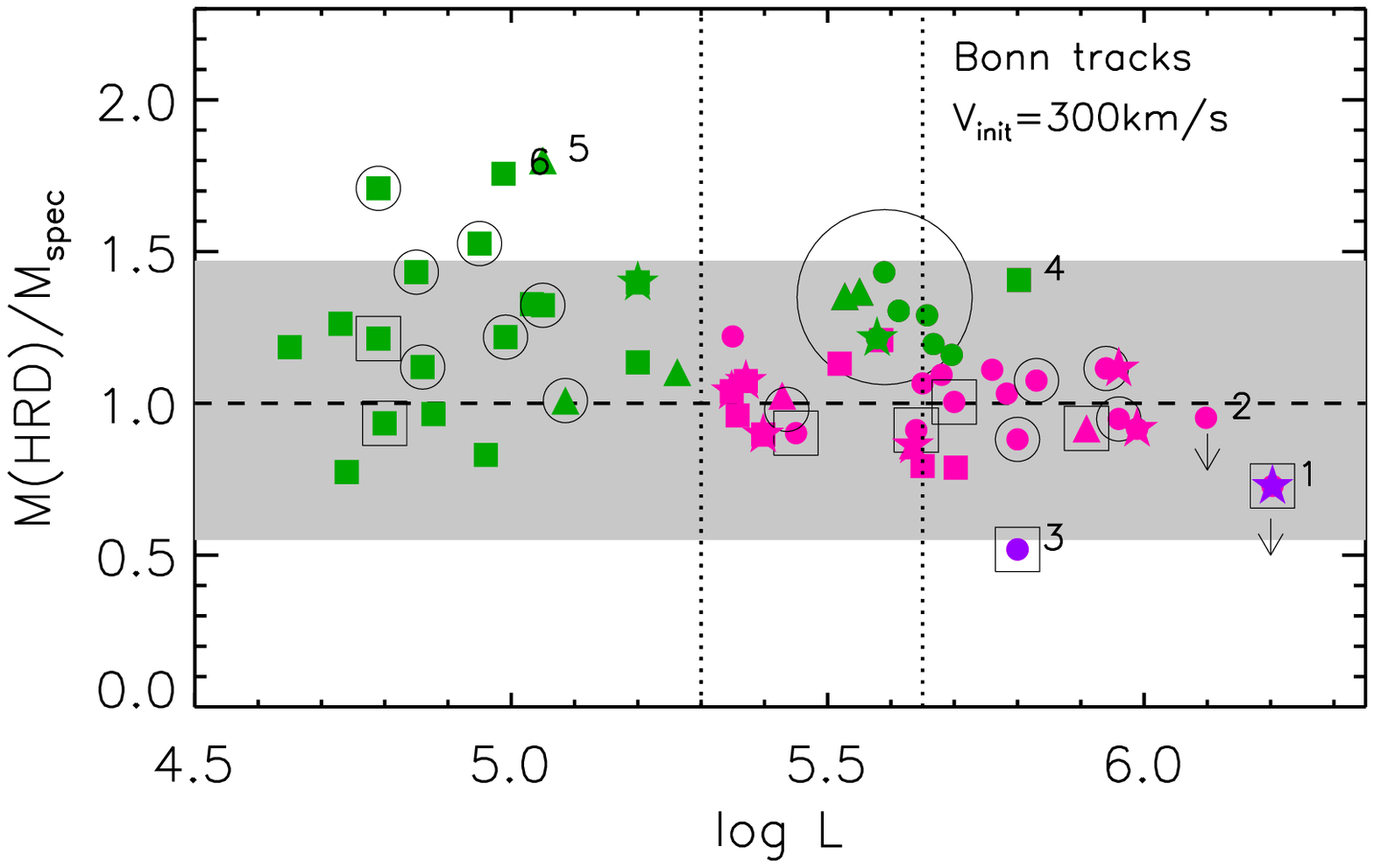}}
{\includegraphics[width=8.5cm,height=5.6cm]{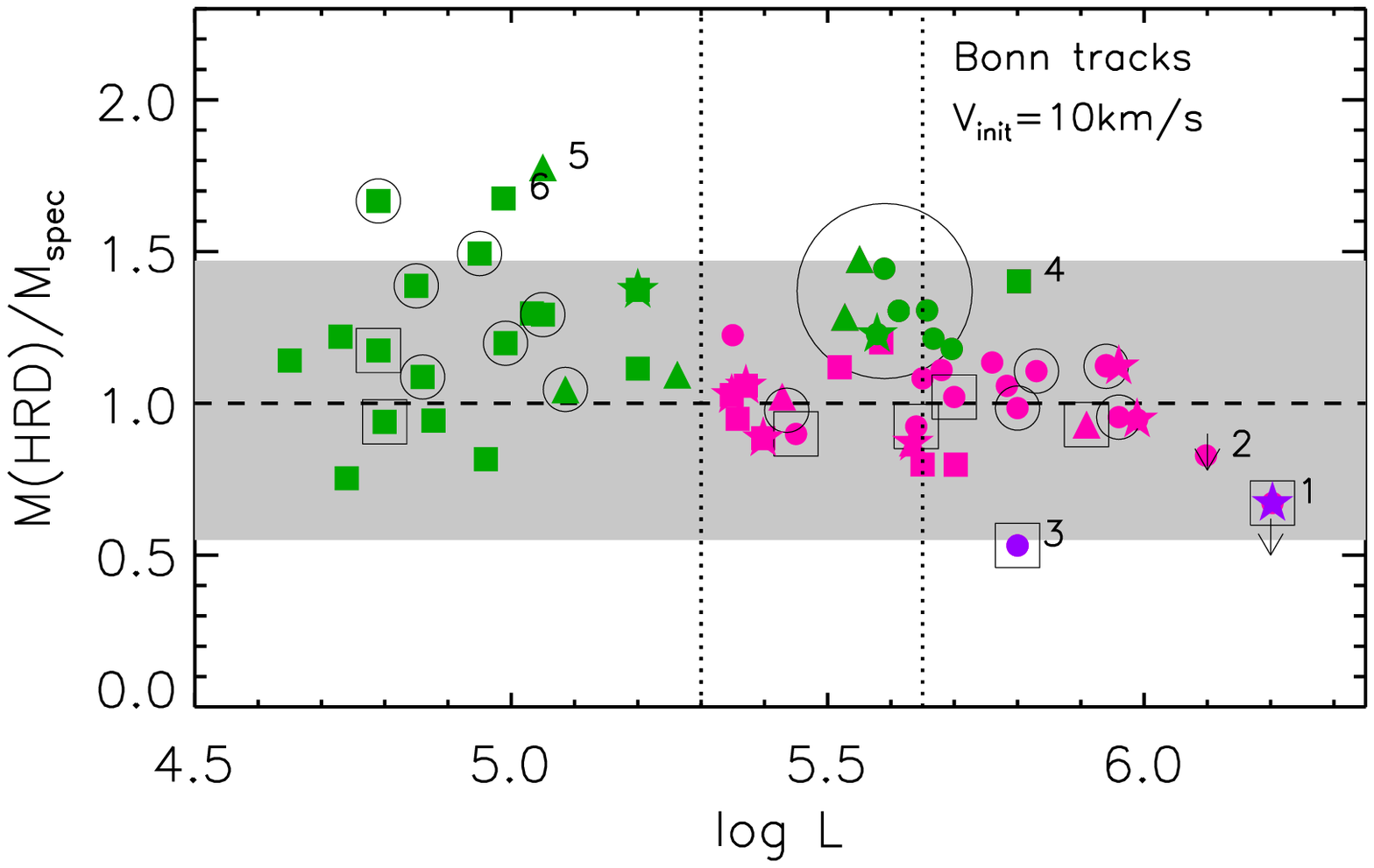}}
\caption{As in Fig.~\ref{fig13} but  for the Bonn tracks with 
\vinit=300~\kms.  Symbols and colour coding as in Figs. 12 and 13. 
 }
\label{fig14}
\end{figure*}

By confronting the Geneva predictions with our observational data, 
and accounting for the loci of each target in the classical HRD, 
we find that the objects that tend to indicate  \Mspec$>$\Minit(HRD) 
(data points in purple) appear to rotate faster, whereas those with 
\Mspec$<$\Minit(HRD) (data  points in green) slower than expected for 
their \Minit(HRD). The exception to this is the fast rotators, which 
deviate showing either consistent (3 out of 7) or significantly larger 
\vsini\ than the corresponding model values. Interestingly, as well 
for the objects with consistent \Mspec\ and \Minit(HRD)\ (data points 
in magenta), the rotational rate is not well reproduced by the models: 
dwarfs and giants indicate \vsini\ lower  than the model values,  
whereas the supergiants (data points enclosed in the large circle) 
rotate faster then expected for their \Minit(HRD).

Further considerations exploiting the \Mevol(HRD)/\Mspec 
ratio for the sample stars, calculated using the rotating and 
the non-rotating  Geneva tracks (lower panels of Fig.~\ref{fig13}),
indicate that for the objects experiencing a negative  mass 
discrepancy (data points in purple), the problem appears as a 
consequence of stellar rotation and/or related processes as 
implemented in the Geneva models; for those demonstrating a 
trend towards a positive mass discordance (data points in green), 
no significant changes in the mass pattern occur due to the 
limited effects of rotation and negligible mass loss in this 
regime (see  Fig.~8 in \citealt{mm00}). We note however 
that in the later case  the  mass discordance  appears 
to be somewhat stronger for the non-rotating compared to the 
rotating tracks: $\sim$23\% versus 20\% for the HRD and  
$\sim$39\% versus $\sim$29\% for the sHRD.

Concerning the stars whose evolutionary mass is  consistent 
with the spectroscopic one (data points in magenta), for 
the dwarfs and giants the mass pattern does not seem to 
depend on \vinit; for the  supergiants (data points enclosed 
by a large circle), an increase in \vinit\, from zero to 
0.4\vcrit\, turns out to be enough to bring  their \Mevol(HRD) in 
perfect agreement with \Mspec.

We  now turn to the Bonn models with \vinit$\approx$300~\kms.
From the upper panels of Fig~\ref{fig14}, it is evident that 
these models rotate generally faster than the sample stars, 
independent of the kind of mass discordance they indicate; this 
is the case with the exception of the four fastest rotators 
whose velocities are 
underestimated, and the three targets with \vsini of about 
200~\kms  whose rotational speeds are well reproduced by 
the models. From the lower panels, on the other hand, we 
find that -- over the whole luminosity range covered by the 
sample -- the mass pattern  does not (or hardly) change when 
non-rotating instead of rotating tracks were used 
to determine \Mevol. Particularly, for the objects in 
the low mass and luminosity regime (data points in green), 
the mass discordance amounts to $\sim$24\% (HRD) and 40\% (sHRD) 
for the rotating  versus $\sim$22 (HRD) and $\sim$38\& (sHRD) 
for the non-rotating Bonn models.
In contrast to the Geneva tracks, also the high luminosity regime 
remains rather unaffected from the inclusion of rotation (at 
least with respect to stellar parameters), which results from 
the consideration of the $\mu$ barrier that almost prohibits the
mixing of helium into the radiative zone (see Sect. 5.3).
Consequently, luminosities (and thus stellar masses) become (almost)
independent on \vinit  unless extreme rotation rates are considered
\citep{brott11}.

Summarising, we conclude that initial rotation that is too fast may 
contribute only for the more massive and evolved objects 
(data points in magenta) to explain the mass discrepancy established 
when comparing to the current Geneva models for \vinit=0.4\vcrit
\footnote{To recapitulate, this is an indirect effect, since in 
these models fast rotation leads, via mixing and an increased mean 
molecular weight in the radiative zone, to higher luminosities and 
larger mass loss, which finally results in masses significantly lower 
than those derived from spectroscopic analyses.}.
Regarding the mass problem  observed  for the less massive 
stars independent of the used grid (data points in green), 
our analysis shows that underestimated \loggc\, from 
spectroscopic analyses may contribute, but  a detailed 
quantitative consideration (accounting for all important parameters 
involved) is required to understand this issue completely. 

Finally in this section, we point out that while both 
model sets are generally inappropriate to represent the rotational 
properties of our sample, the Bonn models rotate generally faster 
than the Geneva models (for the same \Minit\ and \Teff). While 
differences in \vinit\ 
might be an issue, especially in the low mass regime, differences 
in the mass-loss rate and the treatment of angular momentum transport 
(see Sect.~\ref{Geneva_Bonn}) are the main agents dominating 
the process. Indeed, while the Bonn models
account for the extremely efficient angular momentum transport by 
internal magnetic fields (caused by a Spruit-Tayler dynamo, see
\citealt{brott11} and references therein), resulting in a quasi
solid-body rotation, the Geneva models include angular momentum
advection from the meridional circulations, which may transport
angular momentum from the envelope inwards.  As a consequence, 
the Bonn models rotate (at the surface) faster than the Geneva 
models, where the effect increases towards higher initial masses 
owing to the lower mass-loss rates present in the Bonn models (see
Sect.~\ref{Geneva_Bonn}), as already pointed out by \citet{mm05} 
and illustrated in Figs.~12 and 13 of \citet{keszthelyi16}. 
\begin{figure*}
{\includegraphics[width=8.5cm,height=5.2cm]{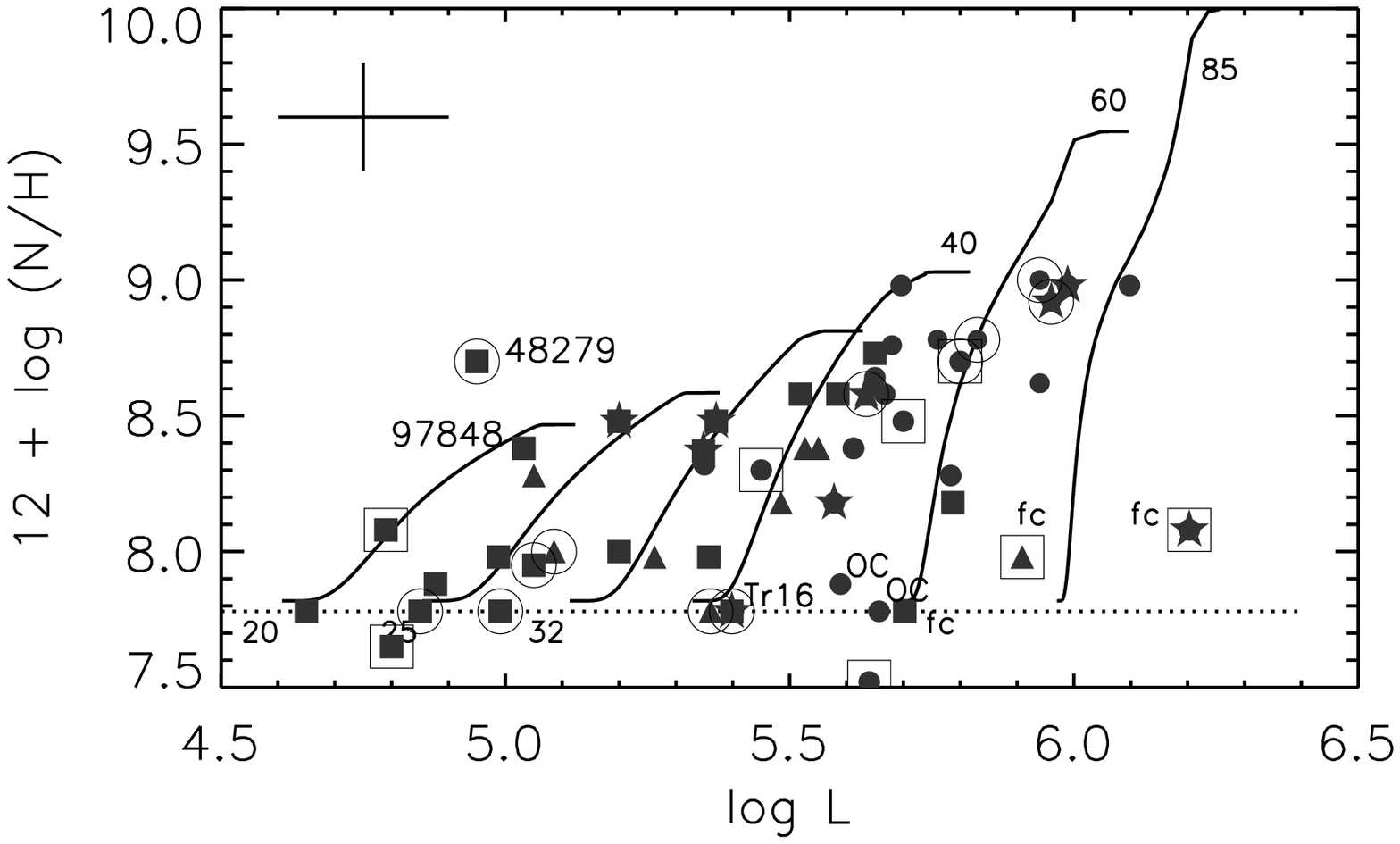}}
{\includegraphics[width=8.5cm,height=5.2cm]{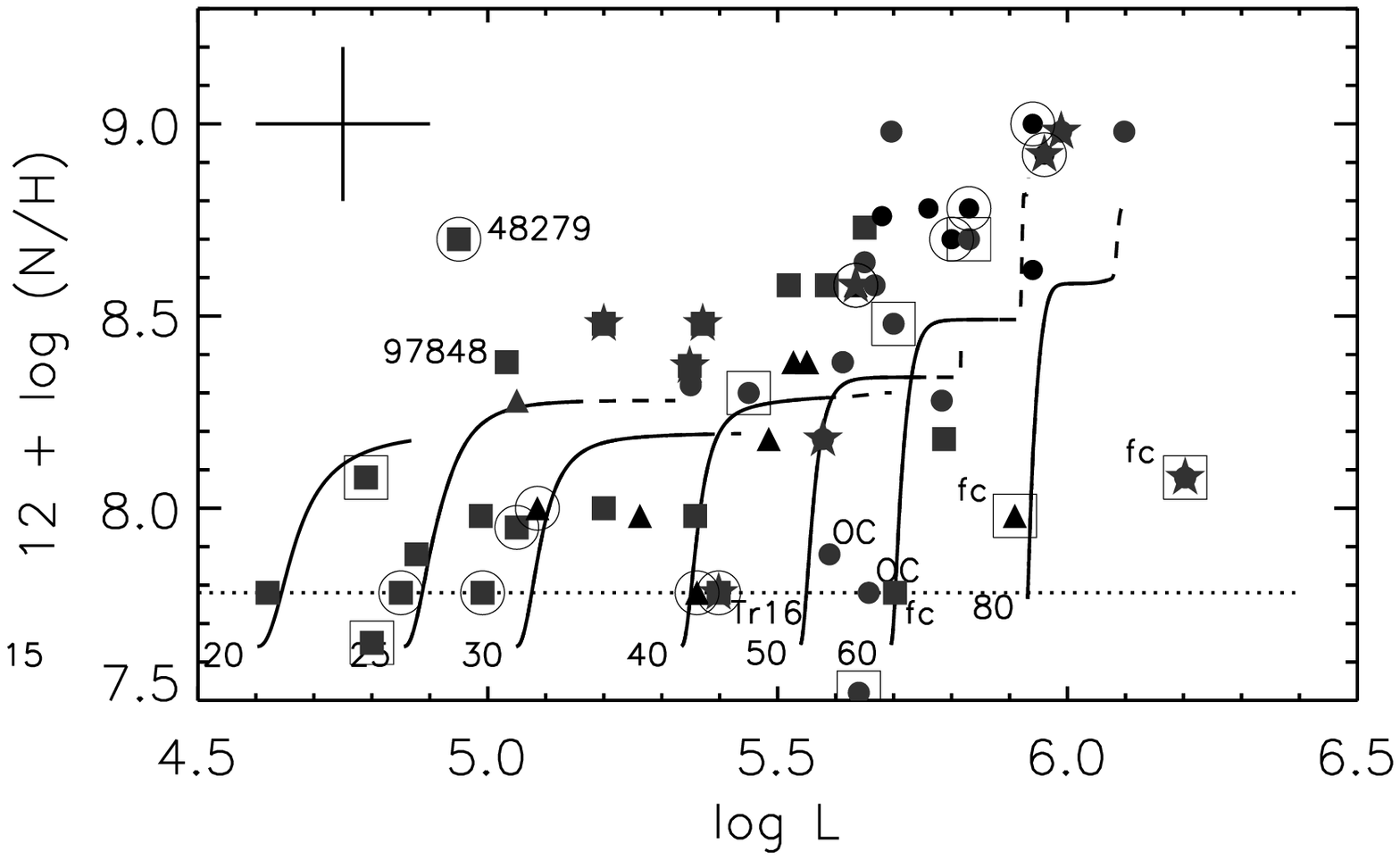}}\\

{\includegraphics[width=8.5cm,height=5.2cm]{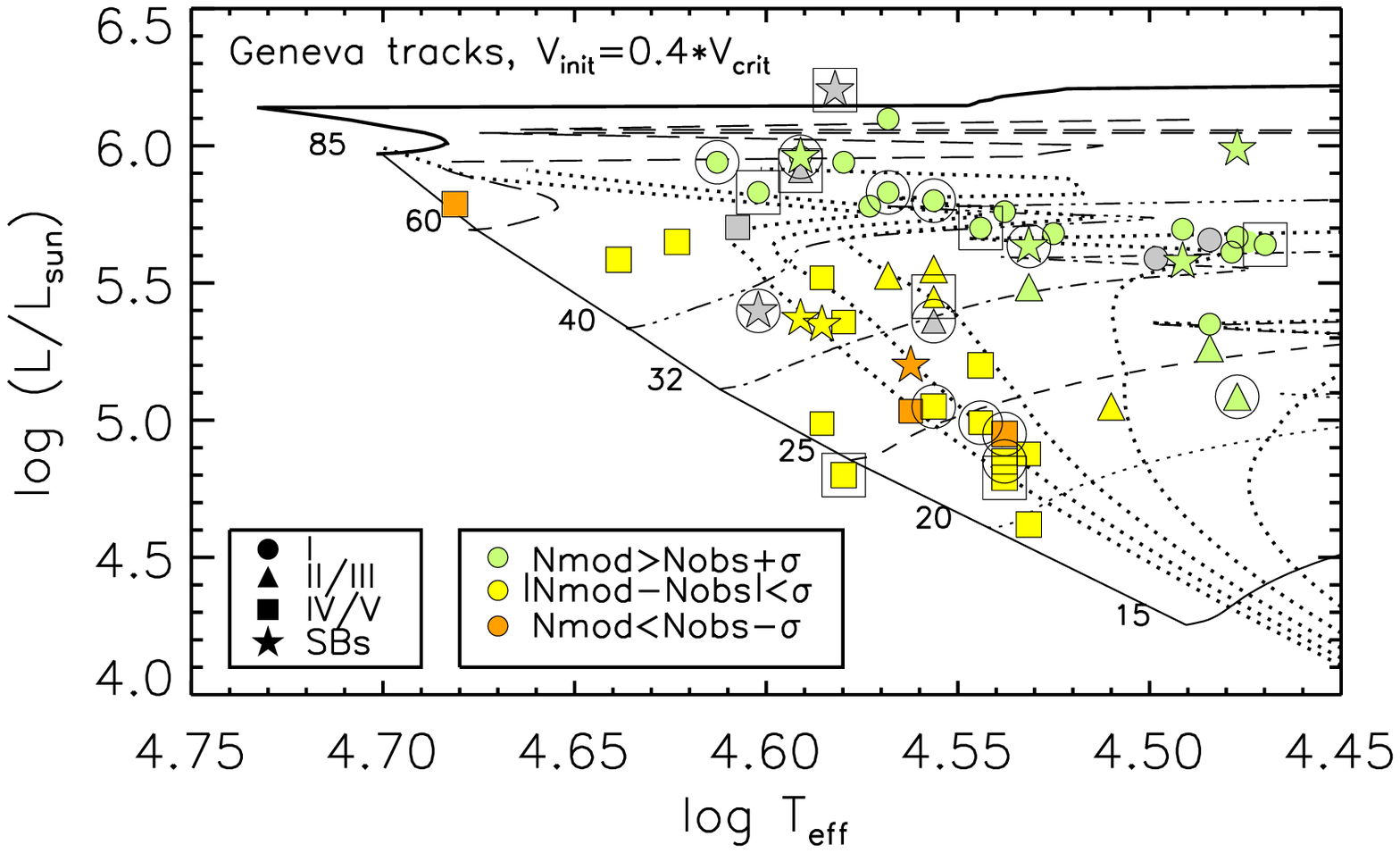}}
{\includegraphics[width=8.5cm,height=5.2cm]{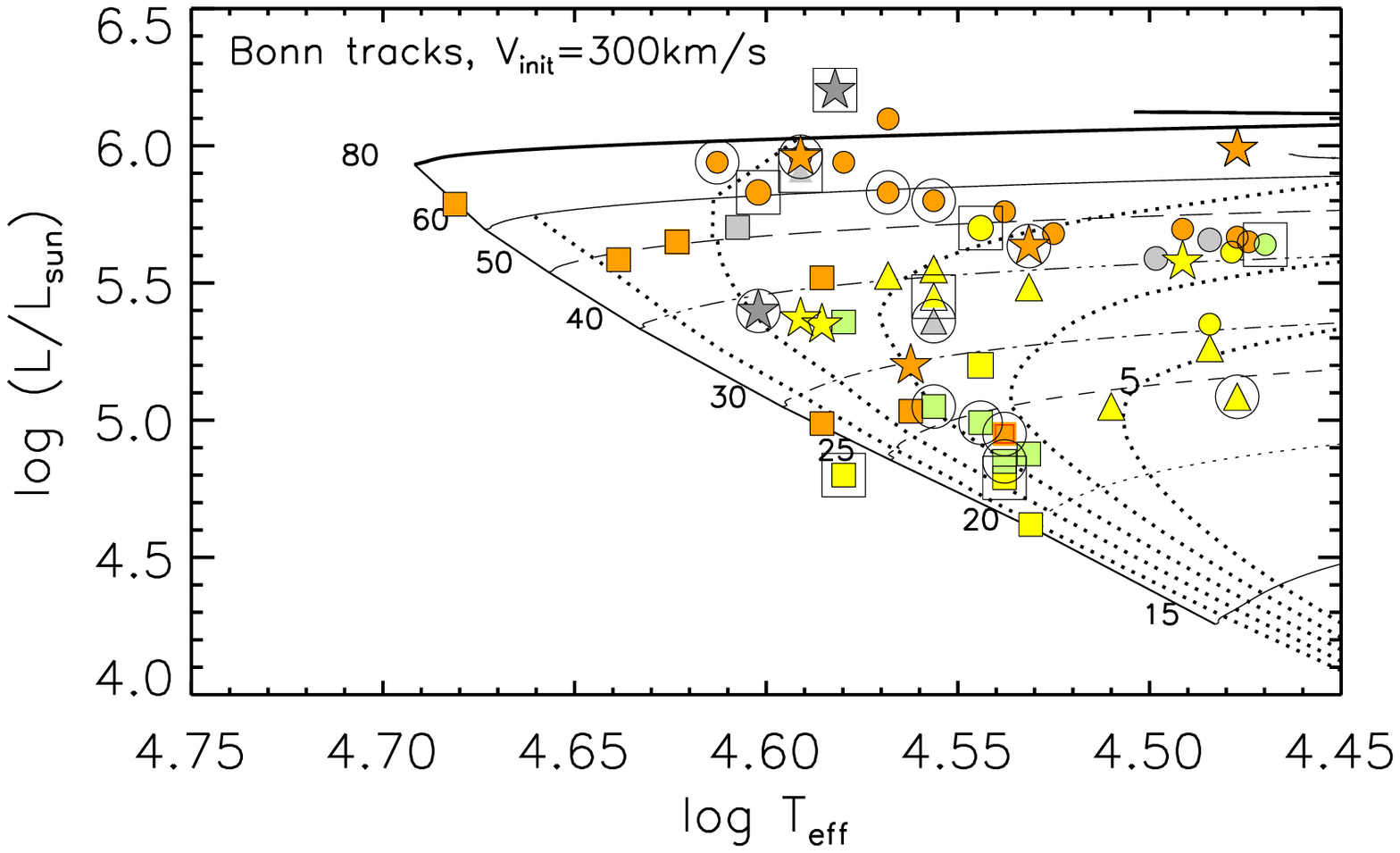}}\\

{\includegraphics[width=8.5cm,height=5.2cm]{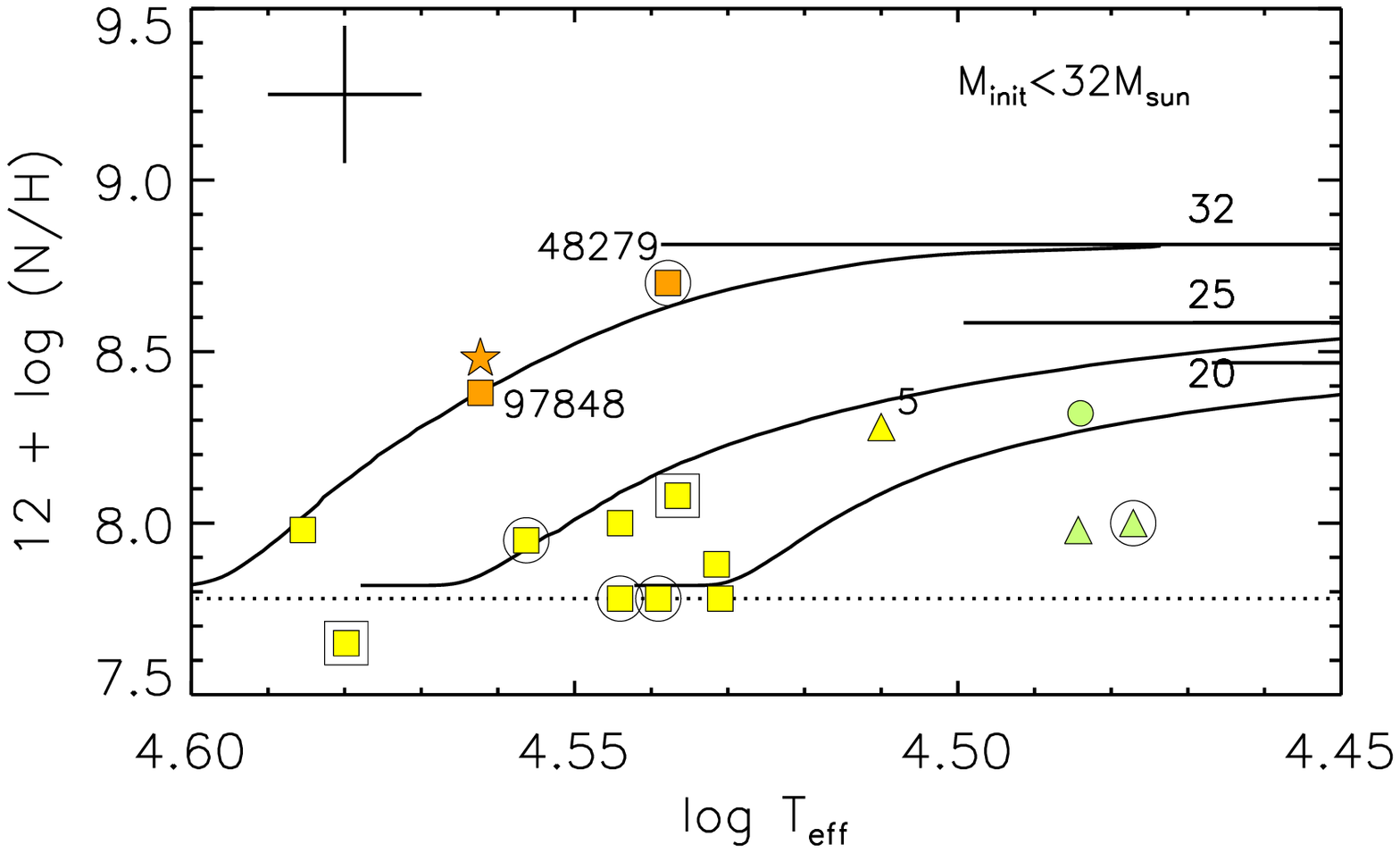}}
{\includegraphics[width=8.5cm,height=5.2cm]{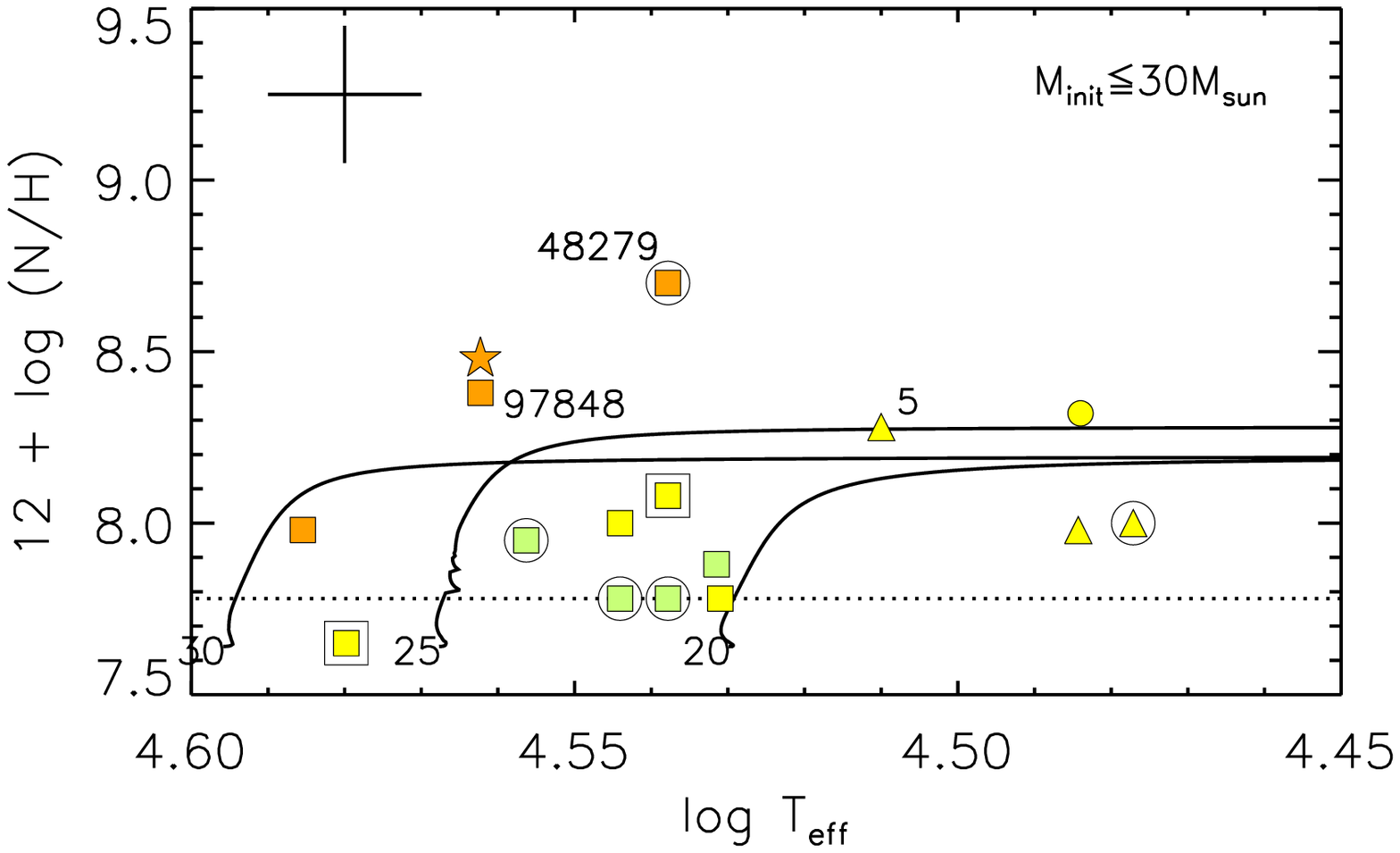}}\\

{\includegraphics[width=8.5cm,height=5.2cm]{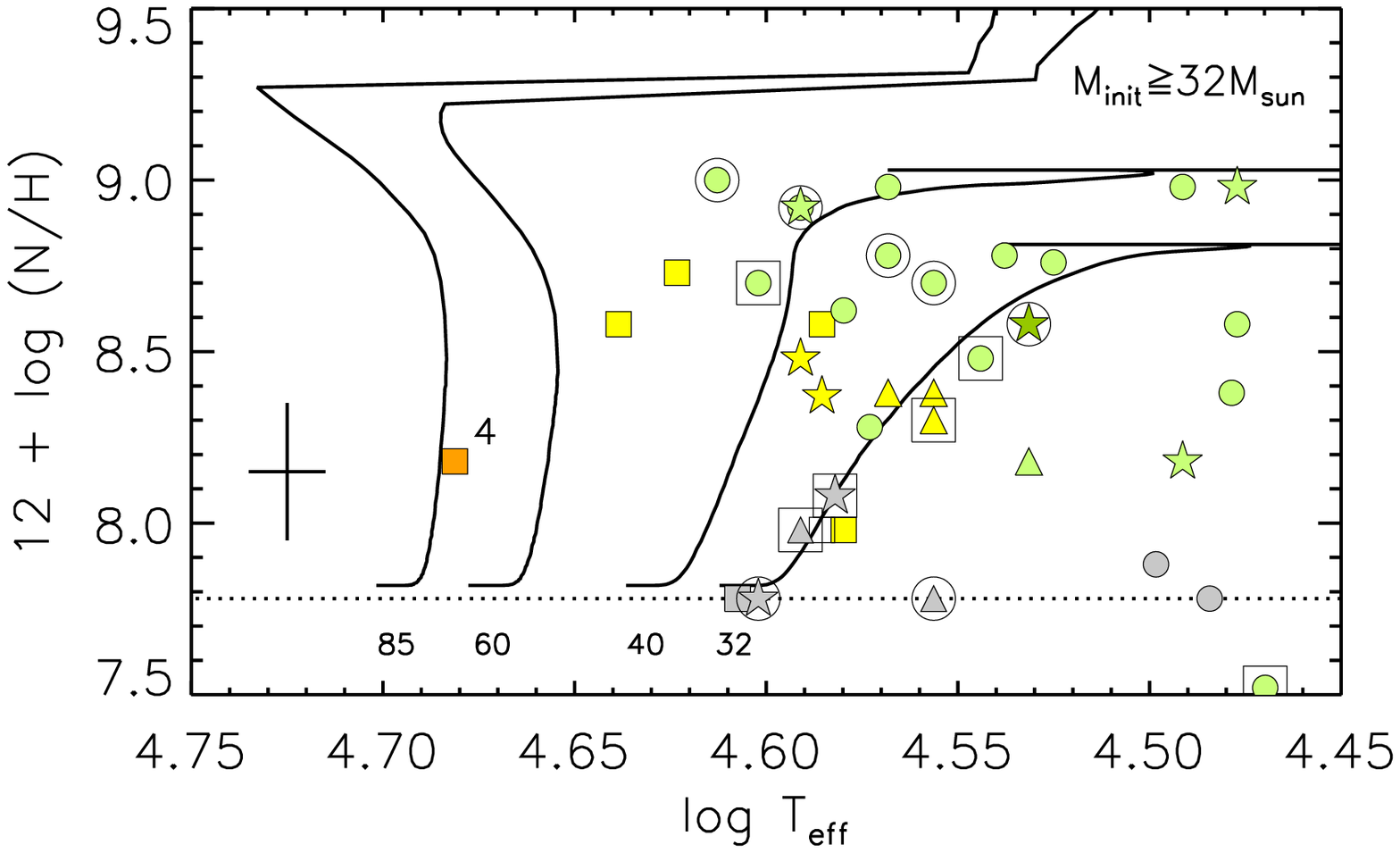}}
{\includegraphics[width=8.5cm,height=5.2cm]{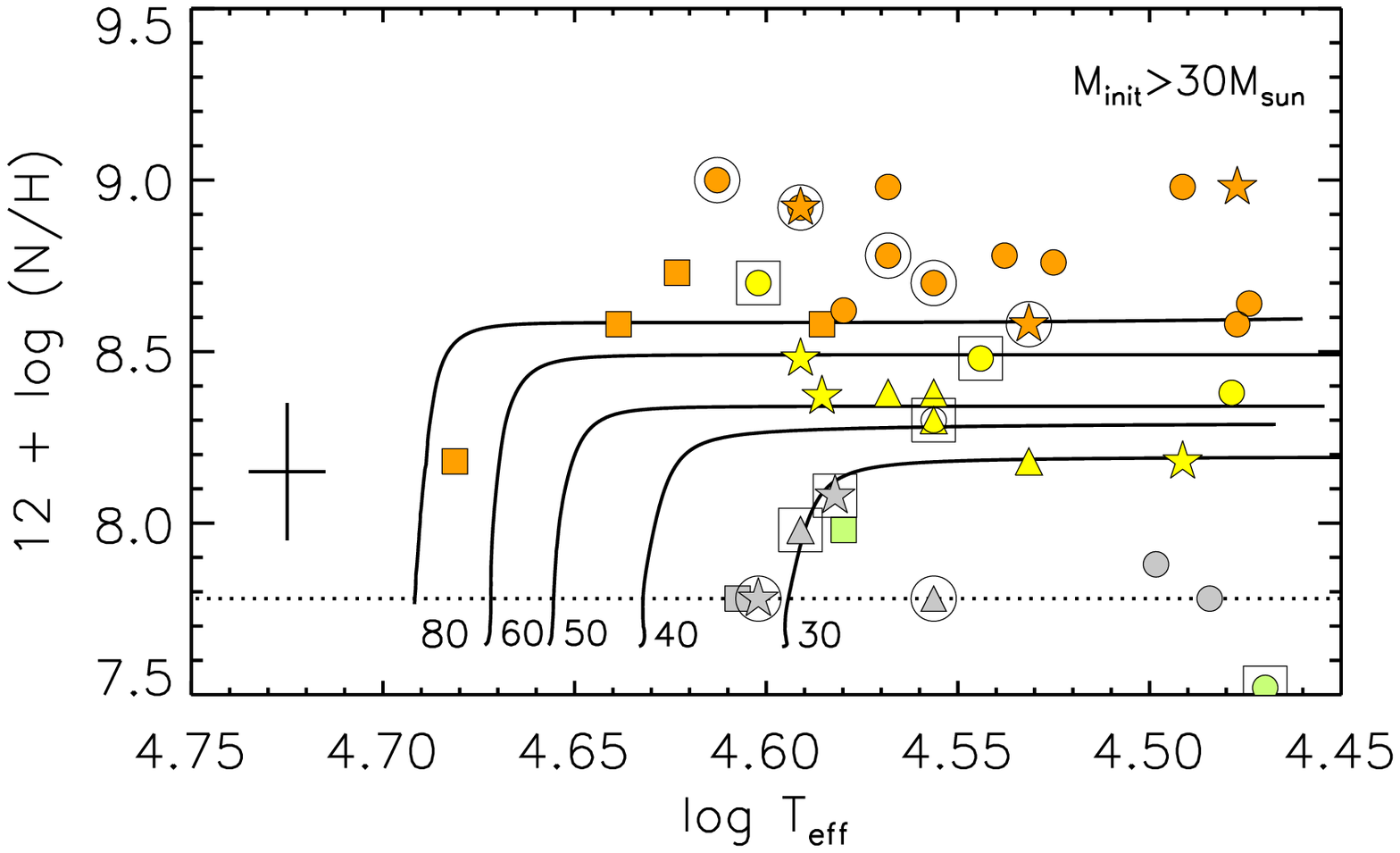}}
\caption{Classical HRD and surface N abundances as a function 
of \logl\ and \Teff\ for the sample stars plotted against the 
current Geneva (left) and the Bonn tracks (right) with rotation. 
Symbol coding is used to distinguish between objects of 
different LC;  magnetic stars and fast 
rotators are  additionally highlighted  by large squares 
and circles, respectively. In the lower three panels, different 
colours denote the morphologically peculiarly objects with very 
weak nitrogen lines (grey), and the stars whose predicted and 
observed [N] values differ by less (yellow) or more (green, 
orange) than 1$\sigma$. In the upper panels, the solid part 
of the tracks corresponds to the O-type phase (\Teff$\ge$29~kK). 
The solar N abundance (from \citealt{asplund05}) is indicated by 
dotted lines. 
}
\label{fig15}
\end{figure*}

\section{Chemical surface enrichment and rotational mixing}\label{rot_mix}
\subsection{Nitrogen enrichment}
\label{obs_mod__Nabn}

For hot massive stars, the surface chemical enrichment is a 
multivariate function of stellar parameters \citep{m09}. Therefore, 
in order to constrain various effects on the derived N abundances, 
in Fig.~\ref{fig15} we show the classical HRD in parallel to the 
[N]--\logl\ and the [N]--log\Teff\, diagrams for the sample stars 
built using the current Geneva tracks with \vinit=0.4\vcrit (left) 
and the Bonn tracks with \vinit=300\kms (right).  To guide the eye 
and facilitate the comparison, different colours have been employed 
to denote those stars whose positions on the \Teff--[N] diagram are 
either consistently (within  1$\sigma$) reproduced by the tracks as 
selected from their position on the classical HRD (yellow) or fail 
to be reproduced  (light green and orange).  

From the  upper left panel  of Fig.~\ref{fig15}, it appears that the 
Geneva models tend to reproduce the N abundances of the sample dwarfs 
generally well  while overpredicting those of the giants and supergiants. 
Further considerations employing the classical HRD in parallel to the 
[N]--log\Teff\ diagram confirm these trends, which indicate that the bulk 
of our dwarfs with \Minit$<$32~\Msun  (in the left panel of the third 
row, yellow squares) are located between the 20 and 25~\Msun model tracks; 
this is  consistent with their position in the classical HRD. Analogous 
findings apply to all but one of the more massive dwarfs with 
\Minit(HRD)$>$32\Msun (yellow squares in the  lower left panel), which 
on both diagrams are distributed  between the 32~\Msun\  and the 50~\Msun 
Geneva tracks. There are only four sample dwarfs   (data points in orange) 
that  stand apart, displaying a nitrogen enrichment larger than predicted 
for their \Minit(HRD) and \Teff by about 0.4 dex: HD~48279, which might 
be a former binary \citep{martins12b}; HD~97848, which is a field star 
with a potentially underestimated luminosity (and hence \Minit(HRD));  
the SB1 system HD~46573;  and the mass outlier No. 4 (HD~64568), whose  
\Teff\, might have been somewhat overestimated (see Appendix~B).

Similar considerations for the sample supergiants indicate that for 
those with \Minit$\ge$32~\Msun, the observed N enrichment  is lower 
then predicted by the models by about 0.3 to 0.4~dex (in the lower 
left panel of Fig.~\ref{fig15}, light green circles). In the classical 
HRD  these objects are distributed between  the 32~\Msun and the 85~\Msun 
tracks, whereas in the [N]--\Teff diagram these objects are generally 
located between the 20 and the 50~\Msun tracks, and the morphologically 
peculiar stars with very weak N lines (Ofc, OC, and Onfp objects, data 
points 
in grey) are the exception. The same behaviour is found for the less 
massive supergiant HD~209975 (light green circle in the left panel of the 
third row), whose  N enrichment is overpredicted by about 1$\sigma$ and 
more. On the classical HRD this star appears either as a 30~\Msun\ MS or 
as a 25~\Msun\ post-MS object, while its N enrichment is well fitted by 
the 20~\Msun\ track. 

For  the sample giants and bright giants, our analysis indicates  good 
agreement (within 1$\sigma$)  for the objects with an intermediate mass 
(yellow triangles in the lower left panel). On both diagrams these stars 
are consistently distributed between the 32~\Msun and 40~\Msun Geneva 
tracks and a severe discrepancy for those with \Minit$<$32~\Msun. Specifically, 
for the two coolest giants (HD~69106 and CD-44\,4865) present in the 
low mass regime (light green triangles in the left panel of the  third 
row of Fig.~\ref{fig15}), the observed N content  is lower by about 0.4 
dex than expected for stars of the same \Teff and \Minit(HRD) of about 
22...25~\Msun. Concerning the bright giant HD~207198 (mass outlier 
No.~5, see Appendix~B), whose enrichment appears to be consistent with 
that predicted by the models, if its  luminosity has been indeed 
underestimated (by about 0.4~dex, as suggested in Appendix B), 
this giant would also indicate a discrepant N enrichment (by about 
0.5~dex). In the classical HRD, it would appear as a 32~\Msun 
object, while its nitrogen content is well reproduced by the 25\Msun 
Geneva track. 

We now turn to the Bonn models for \vinit$\approx$300~\kms. From 
the  upper right panel of Fig.~\ref{fig15}, we find  that for  
\Minit$>$30\Msun,  these models  tend to underpredict the N 
enrichment for the sample stars (objects of OC, fc and nfp categories 
apart, see Sect.~\ref{Nabn}). Exploiting the classical HRD in 
parallel to  the [N]--\Teff\ diagram, we also find that while all 
stars with \Minit(HRD)$>$40~\Msun (save the outliers with peculiarly 
lower N abundances denoted in grey) are more  enriched than 
predicted by the models (by about 0.4~dex and more), a good 
correspondence between model predictions and observations (generally 
within 1$\sigma$)  is present  for those with \Minit(HRD)  between 
30~\Msun and 40~\Msun (in the lower right panel, data points in 
orange and yellow, respectively). 

Similar considerations for the objects in the low mass regime 
(\Minit$\le$30\Msun, right panel of the third row in Fig.~\ref{fig15}) 
show  that the Bonn models can reproduce the N abundance for the sample 
giants and supergiants reasonably well (within 1$\sigma$; data points 
in yellow) but are not able to reproduce that for many of the sample 
dwarfs. On  the HRD and the [N]--log\Teff\, diagram, all (but three) 
of the dwarfs are consistently located between the 20\Msun and 30\Msun 
tracks, yet the N enrichment for half of these (denoted by light green 
squares) is lower (by about 1 to $2\sigma$) than expected for the 
corresponding \Minit(HRD). 
Interestingly, for the bright giant HD~207198 (mass outlier No.5, see 
above), the empirical [N] is also consistent with that predicted by the 
models for the corresponding \Minit(HRD).  
\begin{figure*}
{\includegraphics[width=8.5cm,height=5.6cm]{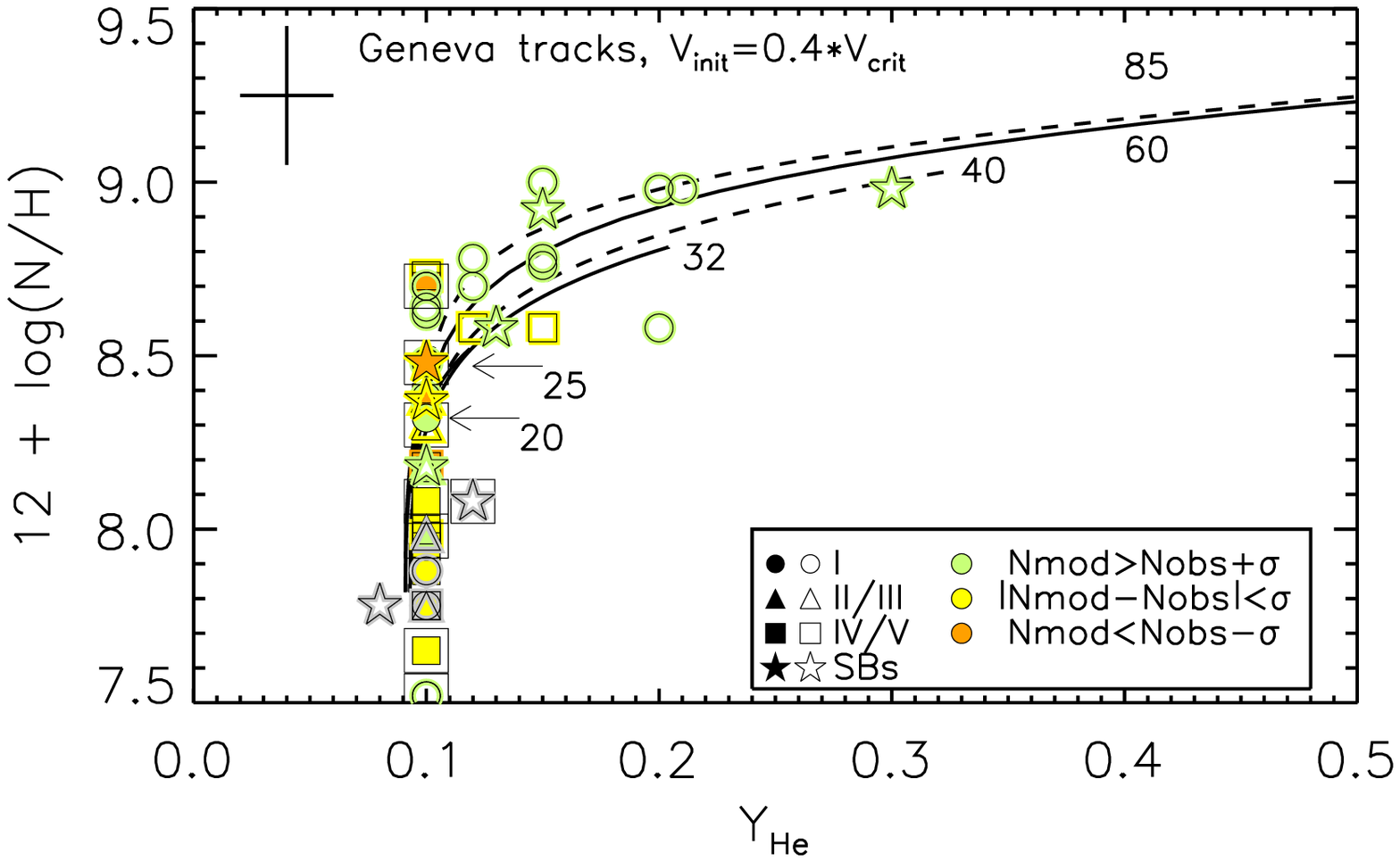}}
{\includegraphics[width=8.5cm,height=5.6cm]{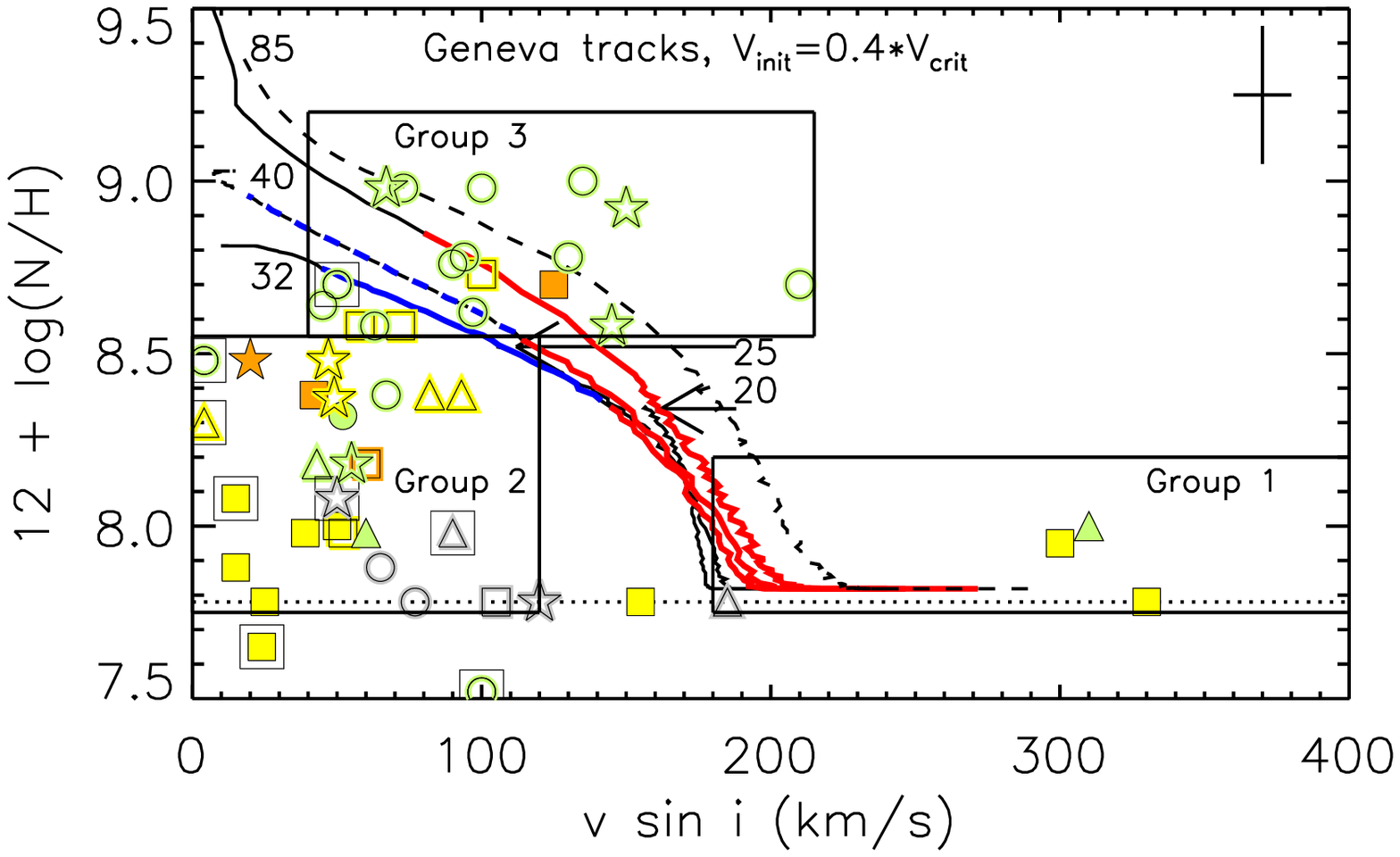}}

{\includegraphics[width=8.5cm,height=5.6cm]{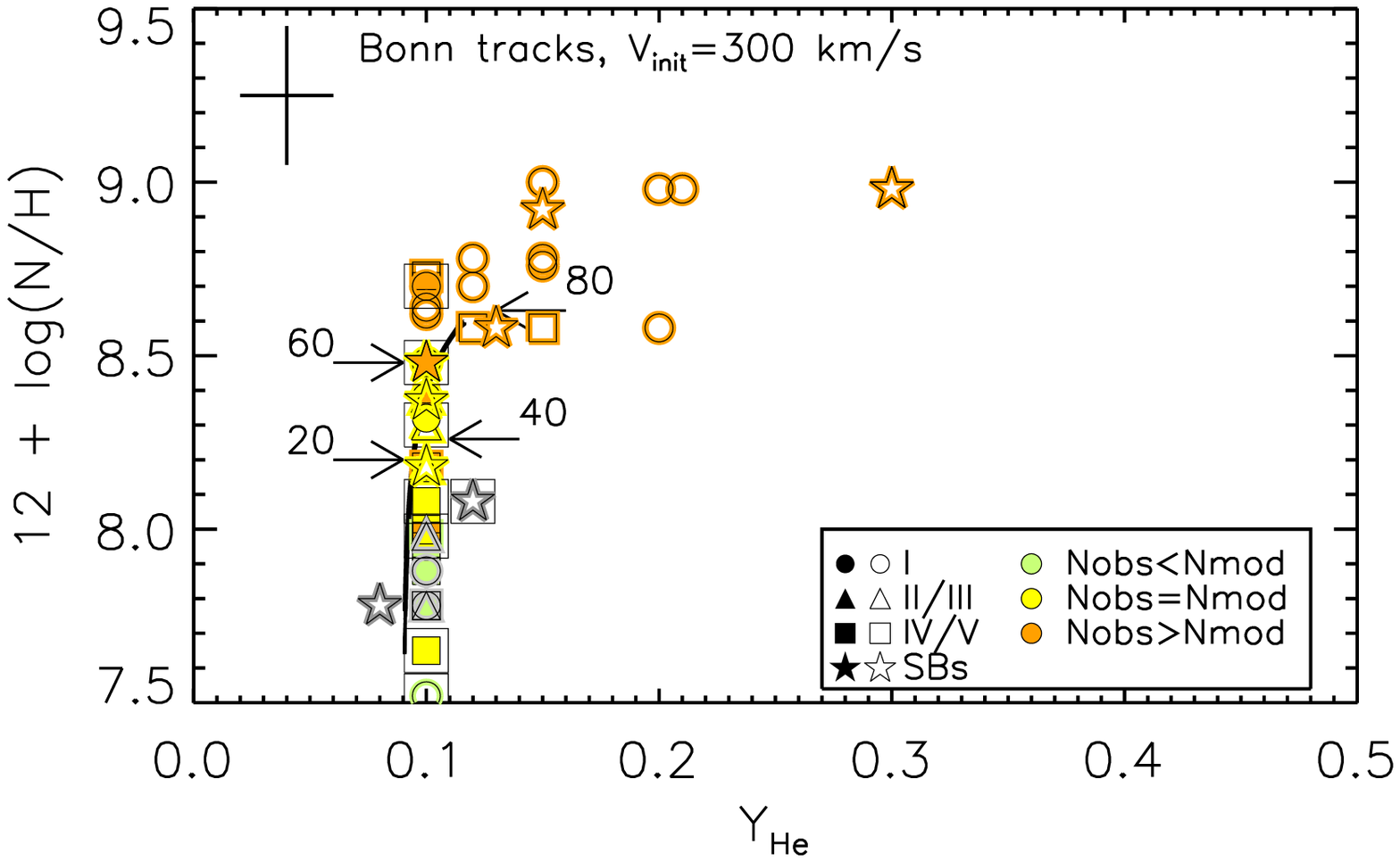}}
{\includegraphics[width=8.5cm,height=5.6cm]{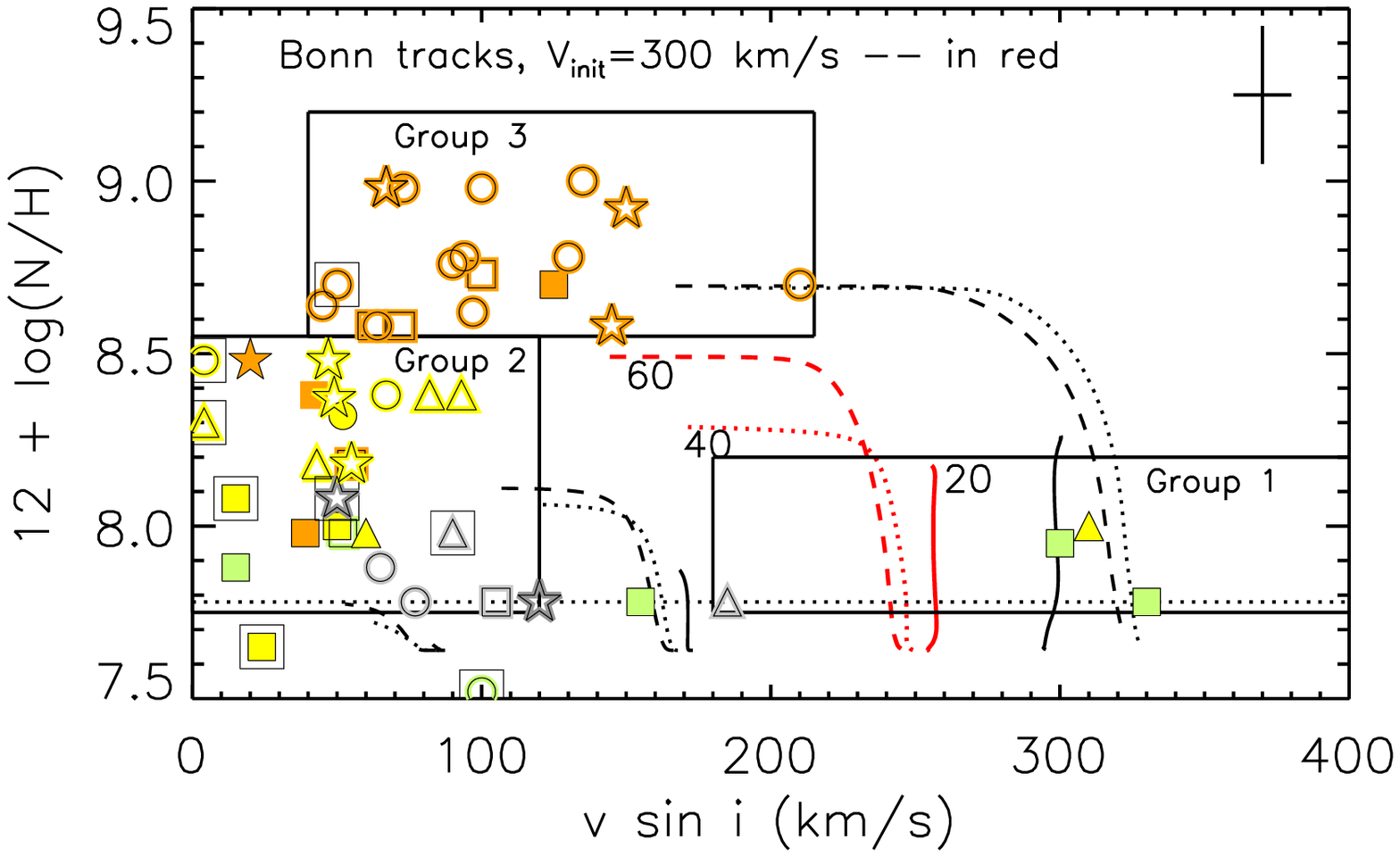}}
\caption{Surface nitrogen abundance as a function of helium 
abundance, \Yhe=N(He)/N(H) (left), and projected rotational 
rate (right). The corresponding predictions for the O-type phase 
(\Teff$\ge$29~kK) are overplotted. Symbols  and colours as in 
Fig.~\ref{fig15}; open/filled symbols denote objects with 
\Minit(HRD) larger/smaller than  32~\Msun (Geneva grids) and 
30~\Msun (Bonn grids), respectively.  For a number of tracks, 
the end of the O-star phase is indicated by horizontal arrows. 
Model \vrot has been multiplied by $\pi$/4 to take the average projection 
effect into account (e.g. \vinit$\approx$300 \kms corresponds to 
\vsini$\approx$240 \kms). In the upper right panel, the coloured 
parts of some tracks represent  the dwarf (\logg$>$3.75, red) and the 
giant (3.75$<$\logg$<$3.6, blue) phases, respectively. 
} 
\label{fig16}
\end{figure*}

Regarding the less massive dwarfs that appear more enriched than 
predicted  by the Bonn models (in the right panel of the third row, 
data points in orange), three of these, HD~48279, HD~97848, and SB1 
HD~46573, were already discussed  as objects with peculiarly high 
enrichment in comparison to the Geneva tracks (see above); the fourth, 
CPD-58\,2620, is a mass outlier (No.6) whose luminosity and gravity 
might be subject to large uncertainties (see Appendix A).

In conclusion,  while our analysis reveals a clear trend of stronger 
N enrichment in more massive and evolved objects, the current Geneva 
models with \vinit=0.4\vcrit appear to reproduce well the enrichment 
of the sample dwarfs  and the less evolved giants, while overpredicting 
that  for the more evolved giants and the supergiants. The Bonn models 
with \vinit$\approx$300~\kms, however seem appropriate to study the 
enrichment for the giants and supergiants with \Minit(HRD)$\le$40~\Msun, 
but tend to overpredict the enrichment for the less massive dwarfs and 
underpredict that for the stars with \Minit(HRD)$>$40~\Msun.

\subsection{Correlation between nitrogen and helium enrichments}

On the left of Fig.~\ref{fig16}, the N enrichment  is shown as a function 
of  He enrichment for the sample stars. For the objects with 
[N]$\le$8.5~dex, no significant He enrichment can be seen; for those with 
[N]$>$8.5~dex,  a clear correlation is observed. While these findings are 
qualitatively consistent with theoretical predictions (from \citealt{ekstroem} 
and from \citealt{brott11}), a closer inspection of the data reveals that 
for the low mass stars (filled symbols; \Minit(HRD) $<$32\Msun and 
$\le$30\Msun for the Geneva and Bonn models, respectively) with [N]$\le$8.5~dex  
the agreement between the observed and predicted N -- He enrichment relation 
is not only qualitative, but also quantitative; for the high mass stars 
(open symbols; \Minit(HRD)$\ge$32\Msun (Geneva tracks) and $>$30\Msun 
(Bonn tracks)), a discrepancy emerges in which the objects appear less evolved, 
and thus less He enriched, compared to the corresponding Geneva models, or 
more He enriched compared to Bonn models  (aside from OC, fc and nfp category 
objects; data points in grey). Indeed, while many of our more massive 
targets are significantly He and N enriched, the Bonn models for  
\Minit$\le$60~\Msun only produce stars with a baseline He abundance, whereas 
the Geneva models become significantly He enriched starting from 
\Minit=32~\Msun on.  As the amount of helium mixed into the stellar surface 
is a direct consequence of the treatment of the mean molecular weight barriers 
(see Sect.~\ref{Geneva_Bonn}), these results might imply that both model grids 
may not adequately account for this physical ingredient.

\subsection{Rotational mixing}\label{obs_mod_Nvsini}

The right part of Fig.~\ref{fig16} shows the \vsini--[N] diagram for the 
sample stars. In observational context, these data indicate that for the 
objects in the low mass regime (data points denoted  by filled symbols; 
\Minit(HRD)$<$32~\Msun (Geneva tracks) and $\le$30~\Msun (Bonn tracks)), 
no clear trend of [N] with \vsini\ can be seen. For those  objects in the 
high mass regime (open symbols), except for the magnetic stars and the 
data points in grey, a weak trend towards more enrichment for larger 
equatorial rotational rate seems to emerge. We note that this trend would 
lose significance, however, if the five fastest rotators in this mass 
subgroup were discarded. Additionally, we also see that at same/similar 
\vsini, the more massive dwarfs are more N-rich than the less massive dwarfs,  
and less enriched than the supergiants in the same mass regime. These results 
confirm  the idea that rotational mixing  is more efficient  at higher masses  
and later evolutionary phases \citep{mm00}. 

Interestingly, the SBs and magnetic stars do not seem to depart from the main 
trends, although for the latter a clear tendency -- independent of mass and 
LC -- to rotate slowly while being relatively unenriched can be observed. 

Another point to be noted here is the clear deficit of more massive 
N-rich objects with very fast rotation (\vsini$>$250~\kms). This 
feature however is most likely a result of observational selection, as 
pointed out in Sect.~\ref{obs_mod_vrot}.

In comparison to the model predictions, the data shown in the right of 
Fig.~\ref{fig16} reveal that  while both the current Geneva models with 
\vinit=0.4\vcrit\ and the Bonn models with \vinit$\approx$300~\kms\ can 
produce fast rotators that are basically unenriched, neither of these models 
can account for (within the corresponding uncertainties) the projected 
rotational rates  of our less massive, rapid rotators with \vsini$\ga$300~\kms, 
as shown in  the upper right panels of Figs.~\ref{fig13} and \ref{fig14}. 
(See also Fig. C~1, where the corresponding vsini--[N] diagram is shown in 
two portions for each model grid: one for the low and another for the 
high mass regime.) Neither do these models predict the existence of stars 
with normal and slow rotation (\vsini$<$120~\kms) that are not or moderately 
nitrogen enriched ([N] below roughly 8.5~dex). From now on we refer to 
these two groups of stars as Group 1 and Group 2, respectively.

Regarding the more massive, N-rich objects with normal and rapid
rotation (Group~3), it is evident that the Bonn grids with 
\vinit$\approx$300~km/s are generally incapable of representing the 
position of these targets, producing only very fast  rotating O stars 
that are not or weakly chemically enriched ([N]$\le$8.5~dex). For the 
Geneva grids, we find that while the models  cover the range of observed 
\vsini\, and [N] for the Group~3 stars, all dwarfs rotate at lower 
velocity, yet indicating  an enrichment consistent with that expected 
for their \Teff\ and \Minit(HRD). Regarding the supergiants (also 
including the three SBs), all of these appear less enriched but with 
faster rotation than predicted by the models: in our  diagram, these 
stars populate the area covered by the red and blue parts of those 
tracks where the dwarfs and giants should reside. 

Overall, the main implication of the above results is similar 
to that already formulated in Sect.~\ref{obs_mod_vrot}, namely 
that additional models, particularly with lower \vinit, might be 
needed to solve the N--\vsini\ problem established when comparing 
to the current predictions. Regarding the Bonn grids, such a test 
is already possible, and we performed this via the BONNSAI tool
\footnote{The BONNSAI analysis was performed using \Teff, \logl, 
\vsini, [N], and \Yhe\ as observables, and a Salpeter initial mass 
function and a uniform initial rotational velocity distribution 
in the range from zero to 600~\kms as independent priors.}. 
The obtained results show that  for each  of the Group~1 and half of 
the  Group~2 objects (about 40\% of the total sample) an acceptable 
or at least compromise solution  for all important parameters 
(\logl, \Teff, \vsini, [N], and \Yhe) can be obtained when \vinit\ 
is allowed to be lower or higher than 300~\kms. For the rest of 
stars in Group~2 and those in Group ~3 however, such solutions cannot 
be reached unless the majority (about 2/3) of their members were fast 
rotators seen pole-on\footnote{ Warning: Since our definition of 
Groups~2 and 3 within the \vsini--[N] diagram is different from that 
used by \citet{hunter08}, any comparison between results derived in 
these two studies should be considered  with  caution.}. Since the 
latter possibility seems rather unlikely, given the large number of 
stars involved, and excluding the seven known binaries, we conclude 
that the N pattern of about 50\% of the single stars in the sample  
cannot be explained in the current framework of rotational mixing, as 
implemented in the Bonn grids for solar metallicity. 

Finally in this section, we point out that mostly as a consequence 
of the smaller initial carbon and oxygen mass fractions in the Bonn 
models ($X_{\rm C}=0.00118$ and $X_{\rm O}=0.00413$, respectively) 
compared to those in the Geneva models ($X_{\rm C}=0.00231$ and 
$X_{\rm O}=0.00573$), CNO equilibium leads to a smaller nitrogen 
abundance in the Bonn models (12+log(N/H)$_e\simeq 8.6$) than in 
the Geneva models (12+log(N/H)$_e \simeq8.8$)  assuming no 
hydrogen depletion. This should be taken into account when both 
model results are compared in Figs.\,15 and\,16.

\section{Comparison to other studies}\label{discussion}

One of the main outcomes of our analysis is that the results of the
comparison between model predictions and observations depend crucially
on the specific choice of the model grid and the kind of diagram used.
With this in mind, and given that the model predictions for B
supergiants are strongly influenced by the behaviour of \Mdot\ at the
bistability jump---an issue that may be problematic for current model
evolutionary calculations (e.g. \citet{keszthelyi16})---we limit our
discussion, when comparing with other studies, to those that either
employ the Ekstr{\"o}m et al.  or Brott et al. grids (or both), and
are furthermore focussed on O stars.

\subsection{Mass discrepancy}\label{mass_discrp_comp}

Based on the analysis of detached eclipsing binaries in the LMC and 
the MW,  \citet{massey12, morrell14}, and \citet{kourniotis15} have 
concluded that in comparison to the Geneva models  with \vinit=0.4\vcrit 
all involved stars, which are late O-type dwarfs and thus fall into 
the less luminosity regime as defined by us, appear undermassive by 
typically 11\% to 20\%.

Regarding the Bonn grids, evidence of a positive mass discrepancy 
(by up to 20\% on average) have been reported by \citet{mahy15} and 
by \citet{R16} for  samples of O stars in the MW and the LMC, 
respectively, with \Mevol(HRD)$<$30\Msun.  In contrast, from the  
analysis of a large sample of O dwarfs in the LMC, \citet{carolina17} 
have concluded that there is no compelling evidence of a systematic 
mass discrepancy. Taken at face values however the data illustrated 
in their Fig. 14 indicate a mass discordance by up to 58\% for masses 
below 20\Msun.

The results outlined above are in good qualitative, and in most of 
the cases  quantitative, agreement with similar findings from the 
present work. Our results indicate that in comparison to the Geneva 
models with \vinit=0.4\vcrit and the Bonn models  with \vinit=300~\kms, 
our stars with \Minit(HRD lower than $\sim$32~\Msun (Geneva) and 30\Msun 
(Bonn)) appear undermassive  by average values of $\sim$20\% and 
$\sim$24\%, respectively.

Regarding the trend towards a negative mass discordance established 
for our sample stars with \Minit(HRD)$\ga$35\Msun\ in comparison to 
the Geneva grids with \vinit=0.4\vcrit, these results are so far first 
and thus unique.

\paragraph{Possible interpretation}
Since the studies referred to above have been performed applying 
either the FASTWIND or CMFGEN code, it does not seem likely that 
he mass problem established in the low mass O star regime might be 
explained by potential shortcomings in the FASTWIND modelling, as 
suggested by \citet{massey13}. 

Rotation of the considered models that is too fast or too slow cannot 
be responsible either because, firstly, evidence of  a mass discordance 
has been obtained using the  BONNSAI tool and accounting for the \vsini\, 
distribution of the corresponding targets (\citet{R16} and  \citet{carolina17}, 
see above) and, secondly, at least for our sample, lowering the value of 
\vinit\ down to zero does not significantly affect the outcome of the mass 
discrepancy analysis (see lower panels of Figs.~\ref{fig13} and \ref{fig14}). 

Underestimated \loggc\, derived from the FASTWIND and CMFGEN  analyses, 
owing to the neglect of turbulent pressure term in hydrodynamic and 
quasi-hydrostatic equations, might help to cure the problem, but this 
possibility has to be carefully investigated in the future accounting 
for all important parameters involved (see Sect.~\ref{obs_mod_logg}).

On the other hand, to explain the discordant masses derived for their targets in 
comparison to the Geneva models  with \vinit=0.4, \vcrit\citet{massey12} 
have suggested that either \vinit\ larger than 0.4*\vcrit or a convective 
overshoot parameter larger than 0.1 might be needed to reconcile observed 
and predicted luminosities. While both hypotheses by Massey et al. are 
legitimate, our analysis indicates that the former hypothesis is not 
applicable to our less massive dwarfs, which generally rotate more slowly 
than predicted by the models; the latter hypotheses might be applicable, 
although an even larger overshoot parameter (larger than 0.3, as adopted 
in the Bonn models) would to be required.

Interestingly, an overshoot parameter larger than 0.3 in the Bonn 
models for \Minit$\ge$15~\Msun was suggested by \citet{castro14} 
investigating the sHRD for about 600 massive stars in the MW, and 
by \citet{mcevoy15}, based on the model atmosphere analysis of a 
sample of single and binary late-O stars and B Sgs in the LMC. 
These findings are somewhat different from similar results from 
\citet{martins14} who argued that an overshoot parameter between 
0.1 and 0.2 in models with \vinit\, between 250 and 300~\kms is 
needed to reproduce the sHRD for hot massive stars with solar 
metallicity.

 With the above outline in mind, we conclude that although 
the reason(s) for the mass discordance observed in the low mass 
O star regime (\Mevol$<$30\Msun) is not presently clear,  there 
is compelling evidence  that stellar masses derived for objects 
in this regime (from stellar luminosity  or surface equatorial 
gravity) are larger than the spectroscopic masses by typically 20 
to 25\% for the HRD and 30 to 40\% for the sHRD. This finding 
does not seem to depend on metallicity (solar or half solar) and is 
also independent of the model grids used: the current Geneva 
and the Bonn grids  both lead to qualitatively similar 
results.

Concerning the negative mass discordance established for our 
more massive stars (\Minit(HRD)$\ga$35\Msun) in comparison to 
the Geneva grids with \vinit=0.4\vcrit,  from our analysis it appears 
that better agreement between model predictions and observations 
might be obtained using models with \vinit\ lower than 0.4~\vcrit 
(see Sects.~\ref{mass_discrepancy} and \ref{rot_mix}).  But an 
alternative explanation in terms of problematic efficiency of 
rotational mixing also seems possible (see Sect.~\ref{Nabn_rot_discussion}). 

\subsection{Chemical surface enrichment and rotational 
mixing}\label{Nabn_discussion}

Studies of chemical surface abundances (apart from H and He)
in O stars compared with evolutionary model predictions 
including rotation have so far been performed by \citet{bouret13} 
in the SMC, by \citet{gonzalez12a, gonzalez12b}, and \citet{grin16} 
in the LMC, and by \citet{bouret12}, Martins et al. (2012a, 2012b, 
2015b, 2015c, 2016 and 2017), \citet{mahy15}, and \citet{cazorla17} 
in the MW. Because \citet{bouret13} used model grids from 
\citet{mm03}, while many of the targets analysed by \citet{bouret12} 
and Martins et al. (2012a, 2012b) are part of our sample, in the 
following we constrain our discussion and comparison to the works by 
Martins et al. (2015b, 2015c, 2016 and 2017),  \citet{mahy15}
\citet{gonzalez12a}, and \citet{grin16}.

\subsubsection{Chemical enrichment as a function of stellar 
mass and evolutionary stage}\label{rotmix_discussion}

Our analysis confirms previous findings from \citet{martins15b} and 
\citet{mahy15} that within each of the subgroups of LC~I, LC~II/III, 
and LC~IV/V objects, more massive targets show, on average, a higher 
degree of N enrichment than the less massive targets (Sect.~\ref{Nabn} 
and Fig.~\ref{fig6}), and   that at the same/similar \Minit\, and 
\vsini, more evolved objects appear to be stronger mixed than the 
less evolved objects (Sect.~\ref{obs_mod__Nabn} and Fig.~\ref{fig16}). 
These findings are qualitatively consistent with the idea that 
rotational mixing is more efficient at higher masses (for a population 
of stars of similar age) and later evolutionary stages.

Regarding the correspondence between predicted and observed N enrichment 
as a function of \Teff, we found that for \Minit(HRD)$<$~40~\Msun the 
current Geneva models with \vinit=0.4\vcrit  do a better job than the 
Bonn models with \vinit$\approx$300~\kms\ for the objects relatively close 
to the ZAMS (dwarfs and less evolved giants), and vice versa for the 
more evolved giants and supergiants.  For the objects in the more
massive regime, both model grids appear to be equally inadequate, 
giving rise to significantly larger (Geneva models) or smaller (Bonn 
models ) enrichments than observed (Sect.~\ref{obs_mod__Nabn} and 
Fig.~\ref{fig15}).

For O-type dwarfs, our results are qualitatively consistent with similar 
findings from \citet{martins15b} for O-type giants and supergiants, 
however, they disagree. Particularly, Martins et al. claimed a good 
correspondence between the current Geneva predictions and the N/C ratio 
derived for about 80\% of their targets, independent of LC, while 
reporting that Bonn predictions are either lower than (for the majority 
of the giants) or consistent with (for all supergiants and half of the 
dwarfs) those observed. As Martins et al. have considered the same models 
grids as we do in the present study, there are (at least) two  possibly 
explanations that could contribute to these discordant findings:
\begin{itemize}
\item[1)] 
Since we used the classical HRD, whereas \citet{martins15b} made use of 
the KD, this might help to (partly) understand the inconsistent results 
(see Sect.~\ref{obs_mod_logg}). At least regarding the current Geneva 
tracks, this possibility can be easily confirmed if one considers the 
KD for the sample stars  to fix the comparison tracks and then compares 
these to results from \citet{martins15b}. 
\item[2)]
Since N enrichment is predicted to depend on stellar mass and age and 
since \citet{martins15b} lacked giants and supergiants with 
\Minit(KD)$\ga$32~\Msun\ (Geneva tracks) or $\ga$40~\Msun\ (Bonn 
tracks) (We found this  by confronting the KDs shown in the right side 
of Fig.~\ref{fig12} and the left upper and lower panels of Fig.~7 of 
Martins et al.) this can explain why Martins et al. claimed a generally 
good correspondence for the majority of their stars independent of LC, 
while our results indicate problems in the corresponding mass and 
temperature regime.
\end{itemize}

Thus, it appears that the lack of consistency between our results 
and those reported by \citet{martins15b} can be, to a large extent, 
understood in terms of differences in the mass range probed by the 
two samples, as well as in terms of the different approaches used 
to fix \Minit\ of the track to which empirical stellar properties 
have to be compared (classical HRD versus KD).

\subsubsection{Chemical enrichment and rotation}\label{Nabn_rot_discussion}

{\bf Ekstr{\"o}m et al. (2012) grids. }\newline
From the analysis of 15 O7-O8 giants with \vsini\ between 50 and 300~\kms, 
\citet{martins17} have found that while the Geneva models  for 
25~\Msun$\ge$\Minit$\le$40~\Msun\ and \vinit=0.4\vcrit can consistently 
account for the \Teff, \loggc\ and log~(N/C)-ratio of their targets, 
these models are not able to correctly reproduced the rotational properties 
of many of the targets (9 out of 15, i.e. $\sim$60\%, as can be seen from 
the left panel of their Fig.~7). There are objects that rotate significantly 
higher or slower than expected for their observed N enrichment and \Minit\ 
as derived from the KD.

The results obtained by Martins et al. are in good qualitative agreement 
with similar findings from the present work which also indicate that for a 
large percentage of Galactic O stars, the [N]-\vsini pattern cannot be 
correctly reproduced by the current Geneva models with \vinit=0.4\vcrit. 

\citet{martins17} have shown that the [N]--\vsini\ problem for their sample 
stars can be successfully solved using Ekstr{\"o}m et al. models with \vinit\ 
higher and lower than 0.4\vcrit.  Such a solution seems possible  for many of 
our sample stars, especially those in Group 1 and Group 2 (see Sect.
~\ref{obs_mod_Nvsini}).  But we point out that to get conclusive results,  
the [N]--\vsini\ problem should be considered in parallel with  the mass 
discrepancy issue accounting for all masses appropriate for O stars,
from 9\Msun to about 80\Msun.

With this in mind an althernative scenario in terms of problematic 
efficiency of rational mixing seems also possible. Particularly, from a 
simple quantitative considerations it follows that if  the efficiency 
of rotationally induced mixing was  lower than that adopted in the Geneva 
models,  this would lead to less luminous tracks and lower mass-loss rates. 
This, in turn, would result in a smaller reduction of current masses and 
surface gravities and in faster rotation compared to the current predictions. 
Also, the He and N enrichment of the models would decrease. The effect would 
be weaker close to the ZAMS, increasing towards higher \Minit\ and later 
evolutionary stages, i.e. just into the direction we need to reconcile model 
predictions and observations, as illustrated in our Figs.~\ref{fig9}, 
\ref{fig10}, and \ref{fig11}. Additionally, as surface gravities (compared to 
stellar masses) are expected to be stronger affected by the process of mass 
loss, \Mevol(sHRD) would be lower than \Mevol(HRD), thus explaining the results 
shown in the upper panels of Fig.~\ref{fig6} and \ref{fig9}.  
\newline\newline
{\bf Brott et al (2011) grids.}\newline
\citet{mahy15} and \citet{cazorla17} have studied the [N]--\vsini\ diagram  
for O stars in the MW in comparison to Bonn predictions. The rotational 
properties of the corresponding samples are significantly different; the 
former is dominated by objects with \vsini$<$200\kms while the latter is 
focussed on the very fast rotators with \vsini\ between 200\kms and 400\kms. 
Both studies, however come to similar conclusions: there are a large number 
of objects whose [N]--\vsini\ properties cannot be accounted for by the 
models and there are fast rotators with a lack of nitrogen enrichment (our 
Group 1) and slow rotators that are highly N rich (our Group 3).

Based on their own \vsini\ and N abundance determinations for a sample of 
25 OB stars in the LMC, \citet{gonzalez12a} have found that about two-thirds 
of their O-type targets are N-rich slow rotators whose properties cannot 
be reproduced by models from  Brott et al.  Analysing a sample 67  O-type 
giants and supergiants, \citet{grin16} have come to similar conclusions. 
For at least 30-40 percent of their targets, the observed [N]--\vsini\ 
pattern cannot be understood in the current framework of rotational mixing, 
as implemented in the Bonn grids for LMC metallicity. 

The results derived in the present study are consistent with those outlined 
above, indicating that -- when all important parameters are accounted for -- 
the Bonn models seem inappropriate to represent the N enrichment in parallel 
with \vsini\ for a large percentage of the sample  (about 50~\%, mainly 
objects from Group~3; see Sect.~\ref{obs_mod_Nvsini}), unless the majority 
were fast rotators seen pole-on. Because of the large number of objects 
involved, this possibility does not seem likely.

To explain the presence of N-rich, slowly and moderately rotating O 
stars (Group 3 in our case) in comparison to the Bonn grids, 
\citet{grin16} have considered 
four scenarios: a rotational mixing efficiency that is larger 
than presently implemented, strong stellar winds leading to 
envelope stripping, an evolution in binary systems, and the 
presence of magnetic fields. \newline \newline
{\it Inadequate efficiency of rotational mixing}\newline
There are at least three observational findings that suggest that the 
efficiency of rotational mixing as implemented in 
the Bonn models might be lower than required by the majority of our more massive 
targets: i) the [N]--\vsini\ problem cannot be solved by varying \vinit, 
unless the majority of the  N-rich objects were seen (almost) pole-on 
(Sect.~\ref{obs_mod_Nvsini}); ii) the observed correlation between N enrichment 
and \vsini\ is steeper than proposed by the models for various \vinit\ 
(right panel of Fig.~\ref{fig16}), and  iii) for \Minit$\ge$40~\Msun\ and 
\vinit$\approx$300~\kms, the models at the end of the O-star phase are 
significantly less mixed than derived from observations (lower right panel 
of Fig.~\ref{fig15}).
\newline
\newline
{\it Mass loss and envelope stripping} \newline
Severe mass loss can reveal the deeper layers of a massive star, which are 
more enriched by nuclear-processed material than the outer layers, thereby 
leading to more chemical surface enrichment compared to a star of similar 
\Minit\ and \Teff, but with a weaker wind. Investigating the run of N 
enrichment as a function of (unclumped) mass-loss rate, we found that only 
for the most luminous supergiants with strongest winds (\logl$\ge$5.8 and 
log~\Mdot[\Msun/yr]$\ge$-5.4) envelope stripping might be an issue (leaving 
CD$-$47\,4551 apart, see Sect.~\ref{Nabn}).  For hot massive stars in the 
LMC, \citet{bestenlehner14} derived  a luminosity threshold of \logl$\ge$6.1. 
\newline 
\newline
{\it Binary evolution}\newline
We have seven binaries in our sample. Our analysis indicates that, regarding 
their main properties, these do not significantly depart from single stars 
with similar characteristics,  no matter which of the two grids were considered. 
Similar results have been reported by \citet{mahy15} in comparison to the 
Bonn grid. However, the identified binaries are likely pre-interaction, whereas 
a significant number of binary interaction products may appear as single 
stars in our sample \citep{deMink14}. The latter may show a significant nitrogen 
and helium enrichment and overluminosity \citep{langer12} which is not accounted 
for by single star evolutionary models. Therefore, an underprediction of the surface 
enrichment compared to the observed stars, as found for the Brott et al. 
models in Sect.\,7, does not necessary imply a problem in the single star 
models.
\newline 
\newline
{\it Effects of magnetic fields}\newline
The presence of relatively slowly rotating, N-enriched objects might be 
explained in terms of magnetically augmented mixing in initially faster 
rotators that have been spun down by angular momentum losses through a 
magnetically confined stellar wind (see e.g. \citealt{udDoula09, meynet11}). 
But this cannot be an effective channel because of the low number of O stars 
with relatively strong magnetic fields (\citealt{wade16}, \citealt{castro15} 
and references therein). Even more, after analysing O stars observed within 
the context of the MiMeS project, \citet{martins15b} failed to relate N 
enrichment with magnetic fields. Our analysis confirms these findings.

\subsubsection{Morphologically peculiar stars with very weak 
N lines.}

\citet{martins16} have recently investigated the properties of four 
Galactic supergiants classified as OC stars, and found that, while their 
\Teff\ and \logg\ are fully consistent with morphologically normal O 
supergiants, they show little, if any, nitrogen enrichment, and carbon 
surface abundances consistent with the initial composition.

Results from the present study confirm the conclusion  by Martins et. al.
about the OC star HD~152249,  and furthermore indicate that a similar 
pattern is present for the more luminous O-type giants and supergiants 
from our sample, classified as Ofc, Onfp, and Nwk objects. Since some 
of these are confirmed or suspected binaries, one might argue that our 
results might be biased by the presence of a companion. Such a possibility, 
however, seems unlikely because the evolution in a binary system is not 
expected to lead to lower N enrichment unless the two components have 
experienced a tidal interaction \citep{deMink13}. Thus we conclude that 
while it is not presently clear whether the OC, Ofc, Onfp, and Nwk stars 
might form a specific class of objects, results derived in the present 
study and in \citet{martins16} clearly indicate that the chemical surface 
enrichment observed in these objects cannot be accounted for in the 
context of rotational mixing in single massive stars unless they all possess 
strong magnetic fields (see \citealt{meynet11}).

\section{Summary}\label{summary}

We have analysed the main properties of 53 Galactic O stars, including  
seven binaries and seven objects with detected surface magnetic fields. 
For 30 of these stars, our own determinations of the main physical parameters 
were derived, using the FASTWIND code; for the remaining stars, literature 
data, obtained by means of the CMFGEN code, were used. The observed 
properties of the sample were compared to evolutionary model predictions for 
single massive stars of solar metallicity from \citet{ekstroem} and \citet{brott11}. 

The main outcome of our analysis can be summarised as follows.
\newline\newline
--  Spectroscopic masses in the low mass O star regime (\Mevol(HRD)$<$30..32~\Msun)  
tend to be smaller than the evolutionary masses (typically by $\sim$20 to 25\% for 
the case of the HRD and 30\% to 40\% for the sHRD), no matter if the Bonn models 
with \vinit$\sim$300~\kms or the current Geneva models with \vinit=0.4\vcrit\ are 
used as a reference. While some weaknesses in the treatment of turbulent pressure 
in the model atmosphere codes might contribute, the problem cannot be fully 
understood in terms of problematic parameters derived from observations. Inadequate 
model values of \vrot\ do not seem to be responsible either. \newline
-- Within each of the two considered grids, inconsistent evolutionary masses 
are derived when using either the stellar luminosity and from the surface equatorial 
gravity (Sect.~\ref{hrd_shrd} and Fig.~\ref{fig8}). The differences are generally 
small, that is smaller than 20\%, but due to their systematic character should 
be considered as important. This finding warns about potential, non-negligible 
differences between results derived using different diagrams.\newline 
-- Evolutionary masses given by Geneva tracks with \vinit=0.4\vcrit are generally 
lower than those inferred from the Bonn tracks for \vinit$\approx$300~\kms, 
that is by up to 50\% for the HRD and up to 70\% for the sHRD. The discordance 
strengthens towards larger masses and later evolutionary stages, leaving O dwarfs 
almost unaffected. While other reasons may contribute, differences in the 
rotational mixing of helium that result in models with substantially different 
luminosities and mass-loss rates, appear to be the main reason for this inconsistency (Sect.~\ref{Geneva_Bonn}).\newline
-- We confirm previous findings from \citet{martins12b} and \citet{mahy15} about 
a clear trend of stronger N enrichment in more massive and evolved O stars  
(Sect.~\ref{Nabn}). This finding is qualitatively consistent with the predictions 
of rotational mixing in single star evolutionary models. In fact, for \logl$>$ 5.3, 
the vast majority of our sample stars are nitrogen enriched, which is difficult 
to understand in any other way.\newline
-- O stars with peculiarly weak N lines, classified as Ofc, OC, or Onfp stars, 
show their own N enrichment trend with luminosity, which runs in parallel to 
the main trend, but at significantly lower N abundances. On the other hand, 
spectroscopic binaries and objects with magnetic fields do not depart (in this 
respect) from the rest of the sample stars.\newline  
-- The empirical N and He surface abundances of our more massive  stars are 
fairly well correlated. Also here, spectroscopic binaries and objects with 
magnetic fields do not depart. However, none of the considered model grids 
can match the observed trend correctly, producing either more (the Geneva 
grids) or less (the Bonn grids) He enriched objects for a given N abundance 
and \Minit. In line with the mismatch in the evolutionary masses as stated 
above, this result may imply that the mixing of helium in the Geneva models 
is too strong, while it might be too weak in the Bonn models.\newline
-- If different \vinit\ were considered, the current Geneva models appear 
well suited to study the properties of O stars with  \Minit(HRD)$\la$40~\Msun. 
These models, however, seem generally incapable of representing the properties 
of more massive and luminous objects with \Minit(HRD)$>$40\Msun. These models 
underpredict stellar masses, surface equatorial gravities, and projected equatorial 
rotational rates and overpredict surface N enrichment. While  \vinit\ that is 
too high or too low can be an issue, our results imply that the efficiency of 
rotational mixing implemented in the current geneva models for the corresponding 
mass regime might be problematic.\newline
-- The Bonn model grids can reasonably well reproduce the main parameters 
(e.g. \logl, \Teff, and \logg ) of O stars in the mass range from 15 to about 
80~\Msun. However, there is a large percentage of objects (about 50\% of our 
sample stars) whose [N]--\vsini\ pattern cannot be understood with the current 
efficiency of rotational mixing in these models, provided the inclination of 
rotational axes is randomly orientated. A problematic transport of angular 
momentum as adopted in these models might contribute as well.

Finally, we end this study with two important remarks. First, the main lesson 
we learned from our investigation  is that  to obtain conclusive results about 
the ability of present day evolutionary models to correctly reproduce the 
physics of (single) massive O stars, one needs to use $all$ parameters derived 
from observations as constraints for the models  in parallel. Second, the large 
percentage of nitrogen enriched massive O stars supports the idea that trace 
elements such as nitrogen can be effectively mixed throughout the star by 
rotationally induced turbulence.  The fraction of helium enriched massive O stars 
is smaller, such that we cannot exclude the suggestion that a binary history is 
responsible for the helium enrichment in these stars. The investigation of a larger 
sample of stars and a comparison to detailed grids of massive binary evolution models 
is desirable in order to further constrain the key uncertainties in the
theory of massive star evolution.

\begin{acknowledgements} 
We thanks the referee for her/his valuable comments and suggestions.
NM acknowledges financial support from the Bulgarian NSF (grant numbers 
DN08/1/13.12.2016 and DN 18/13/12.12.2017 ), 
and the hospitality of the Argelander-Institut f\"ur Astronomie at the 
Bonn University.\newline
This work has made use of data from the European Space Agency (ESA)
mission {\it Gaia} (\url{http://www.cosmos.esa.int/gaia}), processed by
the {\it Gaia} Data Processing and Analysis Consortium (DPAC,
\url{http://www.cosmos.esa.int/web/gaia/dpac/consortium}). Funding
for the DPAC has been provided by national institutions, in particular
the institutions participating in the {\it Gaia} Multilateral Agreement.
\end{acknowledgements}

\appendix
\section{Stellar masses} Table A.1. provides estimates of current
masses, derived for the stars listed in Table~1, applying the
classical and spectroscopic HR diagrams built using the Bonn
tracks with \vinit$\approx$300~\kms and the current Geneva tracks with
rotation.
\begin{table}[h]
\footnotesize
\caption[]{Current evolutionary masses for the sample stars, derived 
from the classical (M1) and the spectroscopic (M2) HRDs. 
Cluster and association members are listed in the upper part, field 
stars in the lower part. Errors shown in italics are 
not derived but adopted values. ''ul" means upper limit. 
For more information, see Sect.~\ref{hrd_shrd}.}
%\label{evol_mass_list}
\tabcolsep.6mm
\begin{tabular}{lllllll}
 \hline
 \hline
\multicolumn{1}{l}{Object}
&\multicolumn{1}{l}{ST}
&\multicolumn{2}{c}{Bonn tracks}
&\multicolumn{2}{c}{Geneve tracks}
&\multicolumn{1}{l}{Notes}\\
\hline
\multicolumn{1}{l}{}
&\multicolumn{1}{l}{}
&\multicolumn{1}{c}{M1}
&\multicolumn{1}{c}{M2}
&\multicolumn{1}{c}{M1}
&\multicolumn{1}{c}{M2}
&\multicolumn{1}{l}{}\\
\hline
HD~64568a  &O3 V  &68.2$^{+ 16.1}_{-\it 17.0}$ & 82$\it \pm{20.5}$ &68.0$^{+16.0}_{-12.4}$&74.8$^{+17.9}_{-10.9}$&\\ 
HD~46223   &O4 V  &47.0$^{+9.2}_{-7.3}$  &51.8$^{+17.4}_{-9.8}$  &47.3$^{+8.4}_{-7.3}$&46.5$^{+1.9}_{-5.0}$&\\
HD~93843a  &O5 III &58.8$^{\it +15.0}_{-11.6}$&52.8$^{\it +13.0}_{-11.9}$&43.6$^{+16.6}_{-3.9}$&38.9$^{+3.5}_{-6.5}$& SB1?\\
CD~$-$47\,4551 &O5~If  & 88 (ul)    &53.4$^{+\it 13.0}_{-12.3}$ &66.2$^{+17.5}_{-\it 16.6}$& 34.2$^{+8.6}_{-2.6}$&SB2\\  
%CD~$-$47\,4551a &60.7$^{+\it 15.0}_{-12.3}$&53.4$^{+\it 13.0}_{-12.3}$&44.3$^{+17.5}_{-3.9}$&34.2$^{+8.6}_{-2.6}$&\\
HD~93204a &O5.5~V&48.0$^{+10.7}_{-7.7}$ &41.4$^{+17.9}_{-6.3}$   &39.3$^{+3.7}_{-2.3}$&38.5$^{+0.7}_{-4.2}$&\\
CPD$-$59\,2600a &O6~V &36.7$^{+6.4}_{-4.8}$  &35.4$^{+6.1}_{-4.5}$   &36.1$^{+3.1}_{-4.5}$&35.0$^{+4.4}_{-4.2}$ &SB1\\
HD~91572a &O6.5~V &34.0$^{+5.8}_{-4.7}$  &34.6$^{+6.6}_{-5.0}$   &32.8$^{+4.1}_{-3.9}$&32.8$^{+4.0}_{-3.7}$&SB1\\
HD~63005a &O6.5~IV &38.9$^{+7.4}_{-5.7}$  &42.2$^{+10.4}_{-7.3}$  &34.2$^{+4.7}_{-2.8}$&34.2$^{+4.5}_{-2.6}$ &\\ 
HD~91824a &O7~V   &35.1$^{+6.0}_{-4.8}$  &36.1$^{+7.2}_{-5.0}$   &33.9$^{+3.8}_{-4.0}$&34.2$^{+3.5}_{-4.0}$&SB1\\
%CPD~$-$58~2620, O7~V &28.1$^{+3.2}_{-9.6}$  &32.9$^{+5.4}_{-4.4}$   &27.7$^{+3.5}_{-7.7}$&31.8$^{+4.2}_{-3.6}$&\\ 
CPD$-$58\,2620a &O7~V &28.1$^{+3.2}_{-9.6}$  &32.9$^{+5.4}_{-4.4}$   &27.7$^{+3.5}_{-7.7}$&31.8$^{+4.2}_{-3.6}$&\\ 
HD~93222 &O7~V   &33.8$^{+5.9}_{-4.8}$  &33.2$^{+5.9}_{-4.8}$     &32.2$^{+3.4}_{-3.7}$&31.5$^{+3.7}_{-3.6}$&\\
HD~94963a &O7~II &35.5$^{+7.0}_{-5.2}$&49.4$^{+\it 12.0}_{-11.0}$&30.8$^{+1.7}_{-3.6}$&32.4$^{+9.4}_{-2.2}$&SB2?\\
HD~94963b &     &41.0$^{+8.5}_{-6.7}$&49.5$^{+\it 12.0}_{-10.2}$&32.2$^{+9.2}_{-1.6}$ &32.4$^{+9.4}_{-2.2}$&\\
HD~94370a &O7   &32.4$^{+3.6}_{-6.7}$  &33.9$^{+10.6}_{-7.1}$    &29.9$^{+1.7}_{-4.0}$&29.8$^{+2.5}_{-4.1}$&SB2?\\
HD~94370b &    &36.2$^{+9.2}_{-3.4}$  &34.1$^{+10.7}_{-7.1}$    &31.0$^{+3.1}_{-2.1}$&29.9$^{+2.5}_{-4.1}$&\\
HD~151804 &O8 Iaf&56.9$^{+\it 13.0}_{-12.8}$ &52.0$^{+\it 13.0}_{-20.1}$&38.1$^{+12.0}_{-6.2}$&31.0$^{+16.4}_{-6.1}$&SB2\\
HD~92504  &O8.5~V&24.0$^{+3.7}_{-2.5}$  &26.0$^{+6.7}_{-3.9}$    &23.9$^{+3.2}_{-2.6}$&25.2$^{+4.1}_{-3.4}$&\\
HD~75211 &O8.5~II&37.3$^{+8.6}_{-6.6}$  &38.0$^{+16.3}_{-9.4}$   &31.8$^{+4.9}_{-3.3}$&29.1$^{+3.5}_{-3.6}$&SB1\\
HD~152249 &OC9 Iab &36.8$^{+7.9}_{-5.4}$&51.5$^{+\it 13.0}_{-11.3}$ &29.5$^{+3.5}_{-2.8}$ &30.9$^{+6.2}_{-2.7}$&\\
HD~46202  &O9.2~V&22.0$^{+2.7}_{-2.3}$  &21.9$^{+2.4}_{-2.2}$    &21.8$^{+2.8}_{-2.4}$&21.6$^{+2.6}_{-2.2}$&\\ 
CD~$-$44\,4865 &O9.7~III&26.9$^{+5.3}_{-4.0}$  &28.3$^{+6.7}_{-4.7}$    &24.5$^{+2.8}_{-2.8}$&24.6$^{+2.5}_{-2.7}$&\\
HD~152003 &O9.7~Iab &39.6$^{+8.6}_{-6.5}$&49.9$^{+\it 12.0}_{-11.3}$ &30.8$^{+7.7}_{-4.0}$ &30.8$^{+2.4}_{-3.9}$&\\
HD~75222 &O9.7 Iab  &35.4$^{+5.2}_{-7.8}$&47.4$^{+17.6}_{-11.2}$ &28.6$^{+1.7}_{-3.9}$ &29.5$^{+2.2}_{-2.9}$&SB?\\
HD~75222a & &39.8$^{+10.9}_{-4.4}$&47.4$^{+17.6}_{-11.2}$ &30.8$^{+2.8}_{-3.9}$ &29.5$^{+2.2}_{-2.9}$&\\
HD~78344  &O9.7~Iab &39.8$^{+8.7}_{-6.6}$& 46.7$^{+19.3}_{-11.2}$&30.8$^{+2.8}_{-3.9}$ &29.4$^{+2.4}_{-3.0}$&\\
\hline                                  
HD~169582 &O6 Iaf  & 82 (ul)    &58.6$^{+\it 12.0}_{-18.1}$&48.7$^{+\it 12.2}_{-5.0}$& 40.2$^{+21.4}_{-9.2}$&\\
CD~$-$43~4690 &O6.5 III&39.9$^{+7.5}_{-5.6}$& 45.3$^{+15.8}_{-8.7}$&32.1$^{+6.7}_{-2.3}$&32.4$^{+8.1}_{-1.9}$&\\
HD~97848 &O8 V  &26.5$^{+3.8}_{-2.9}$  &28.9$^{+5.2}_{-4.0}$    &25.8$^{+3.5}_{-2.8}$&28.0$^{+3.5}_{-3.4}$&\\
HD~69464 &O7 Ib &48.4$^{+11.9}_{-8.4}$ &49.7$^{+\it 20.3}_{-11.1}$ & 36.8$^{+7.6}_{-4.9}$&32.4$^{+9.5}_{-2.2}$&\\
HD~302505 &O8.5~III &33.2$^{+6.2}_{-5.3}$  &33.6$^{+7.5}_{-6.0}$    &28.7$^{+3.0}_{-3.0}$&27.8$^{+2.8}_{-2.8}$&\\
HD~148546 &O9 Iab &41.4$^{+9.1}_{-7.0}$&47.5$^{+ 17.9}_{-11.1}$ &31.1$^{+6.1}_{-3.0}$ &29.8$^{+3.2}_{-3.3}$&\\
HD~76968a &O9.2 Ib &36.2$^{+7.8}_{-5.4}$&43.1$^{+ 15.9}_{-8.6}$ &29.2$^{+3.4}_{-3.2}$ &30.2$^{+2.1}_{-4.2}$&SB1\\
HD~69106 &O9.7~II&22.0$^{+3.9}_{-3.0}$  &22.4$^{+7.6}_{-3.9}$    &21.8$^{+2.7}_{-2.6}$&21.8$^{+3.7}_{-3.2}$&\\
\hline
\end{tabular}
% \end{center}
\end{table}
\section{Individual stars with highly discrepant spectroscopic 
and evolutionary masses}

{\bf CP$-$47\,4551 (No. 1).} In Sect.~\ref{gen_comments} we 
noted that this star is a SB2, most likely with a colliding 
wind, and that it also possesses a magnetic field. Given 
these features, its appearance as a mass outlier is easy to 
understand.  

{\bf HD~169582 (No.~2)} is an object from the field, and a 
suspected SB (see Sect.~\ref{gen_comments}). Hence, an 
underestimated luminosity caused by unknown distance and/or 
an overestimated gravity due to the  presence of a companion 
might both contribute to explain the discordance between its 
\Mevol\ and \Mspec. For the Geneva tracks, the discrepancy 
is obviously stronger, but this  is to be expected given 
the comments provided in Sect.~\ref{Geneva_Bonn}.

{\bf HD~148937 (No.~3)} is a magnetic star that rotates relatively 
fast for its luminosity class and the Of?p %\LEt{Please check for typo.}
category it belongs to 
\citep{naze10}. Additionally, this star exhibits a massive,
nitrogen-rich circumstellar nebula, with an expansion age of only
3000\,yr \citep{lch87}, and its \Mspec\ (about 100~\Msun, see
\citealt{martins12a}) appears as too large.  To explain these peculiar
features, \citet{langer12} suggested that HD~148937 might be the 
product of a very recent stellar merger. Thus, a
former binarity could be responsible for this star to appear as a mass
outlier in Figs.~\ref{fig8} and ~\ref{fig11}. In this case then, its
overluminosity on the HRD compared to the sHRD might be explained as a
consequence of energy released inside the star during the merger
event, implying that the star is not yet in its thermal equilibrium
state \citep{glebbeek13}.

{\bf HD~64568 (No.~4).} The \logl\ and \loggc\ values derived by us 
are consistent with those proposed in the calibrations by Martins et. al. for an
early O3 dwarf, whereas its \Teff\ and \Rstar\ are significantly
higher and smaller, respectively. This could lead to the assumption 
that an overestimated \Teff\ might be responsible for HD~64568 to
display \Mevol$>$\Mspec. While such possibility cannot be 
excluded (see Sect.~\ref{mod_res}), we note that the
location of HD~64568 exactly on the ZAMS (see Fig.~\ref{fig6}) is
fully consistent with a very young age, as suggested by its morphological
Vz designation. 

{\bf HD~207198 (No.~5)} has been analysed in terms of fundamental 
stellar and wind properties by \citet{repo} and by \citet{martins15}. 
While the derived \Teff\, and \loggc\, agree perfectly, the 
\logl--estimates are significantly different: 5.05~dex in Martins 
et al. (which has been used here)
and 5.47~dex in Repolust et al. Thus, large uncertainties 
in this parameter might be resposible for this star to appear as 
mass outlier in Figs.~\ref{fig8} and \ref{fig11}.

{\bf CPD~-58\,2620 (No.~6).} As  a member of the Tr~14 cluster ,
which is known for its anomalous reddening law (see Sect.~\ref{absmag}),
this star might suffer from a highly uncertain luminosity. This 
possibility might be supported by GAIA measurements, which 
indicate a distance that is by  about a factor of four larger 
than the photometric distance adopted here. Further considerations, however, 
revealed that a luminosity larger than that derived and used 
by us would result in an even stronger mass discrepancy, whereas a 
lower luminosity would place the star to the left of the ZAMS. On the other hand, 
an underestimated surface gravity might help to solve the mass 
problem, which is a possibility that seems to be additionally supported by 
the fact that CPD~-58\,2620 appears overluminous in the sHRD, compared 
to the HRD, independent of the used model grid (see Fig.~\ref{fig8}).

\section{Rotational mixing}
\begin{figure*}
{\includegraphics[width=8.5cm,height=5.6cm]{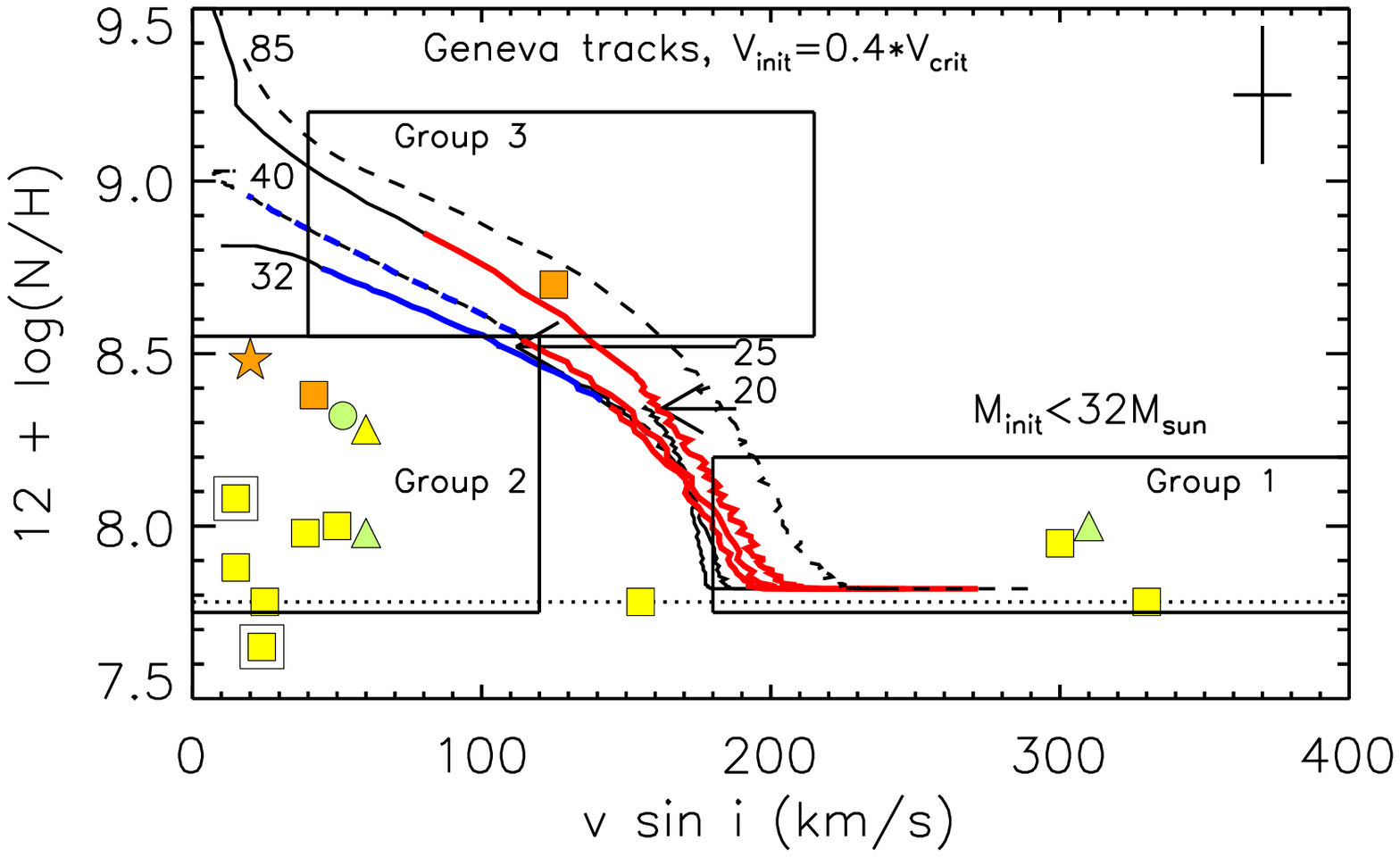}}
{\includegraphics[width=8.5cm,height=5.6cm]{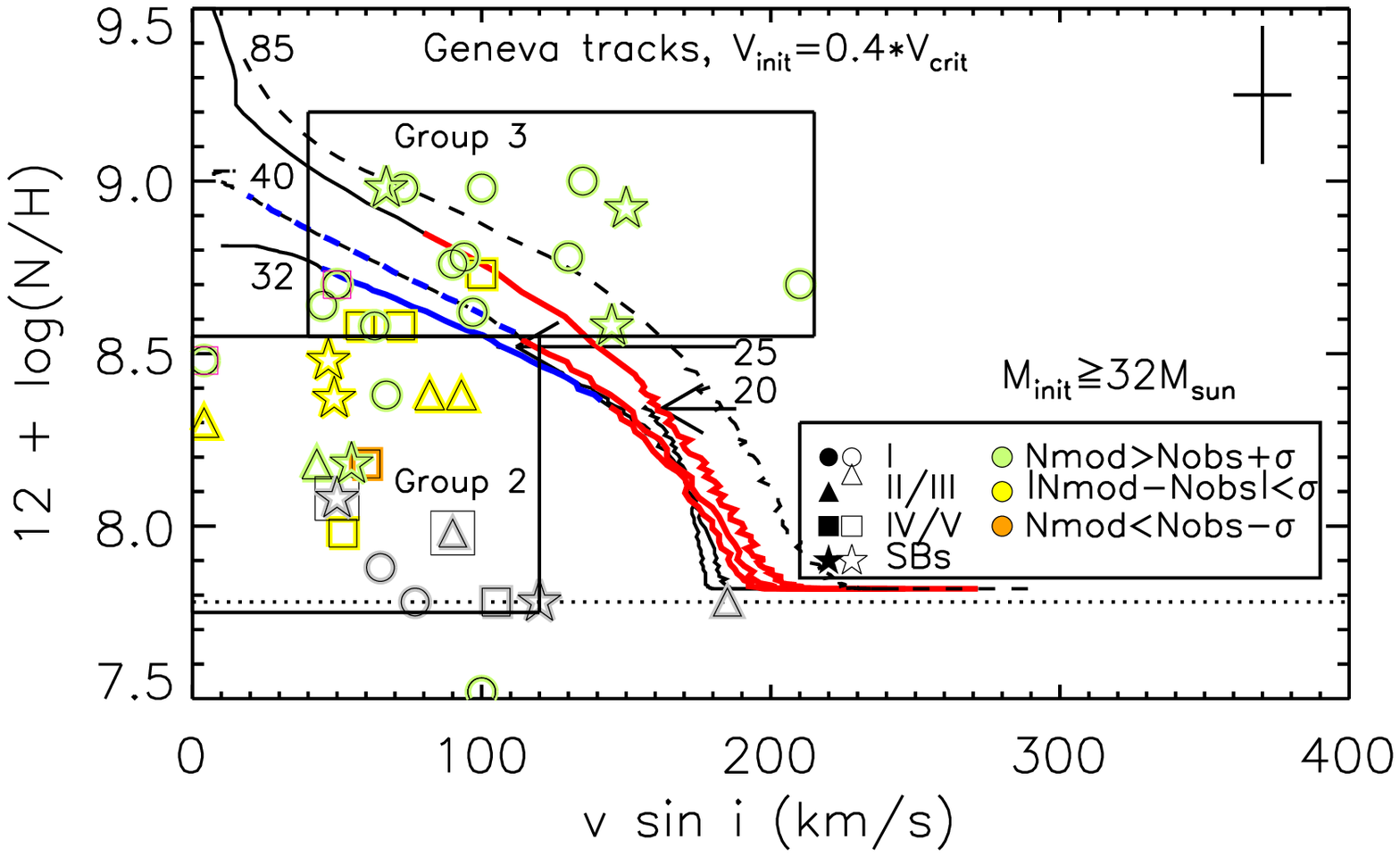}}

{\includegraphics[width=8.5cm,height=5.6cm]{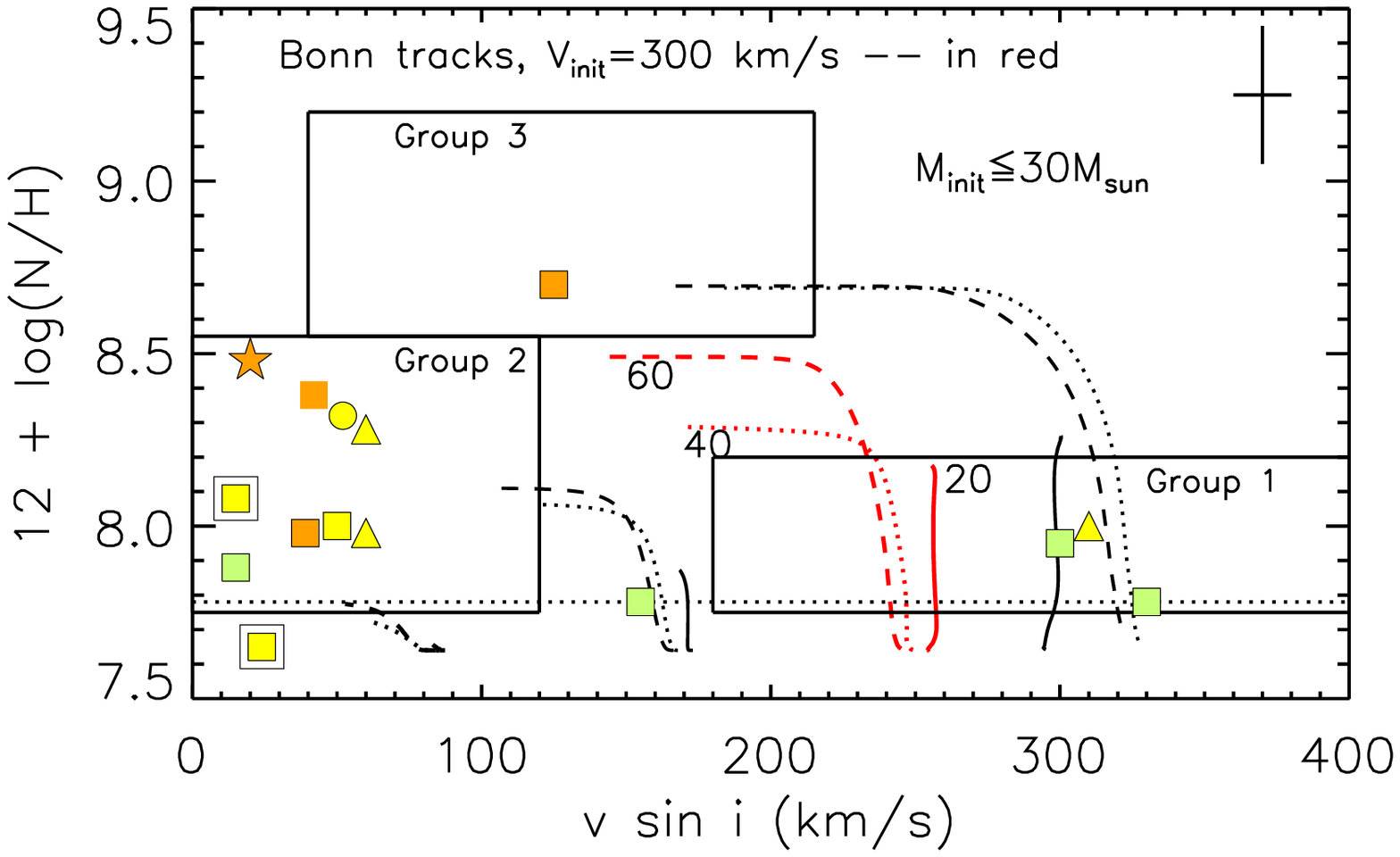}}
{\includegraphics[width=8.5cm,height=5.6cm]{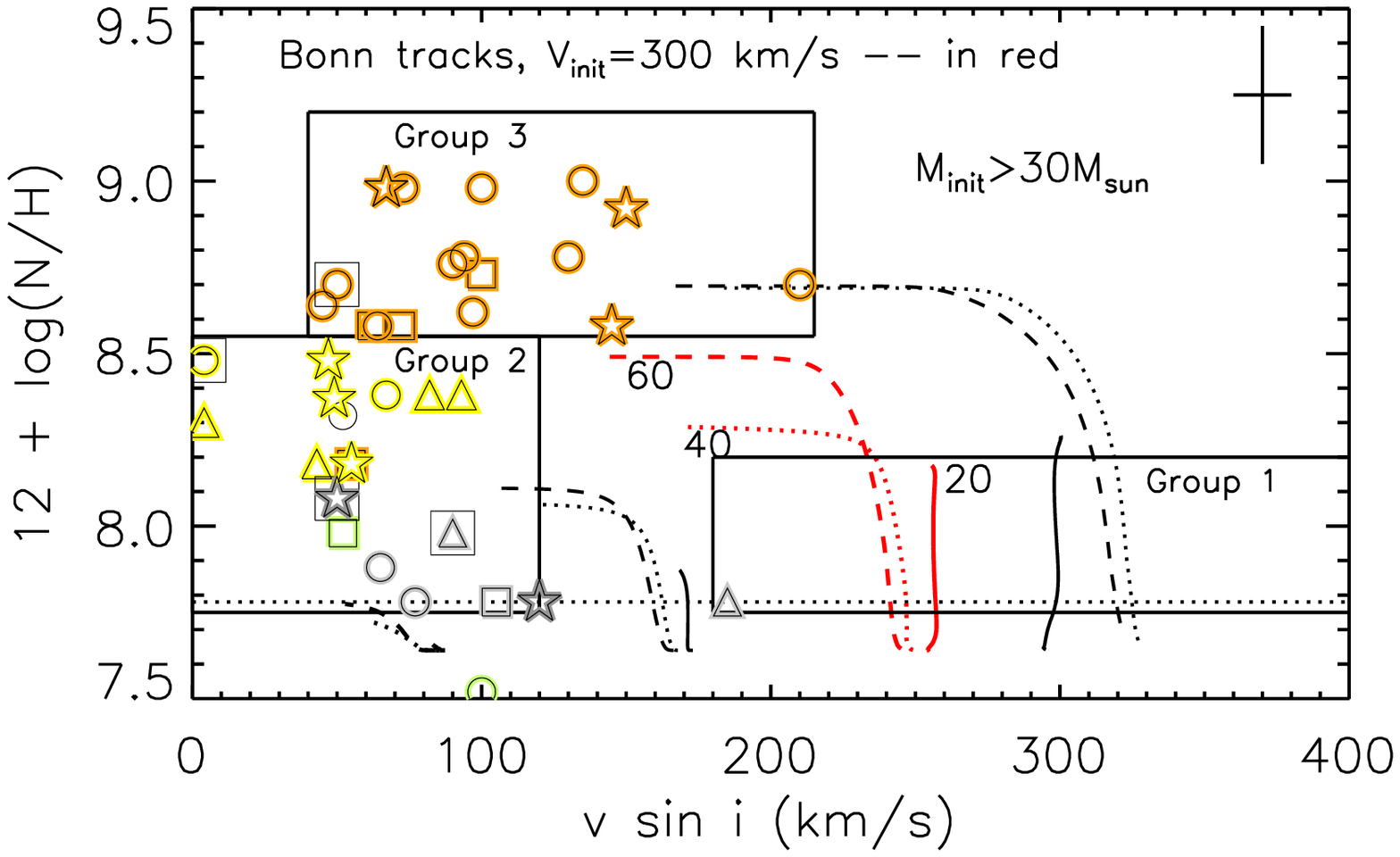}}
\caption{Surface nitrogen abundance for the sample stars divided 
into two mass bins (lower masses, left; higher masses, right) to 
enable better visibility, as a function of  projected rotational 
rate and plotted against the current Geneva and 
Bonn tracks with rotation (upper and lower panels, respectively).
Symbol coding distinguishes between objects of different LC;  
magnetic stars  are  additionally highlighted  by 
large squares. Colour coding 
is used to denote the morphologically peculiarly objects with very 
weak nitrogen lines (grey), and the stars whose predicted and 
observed [N] values differ by less (yellow) or more (green, 
orange) than 1$\sigma$. 
The observed \vsini\ accounts for the effects of macroturbulence; 
the model \vrot has been multiplied by $\pi$/4 to take the average 
projection effect into account (e.g. \vinit$\approx$300~\kms 
corresponds to \vsini$\approx$240 \kms). In the upper panels, the 
coloured parts of some tracks represent  the dwarf (\logg$>$3.75, 
red) and the giant (3.75$<$\logg$<$3.6, blue) phases, respectively. 
} 
\label{fig17}
\end{figure*}


\begin{thebibliography}{80}
\bibitem[Arenou et al.(2016)]{arenou16}
Arenou, F., Luri, X.,  Babusiaux, C., et al. (2016) {\it 
Gaia Data Release 1: Catalogue data validation: procedures, statistics 
and conclusions.} A\&A special Gaia volume. 
%
\bibitem[Asplund et al.(2005)]{asplund05}
Asplund, M., Grevesse, N., \& Sauval, A. J. 2005, in Cosmic Abundances as
Records of Stellar Evolution and Nucleosynthesis, ed. T. G. Barnes III, 
\& F. N. Bash, ASP Conf. Ser., 336, 25
%
\bibitem[Asplund et al.(2009)]{asplund09}
Asplund, M., Grevesse, N., Sauval, A.J. \& Scott, P. 2009, ARA\&A, 47, 481
%
\bibitem[Barbier-Brossat and Figon(2000)]{barbier}
Barbier-Brossat, M. \& Figon, P. 2000, A\&AS, 142, 217

\bibitem[Bestenlehner et al.(2014)]{bestenlehner14}
Bestenlehner, J.M., Gräfener, G, Vink, J.S., et al. 2014, A\&A, 570, 38
%
\bibitem[Bouret et al.(2012)]{bouret12}
Bouret, J.-C., Hillier, D.~J., Lanz, T., \& Fullerton, A.~W. 2012, A\&A, 544, A67

\bibitem[Bouret et al.(2013)]{bouret13}
Bouret, J.-C., Lanz, T., Martins, F., et al. 2013, A\&A, 555, A1

\bibitem[Brott et al.(2011)]{brott11}
Brott, I., de Mink, S. E., Cantiello, M., et al. 2011, A\&A, 530, 115   

\bibitem[Cantiello et al.(2009)]{cantiello09}
Cantiello, M., Langer, N., Brott, I., et~al. 2009, A\&A, 499, 279

\bibitem[Carraro \& Patat(2001)]{CP01}
Carraro, G. \& Patat, F. 2001, A\&A, 379, 136

\bibitem[Castro et al.(2014)]{castro14}
Castro, N., Fossati, L., Langer, N.,  et al. 2014, A\&A, 570, 13

\bibitem[Castro et al.(2015)]{castro15}
Castro, N., Fossati, L., Hubrig, S., et al. 2015, A\&A, 581, A81

\bibitem[Cazorla et al(2017)]{cazorla17}
Cazorla, C., Nazé, Y., Morel, Th. et al. 2017, A\&A, 604, 123
%
\bibitem[Chieffi \& Limongi(2013)]{chieffi13}
Chieffi, A. \& Limongi, M. 2013, ApJ, 764, 21
  
\bibitem[Crowther \& Evans(2009)]{ce09}
  Crowther, P. \& Evans, C. 2009, A\&A, 503, 85 

  
\bibitem[de Mink et al.(2009)]{deMink09}
de Mink, S. E., Cantiello, M., Langer, N., et al. 2009, A\&A, 497, 243

\bibitem[de Mink et al.(2013)]{deMink13}
de Mink, S., Langer, N., Izzard, R. G., et al. 2013, ApJ 764, 166
     
 \bibitem[de Mink et al.(2014)]{deMink14}
de Mink, S., Sana, H., Langer, N. Izzard, R. G. et al 2014, ApJ 782, 7
    
     
\bibitem[Ekstr{\"o}m et al.(2012)]{ekstroem}
Ekstr{\"o}m, S., Georgy, C., Eggenberger, P., et~al. 2012, A\&A, 537, 146   

\bibitem[Evans et al.(2006)]{evans06}
Evans, C. J., Lennon, D. J., Smartt, S. J., \& Trundle, C. 2006, A\&A, 456, 623

\bibitem[Evans et al.(2011)]{evans11}
Evans, C. J., Taylor, W. D., Henault-Brunet, V., et al. 2011, A\&A, 530, A108

\bibitem[GAIA collaboration et al.(2016)]{GAIA16}
Gaia Collaboration, Brown, A., Vallenari, A., Prusti, T.  et al. (2016) 
{\it Gaia Data Release 1: Summary of the astrometric, photometric, and survey 
properties.} A\&A special Gaia volume. 

\bibitem[Gies(1987)]{gies87}
Gies, D. 1987, ApJSS, 64, 545
%
\bibitem[Glebbeek et al.(2013)]{glebbeek13}
Glebbeek, E., Gaburov, E., Portegies Zwart, S., et al. 2013, MNRAS, 434, 349 

\bibitem[Grassitelli et al.(2015a)]{grassitelli15a}
Grassitelli, T., Fossati, L., Sim{\'o}n-Diáz, S., et al. 2015a, ApJ, 808, 31

\bibitem[Grevesse et al.(1996)]{grevesse96}
Grevesse, N., Noels, A., \& Sauval, A. J. 1996, in Astronomical Society of
the Pacific Conference Series, held in: Astrophysics Conference in College
Park; Maryland; 9-11 October 1995; San Francisco., Vol. 99, Proceedings of
the sixth annual October Astrophysics Conference, ed.: by S.S. Holt and G.
Sonneborn, 117

\bibitem[Grin et al.(2016)]{grin16}
Grin, N., Ram{\'{\i}}rez-Agudelo, O.H., de Koter, A., et al. 2016, A\&A, 600, 82

\bibitem[Herrero et al.(1992)]{herrero92} 
Herrero, A., Kudritzki, R.-P., Vilchez, J. M., et al.  1992, A\&A, 261, 209
%
\bibitem[Hillier \& Miller(1998)]{HM98}
Hillier, D.J. \& Miller, D.L. 1998, ApJ, 496, 407 
%
\bibitem[Hohle et al.(2010)]{hohle10}
Hohle, M.~M., Neuh{\"a}user, R., \& Schutz, B.~F. 2010, Astronomische 
Nachrichten 331, 349
%
\bibitem[Howarth et al.(1997)]{howarth97} 
Howarth, I.D., Siebert, K.W., Hussain, G.A.J., et al. 1997, MNRAS, 284, 265

\bibitem[Hubrig et al.(2011)]{hubrig11}
Hubrig, S., Sch{\"o}ller, M., Kharchenko, N.V., et al. 2011, A\&A, 258, 151

\bibitem[Humphreys(1978)]{hump78}
Humphreys, R. 1978, ApJS, 38, 309

\bibitem[Hunter et al.(2008)]{hunter08} 
         Hunter, I., Brott, I., Lennon, D., et al. 2008, ApJ, 676, L29       

\bibitem[Hunter et al.(2009)]{hunter09} 
         Hunter, I., Brott, I., Langer, N., et al. 2009, A\&A, 496, 841
       
\bibitem[Hur et al.(2012)]{HSB12}
Hur, H., Sung, H. \& Bessell, M. 2012,  AJ, 143, 41

\bibitem[Feinstein(1995)]{F95}
Feinstein, A. 1995, RMxAC, 2, 57

\bibitem[Kaltcheva \& Hilditch(2000)]{KH00}
Kaltcheva, N. T. \& Hilditch, R. W. 2000, MNRAS, 312, 753

\bibitem[Kaltcheva \& Scorcio(2010)]{KS10}
Kaltcheva, N. T. \& Scorcio, M. 2010, A\&A, 514, 59

\bibitem[Kaufer et al.(1999)]{kaufer99}  
Kaufer, A., Stahl, O., Tubbesing, S., et al. 1999, The Messenger 95, 8.

\bibitem[Keszthelyi et al.(2016)]{keszthelyi16}
Keszthelyi, Z., Puls, J. \& Wade, Gr. 2017, A\&A, 598, 4

\bibitem[Kippenhahn \& Weigert(1990)]{kw90}
Kippenhahn, R. and Weigert, A. 1990. {\it Stellar Structure and Evolution} 
Berlin: Springer

\bibitem[Kourniotis et al.(2015)]{kourniotis15}
Kourniotis, M., Bonanos, A. Z., Williams, S. J., et al. 2015, A\&A, 582, 42

\bibitem[Kudritzki(1980)]{kudri80} 
Kudritzki, R.-P. 1980, A\&A, 85, 174

\bibitem[Kudritzki(1992)]{kudri92} 
Kudritzki, R.-P. 1992, A\&A, 266, 395
 
\bibitem[Kudritzki \& Puls(2000)]{kudri00} 
Kudritzki, R.-P. \& Puls, J. 2000, ARA\&A, 38, 613
%
\bibitem[Kudritzki et al.(2006)]{KUP06}
Kudritzki, R.-P., Urbaneja, M. \& Puls, J. 2006, IAU 234
%
\bibitem[Langer(2012)]{langer12}
Langer, N. 2012, ARAA, 50, 107

\bibitem[Langer \& Kudritzki(2014)]{LK14}
Langer, N. \& Kudritzki, R.~P. 2014, A\&A, 564, A52

\bibitem[Leitherer \& Chavarria(1987)]{lch87}
Leitherer, C. \& Chavarria-K., C. 1987, A\&A, 175, 208

\bibitem[Maeder \& Meynet(2000)]{mm00}
Maeder, G. \% Meynet, G. 2000, A\&A, 361, 159

\bibitem[Maeder \& Meynet(2015)]{mm14}
 Maeder, A. \& Meynet, G. 2015 Proceedings IAU Symposium No. 307, 9
(eds. G. Meynet, C. Georgy, J.H. Groh \& Ph. Stee)
%

\bibitem[Maeder et al.(2009)]{m09}
Maeder, A., Meynet, G., Ekström, S., \& Georgy, C. 2009, Commun.
Asteroseismol., 158, 72
%
\bibitem[Maeder et al.(2014)]{maeder14}
Maeder, A., Przybilla, N., Nieva, M.F., et al. 2014, A\&A, 565, 39
%
\bibitem[Mahy et al(2015)]{mahy15}
Mahy, L., Rauw, G., De Becker, M. et al. 2015, A\&A, 577, 23

\bibitem[Ma\'iz-Apell\'aniz et al.(2004)]{maiz04}
Ma\'iz-Apell\'aniz, J., Walborn, N.~R., Galu{\'e},
H.~{\'A}., Wei, L.~H., 2004, ApJS, 151, 103

\bibitem[Marcolino et al.(2009)]{marcolino}
Marcolino, W., Bouret, J-C, Martins, F., et al. 2009, A\&A, 498, 837

\bibitem[Markova \& Puls(2015)]{MP15} 
Markova, N. \& Puls, J. 2015, IAUS 307, 117

\bibitem[Markova et al.(2004)]{markova04} 
Markova, N., Puls, J., Repolust, T., et al. 2004, A\&A, 413, 693

\bibitem[Markova et al.(2011)]{markova11} 
Markova, N. , Puls, J. , Scuderi, S., et al. 2011, A\&A,  530, 11 (Paper I)

\bibitem[Markova et al.(2014)]{markova14}
Markova, N., Puls, J., Simon-Di\'az, S., et al. 2014, 
A\&A, 562, 37  (Paper II)

\bibitem[Mart{\'i}nez-N{\'u}nez et al.(2017)]{martinez17}
Mart{\'i}nez-N{\'u}nez, S, Kretschmar, P., Bozzo, E., et al. 2017, SSRv, 13 

\bibitem[Martins(2015)]{martins15}
Martins, F. 2015, ASSL, 412, 9 

\bibitem[Martins \& Plez(2006)]{MP06}
Martins, F. \& Plez, B. 2006, A\&A, 457, 657

\bibitem[Martins \& Palacios(2014)]{martins14}
Martins, F. \& Palacios, A. 2014, A\&A, 560, 16

\bibitem[Martins et al.(2005a)]{martins05a}
Martins, F., Schaerer, D. \& Hillier, D.J. 2005, A\&A, 436, 1049 

\bibitem[Martins et al.(2005b)]{martins05b}
Martins, F., Schaerer, D. \& Hillier, D.J., et al. 2005, A\&A, 441, 735 

\bibitem[Martins et al.(2012a)]{martins12a}
Martins, M., Escolano, C., Wade, G. A., et al. 2012a, A\&A, 538, 29 

\bibitem[Martins et al.(2012b)]{martins12b}
Martins, M., Mahy, L.,  Hillier, D. J., \& Rauw, G. 2012b, A\&A, 538, 39

\bibitem[Martins et al.(2015a)]{martins15a} 
Martins, F., Marcolino, W., Hillier, D.J., et al. 2015a, A\&A, 574, 142 

\bibitem[Martins et al.(2015b)]{martins15b}
Martins, F., Hervé, A., Bouret, J.-C., et al. 2015b, A\&A, 575, A34 

\bibitem[Martins et al.(2015c)]{martins15c}
Martins, F., Sim{\'o}n-D{\'{\i}}az, S., Palacios, A., et al. 2015c, A\&A, 578, 109 

\bibitem[Martins et al.(2016)]{martins16}
Martins, F., Foschino, S., Bouret, J.-C., et al. 2016, A\&A, 588, 64

\bibitem[Martins et al.(2017)]{martins17}
Martins, F., Sim{\'o}n-D{\'{\i}}az, S., Barb{\'a}, R.H., et al. 2017, A\&A 599, 30 

\bibitem[Massey et al.(2012)]{massey12}
Massey, P., Morrell, N., Neugent, K.~F., et al. 2012, ApJ, 748, 96

\bibitem[Massey et al.(2013)]{massey13}
Massey, P., Neugent, K.~F., Hillier, D.~J., \& Puls, J. 2013, ApJ, 768, 6

\bibitem[Mathys(1988)]{M88} 
Mathys, G. 1988, A\&AS, 76, 427

\bibitem[McEvoy et al.(2015)]{mcevoy15}
McEvoy, C. M., Dufton, P.L., Evans, C., et al. 2015, A\&A, 575, 70 

\bibitem[Meynet \& Maeder(2003)]{mm03} 
Meynet, G. \& Maeder, A. 2003, A\&A, 404, 975

\bibitem[Meynet \& Maeder(2005)]{mm05}
Meynet, G. \& Maeder, A.  2005, A\&A, 440, 1041 

\bibitem[Meynet et al.(2011)]{meynet11}
Meynet, G., Eggenberger, P. and Maeder, A. 2011, A\&A, 525, L11

\bibitem[Mokiem et al.(2007)]{mokiem07}
Mokiem, M.~R., de Koter, A., Evans, C.~J., et~al. 2007, A\&A, 465, 1003

\bibitem[Morel et al.(2008)]{morel08}
Morel, T.,Hubrig, S. \& Briquet, M. 2008, A\&A, 481, 453

\bibitem[Morrell et al.(2014)]{morrell14}
Morrell, N., Massey, P., Neugent, K., et al. 2014, ApJ, 789, 139

\bibitem[Naz{\'e} et al.(2010)]{naze10} 
Naz{\'e}, Y., Ud-Doula, A., Spano, M., et al. 2010, A\&A, 520, 59

\bibitem[Naz{\'e} et al.(2016)]{naze16}
Naz{\'e}, Y., ud Doula, A. \& Zhekov, S. 2016, ApJ, 831, 138

\bibitem[Nieva \& Przybilla(2014)]{nieva14}
Nieva, M.-F. \& Przybilla, N. 2014, A\&A, 566, 7

\bibitem[Patriarchi et al.(2001)]{patriarchi01}
Patriarchi, P., Morbidelli, L., Perinotto, M., et al. 2001, A\&A, 372,644 

\bibitem[Penny(1996)]{penny96} 
Penny, L. R. 1996, ApJ, 463, 737

\bibitem[Penny et al.(2004)]{penny04} 
Penny, L. R., Spraguel, A.J., Seago, G \& Gies, D.R. 2004, ApJ, 463, 737

\bibitem[Prinja et al.(1990)]{prinja90} 
Prinja, R.K., Barlow, M.J. \& Howarth, I.D. 1990, ApJ, 361, 607

\bibitem[Puls et al.(1996)]{puls96}
Puls, J., Kudritzki, R.P., Herrero, A., et al. 1996, A\&A, 305, 171

\bibitem[Puls et al.(2005)]{puls05} 
Puls, J., Urbaneja, M.A., Venero, R., et al. 2005, A\&A, 435, 669   

\bibitem[Puls et al.(2006)]{puls06}  
Puls, J., Markova, N., Scuderi, S., et al. 2006, A\&A, 454, 625  

\bibitem[Puls et al.(2008)]{puls08}
Puls, J., Vink, J. \& Najarro, F. 2008, A\&ARv 16, 209

\bibitem[Puls et al.(2015)]{PSM15}  
Puls, J., Sundqvist, J. O. \& Markova, N. 2015, IAUS 307,25

\bibitem[Ram\'irez-Agudelo et al.(2013)]{R13}
Ram\'irez-Agudelo, O. H., Sim\'on-D\'iaz, S., Sana, H., et al. 2013, A\&A, 560, 29

\bibitem[Ram\'irez-Agudelo et al.(2017)]{R16}
Ram\'irez-Agudelo, O. H., Sana, H., de Koter, A., et al. 2017, A\&A, 600, 81
%
\bibitem[Reed(2000)]{reed00}
Reed, B.C. 2000, AJ, 119, 1855

\bibitem[Repolust et al.(2004)]{repo}
Repolust, T., Puls, J. \& Herrero, A. 2004, A\&A, 415, 349

\bibitem[Rivero-Gonz{\'a}lez et al.(2011)]{gonzalez11}
Rivero Gonz{\'a}lez, J. G., Puls, J. \& Najarro, F. 2011, A\&A, 536, 58

\bibitem[Rivero Gonz{\'a}lez et al.(2012a)]{gonzalez12a}
Rivero Gonz{\'a}lez, J.G., Puls, J., Najarro,  J.F. \& Brott, I. 2012a, A\&A, 537, 79

\bibitem[Rivero Gonz{\'a}lez et al.(2012b)]{gonzalez12b}
Rivero Gonz{\'a}lez, J. G., Puls, J., Massey, P. and Najarro, F., 2012b, A\&A, 543, 95

\bibitem[Sab{\'i}n-Sanjuli{\'a}n et al.(2014)]{carolina14}
 Sab{\'i}n-Sanjuli{\'a}n, C., Sim{\'o}n-D{\'{\i}}az, S., Herrero, A., et al. 2014, A\&A, 564, 39
 
 \bibitem[Sab{\'i}n-Sanjuli{\'a}n et al.(2017)]{carolina17}
 Sab{\'i}n-Sanjuli{\'a}n, C., Sim{\'o}n-D{\'{\i}}az, S., Herrero, A., et al. 2017, A\&A, 601, 79

\bibitem[Salpeter(1955)]{salpeter55}
 Salpeter, E.E. 1955, ApJ, 121, 161
 
\bibitem[Sana et al.(2014)]{sana14}
Sana, H., Le Bouquin, J.-B., Lacour, S., et al. 2014, ApJS, 2015, 15 

\bibitem[Sander et al.(2015)]{sander15}
Sander, A.,  Shenar, T., Hainich, R., et al. 2015, A\&A, 577, 13

\bibitem[Schneider et al.(2014)]{schneider14}
Schneider, F., Langer, N., de Koter, A., et al.  2014, A\&A, 570, A66

\bibitem[Schr\"oder et al.(2004)]{Schroeder04} 
         Schr\"oder, S.E., Kaper, L., Lamers, H.J.G.L.M., et al. 2004, A\&A, 428, 149        


\bibitem[Sim{\'o}n-D{\'{\i}}az \& Herrero(2014)]{SH14}
Sim{\'o}n-D{\'{\i}}az, S. \& Herrero, A. 2014, A\&A, 562, 135
  
\bibitem[Sim{\'o}n-D{\'{\i}}az et al(2014)]{S14}
Sim{\'o}n-D{\'{\i}}az, S., Herrero, A., Sab{\'i}n-Sanjuli{\'a}n, C., et al. 2014, A\&A, 570, 6


\bibitem[Smith \& Brooks(2008)]{SB08}  
Smith, N., Brooks, K. 2008,  Handbook of Star 
Forming Regions, Volume II: The Southern Sky ASP 
Monograph Publications, Vol. 5. Edited by Bo Reipurth, p.138    

\bibitem[Sota et al.(2008)]{sota08}
Sota, A., Ma\'iz-Apell\'aniz, J., Walborn, N \& Shida, R. 2008, RevMexAA, 33, 56

\bibitem[Sota et al.(2014)]{sota14}
Sota, A., Ma\'iz-Apell\'aniz, J., Morrell, N.I., et al. 2014, ApJS, 211, 10

\bibitem[Sundqvist et al.(2012)]{jon12}
Sundqvist, J. O., ud-Doula, A., Owocki, S.P., et al. 2012, MNRAS, 423,21

\bibitem[Tapia et al.(2003)]{tapia03}
Tapia, R., Roth, M., Vazquez, R., et al. 2003, MNRAS, 339, 44

\bibitem[Townsend et al.(2007)]{townsend07}  
Townsend, R., Owocki, S.P. \% Ud-Doula, A. 2007, MNRAS, 382, 139

\bibitem[Vazquez et al.(1996)]{vazquez96}
Vazquez, R. A., Baume, G., Feinstein, A. \& Prado, P. 1996, RMxAC, 4, 131

\bibitem[Vink et al.(2000)]{vink00}
Vink, J.~S., de Koter, A., \& Lamers, H.~J.~G.~L.~M. 2000, A\&A, 362, 295

\bibitem[Vink et al.(2001)]{vink01}
Vink, J.-S., de Koter, A. \& Lamers, N. 2001, A\&A, 369, 574

\bibitem[ud-Doula \& Owocki(2002)]{udDoula02}
ud-Doula, A., \& Owocki, S. P. 2002, ApJ, 576, 413

\bibitem[ud-Doula et al.(2009)]{udDoula09}
ud-Doula, A., Owocki, S. P., \& Townsend, R. H. D.,  2009, MNRAS, 392, 1022

\bibitem[Wade et al.(2016)]{wade16}
Wade, G. A., Neiner, C., Alecian, E., et al. 2016, MNRAS, 456, 2

\bibitem[Walborn(1973)]{W73}
Walborn, N. R. 1973, AJ, 78, 1067

\bibitem[Walborn(2009)]{walborn09}
Walborn, N. R. 2009, in Massive stars: from Pop III and GRBs to the Milky Way,
eds. M. Livio, \& E. Villaver, STScI Symp. Ser., 20, 167

\bibitem[Walborn et al.(2002)]{walborn02}
Walborn, N., Howarth, I., Lennon, D., et al. 2002, AJ, 123, 2754

\bibitem[Walborn et al.(2010)]{walborn10}
Walborn, N. R., Sota, A., Ma\'iz-Apell\'aniz, J., et al. 2010, ApJ, 711, L143

\bibitem[Wegner(1994)]{wegner94}
Wegner, W. 1994, MNRAS, 270, 229

\bibitem[Weidner \& Vink(2010)]{WV10}
Weidner, C. \& Vink, J.~S. 2010, A\&A, 524, A98

\bibitem[Wolff et al.(2006)]{wolff06}
Wolff, S. C., Strom, S. E., Drot, D., et al. 2006, AJ, 132, 749

\bibitem[von Zeipel(1924)]{vonZ}
von Zeipel H. 1924, MNRAS, 84, 665

\bibitem[de Zeeuw et al.(1999)]{Z99} 
         de Zeeuw, P.T., Hoogerwerf, R., de Bruijne, J.H.J., et al. 1999, AJ, 117, 354 

\end{thebibliography}
\end{document}